\documentclass[aps,prb,showpacs,floatfix,superscriptaddress,twocolumn]{revtex4-1}

\pdfoutput=1

\usepackage{natbib} 
\usepackage{amsmath} 
\usepackage{amssymb} 
\usepackage{graphicx}

\usepackage{fancyref}

\begin{document}

\title{Pathways to self-organization: crystallization via nucleation and growth}

\author{Swetlana Jungblut}
\affiliation{ Faculty of Physics, University of Vienna, Boltzmanngasse 5, 1090 Wien, Austria}
\author{Christoph Dellago}
\affiliation{ Faculty of Physics, University of Vienna, Boltzmanngasse 5, 1090 Wien, Austria}

\begin{abstract}
Crystallization, a prototypical self-organization process during which a disordered state spontaneously transforms into a crystal characterized by a regular arrangement of its building blocks, usually proceeds by nucleation and growth. In the initial stages of the transformation, a localized nucleus of the new phase forms in the old one due to a random fluctuation. Most of these nuclei disappear after a short time, but rarely a crystalline embryo may reach a critical size after which further growth becomes thermodynamically favorable and the entire system is converted into the new phase. In these lecture notes, we will discuss several theoretical concepts and computational methods to study crystallization. More specifically, we will address the rare event problem arising in the simulation of nucleation processes and explain how to calculate nucleation rates accurately. Particular attention is directed towards discussing statistical tools to analyze crystallization trajectories and identify the transition mechanism.  
\end{abstract}


\maketitle

\section{Introduction}
\label{sec:intro}

If liquid water is cooled below zero degrees Celsius at atmospheric pressure, it suddenly changes into solid ice with physical and chemical properties drastically differing from those of the liquid. On a microscopic level, during this freezing transition a fluid without long range order of its molecules turns into a solid, in which the molecules are arranged on a regular lattice. Due to thermal motion, the molecules fluctuate around their ideal lattice positions even in the solid, but on the average their positions are ordered over macroscopic distances. Similar freezing transitions are observed in all liquids (except helium) provided they are subject to suitable cooling procedures that prevent them from forming a glass, an amorphous solid stabilized by arrested dynamics rather than thermodynamics. Crystals can also form by deposition from the vapor or by precipitation from a supersaturated solution (this is how crystalline sea salt is produced by evaporation). While crystallization was first discovered for atomic and molecular liquids, much larger units such as proteins \cite{McPherson19911,tenwolde:1997}, colloidal particles \cite{gasser:2009} or even dusty plasmas, charged ``dust'' particles levitated in a plasma \cite{PhysRevLett.73.652,RevModPhys.81.1353}, also form crystals and many of these solidification events play an important role in a variety of natural and technological processes ranging from the formation of ice crystals in stratospheric clouds and bone mineralization to the phase decomposition of alloys and the manufacturing of pharmaceuticals. Indeed, crystallization can be viewed as a prototypical self-assembly process, during which building blocks of a few species spontaneously arrange in ordered structures. In this article, we will discuss the mechanism by which crystals of various substances form. In doing so, we will pay particular attention to the theoretical concepts of computational methods and statistical analysis techniques that can be used to investigate and understand self-assembly by crystallization. 

While at temperatures below zero degrees Celsius ice is the thermodynamically stable phase of water, it has been known for centuries that one can cool water down below the freezing point without the liquid actually freezing. (Such water is then said to be in a supercooled state.) In fact, in 1724 the German scientist Daniel Gabriel Fahrenheit found that boiled water in sealed containers remained liquid even when put outside in winter nights at temperatures as cold as -9~$^\circ$C \cite{Fahrenheit1724}. But, when he then dropped small ice crystals into the water, the liquid suddenly froze and a similarly rapid crystallization occurred upon shaking. Subsequently, Fahrenheit's experiments were reproduced and improved by many other scientists \cite{stoekel:2005} and currently the world record for the supercooling of water lies at -46~$^\circ$C as determined by probing water droplets cooled by evaporation with ultrafast x-rays \cite{Sellberg2014}. Deep supercooling as well as the dramatic speedup of crystallization by catalyzing particles was demonstrated for many other substances and it is now clear that all liquids can be supercooled regardless of the type of interaction acting between their basic constituents.

\begin{figure}[t]
\begin{center}
\includegraphics[width=75mm]{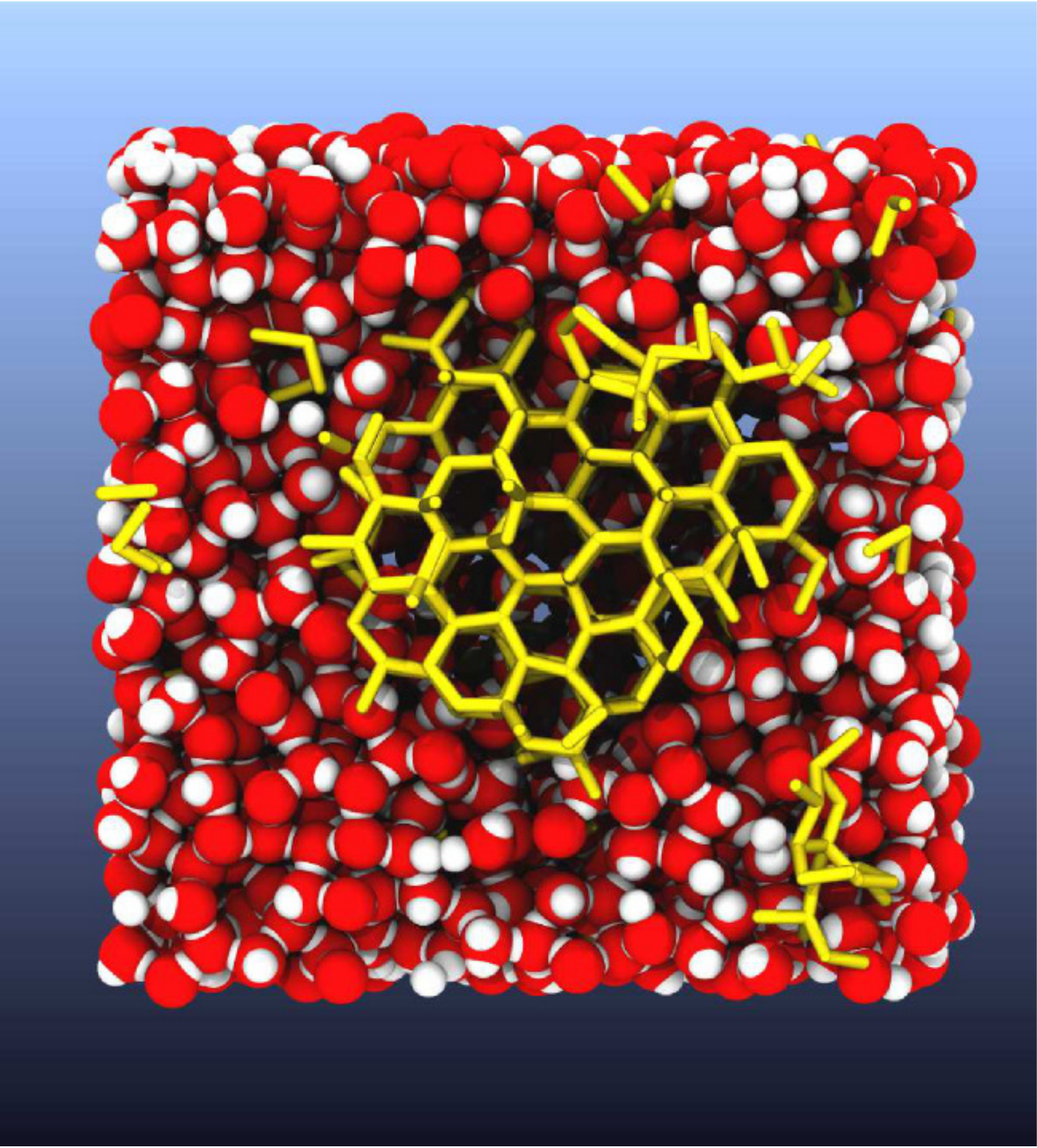}
\end{center}
\caption {\label{fig:ice} Ice nucleus forming in supercooled liquid water (snapshot from a simulation). Molecules belonging to the liquid are represented by the space-filling model. To highlight the regular hexagonal structure of the nucleus, the crystalline part is depicted by sticks corresponding to the respective hydrogen bonds. (Picture courtesy of Philipp Geiger.)}
\end{figure}

The fact that liquid water and other fluids can be kept almost indefinitely in a supercooled state indicates that crystallization is opposed by a free energy barrier that is large compared to typical thermal fluctuations. This barrier prevents the system from rapidly transforming into the thermodynamically stable state such that it can exist in a metastable state for long times. The origin of this barrier can be understood, at least qualitatively, in the framework of classical nucleation theory (CNT), a theory originally developed for droplet condensation from supersaturated vapors by Volmer and Weber \cite{Volmer:1926} and Becker and D{\"o}ring \cite{Becker:1935} and later generalized to the crystallization of supercooled liquids by Turnbull \cite{Turnbull:1949}. Classical nucleation theory, discussed in detail below, posits that the transformation occurs via the formation of a small, localized region of the new phase in the metastable old phase. An example of such a nucleus is given in fig.~\ref{fig:ice} showing an ice crystallite forming in supercooled water as observed in a computer simulation based on the TIP4P/ice model \cite{JCP_Geiger_2013}. The free energy of this nucleus has two main contributions: a negative bulk term, which arises from the formation of the stable phase and lowers the total free energy, and a positive surface term occurring due to the free energetic cost of creating an interface between the nucleus and the metastable phase in which it forms. Since the first term scales with the volume of the nucleus while the second one scales with its surface area, the free energy increases for small nucleus sizes, but for growing nuclei the volume term eventually dominates leading to a decrease of the free energy with nucleus size. The competition of these two terms creates a free energy barrier that prevents the supercooled liquid from immediate crystallization. Nevertheless, small regions of the stable phase randomly appear in the metastable phase due to thermal fluctuations, but usually disappear after a short time due to the increasing free energy. The macroscopic phase transition only takes place if, due to a rare random fluctuation, one of these regions grows beyond the so-called critical size. Once that happens, the crystallite has a strong tendency to grow further driven by the free energy, which now decreases with growing nucleus. This growth process then rapidly leads to the crystallization of the entire system. 

If a nucleation event takes place deep in the bulk of the metastable phase away from surfaces or impurities, one speaks of homogeneous nucleation. In contrast, during heterogeneous nucleation the nucleus forms near foreign objects that reduce the free energetic cost of the interface and, thus, reduce the height of the nucleation barrier, thereby facilitating the phase transition. Most nucleation processes occurring in technology and nature are heterogeneous, for instance, the condensation of water droplets in the atmosphere is catalyzed by nanoparticles. Similarly, crystallization usually starts near surfaces or impurities.

The scenario of (homogeneous or heterogeneous) nucleation followed by growth envisioned by CNT captures the essential mechanism of most first-order phase transitions. (An exception is the spinodal decomposition \cite{Cahn1958,Binder_RepProgPhys_1987}, in which the old phase is unstable against small fluctuations, which are exponentially amplified and lead to a rapid evolution towards the thermodynamically stable phase simultaneously over the entire sample volume.) On a more quantitative level, however, nucleation theory often yields nucleation rates (number of nucleation events per unit time and unit volume) that deviate from experimentally measured nucleation rates by orders of magnitude. The exact reason for such discrepancies may lie in the application of macroscopic concepts such as a well-defined surface and its excess free energy to a microscopic situation, but also in the neglect of certain degrees of freedom which may importantly influence the nucleation mechanism and its dynamics. For instance, the size of the nucleus of the new phase forming in the old one may not be the only relevant variable and also its shape and structure may matter. Due to the limited temporal and spatial resolution of experimental probes, it is usually difficult to infer detailed information about the nucleation mechanism from experiments, which usually detect only the macroscopic consequence of nucleation long after the occurrence of the crucial barrier crossing event. Hence, such experiments can only determine nucleation rates and their dependence on external conditions, but lack the time and space resolution to provide specific information on the nucleation mechanism. However, in the last decades, computer simulations \cite{Frenkel_Smit_book,Allen_Tildesley_book} have played an increasingly important role in the study of nucleation processes, providing a powerful way to obtain microscopic details of the formation of the new phase and to test models that go beyond the simplified assumptions of CNT. In fact, most of what we know today about the microscopic mechanism of nucleation has been learned from computer simulations. (It is also true that simulation studies of nucleation are usually done for systems that are much simpler than those studied in laboratory experiments.) 

Due to the free energy barrier opposing a rapid phase transition, nucleation processes are characterized by rare barrier crossing events. The resulting separation between the basic time scales of atomic motion and the long time scales of nucleation makes straightforward computer simulation methods such as molecular dynamics (MD) and Monte Carlo (MC) simulations impractical and more sophisticated approaches are required for studying phase transformations numerically. In this review article, we will discuss several computational methods that can be used to overcome the rare event problem of nucleation. These approaches include methods for the calculation of free energies and rate constants that require a priori knowledge of the reaction coordinate, but also more advanced path sampling procedures and statistical analysis tools, where such information is not needed in advance. Rather, these methods are designed to find reaction coordinates and identify the important degrees of freedom that characterize the crystallization mechanism. We will illustrate these computational methods with several examples taken from the recent literature. In doing so, we will focus on homogeneous nucleation processes and will not say much about heterogeneous nucleation occurring close to impurities or surfaces. Such nucleation sites can not only speed up the nucleation dramatically by lowering the free energy barrier but also lead to the formation of different crystalline structures that would not form during homogeneous nucleation. We would like to stress, however, that the computational methods considered here can be applied to a variety of processes dominated by events that occur rarely, but proceed rapidly when initiated. In particular, these methods can be applied to study both homogeneous and heterogeneous crystallization, but also other nucleation processes such as melting, condensation, cavitation, phase separation and structural phase transitions. 

While the rate determining step for crystallization is usually the nucleation of a supercritical crystallite, the growth following the nucleation event is equally important in determining the morphology of the resulting crystalline material. For instance, urea crystals of many shapes ranging from needles to regular tetrahedra can be grown from solutions by tuning the relative growth rates of different facets through variation of additive concentrations and supersaturation \cite{salvalaglio:2012,salvalaglio:2013}. Although such surface growth processes may also be dominated by activated events, we will not discuss them in this article but only point out that they can also be studied with the computer simulation methods presented here. Furthermore, we would like to mention that simple nucleation and growth is not the only mechanism for spontaneous self-assembly and a variety of other dynamical pathways to self-assembly exist. These range from non-classical nucleation, in which one or more crystalline or amorphous metastable intermediates form on the way to the final thermodynamically stable state \cite{baumgartner:2013,gebauer:2014,Jacobs19052015}, to far-from-equilibrium processes, in which kinetic traps or the active motion of the building blocks determine the shape and structure of the assemblies emerging at the end of the process \cite{Whitelam_Jack_2015}. Here, we do not consider such processes, but limit ourselves to the discussion of near-equilibrium self-assembly through crystallization over a single barrier.

\section{Classical nucleation theory}
\label{sec:cnt}

\begin{figure}
\begin{center}
\includegraphics[clip=,width=0.95\columnwidth]{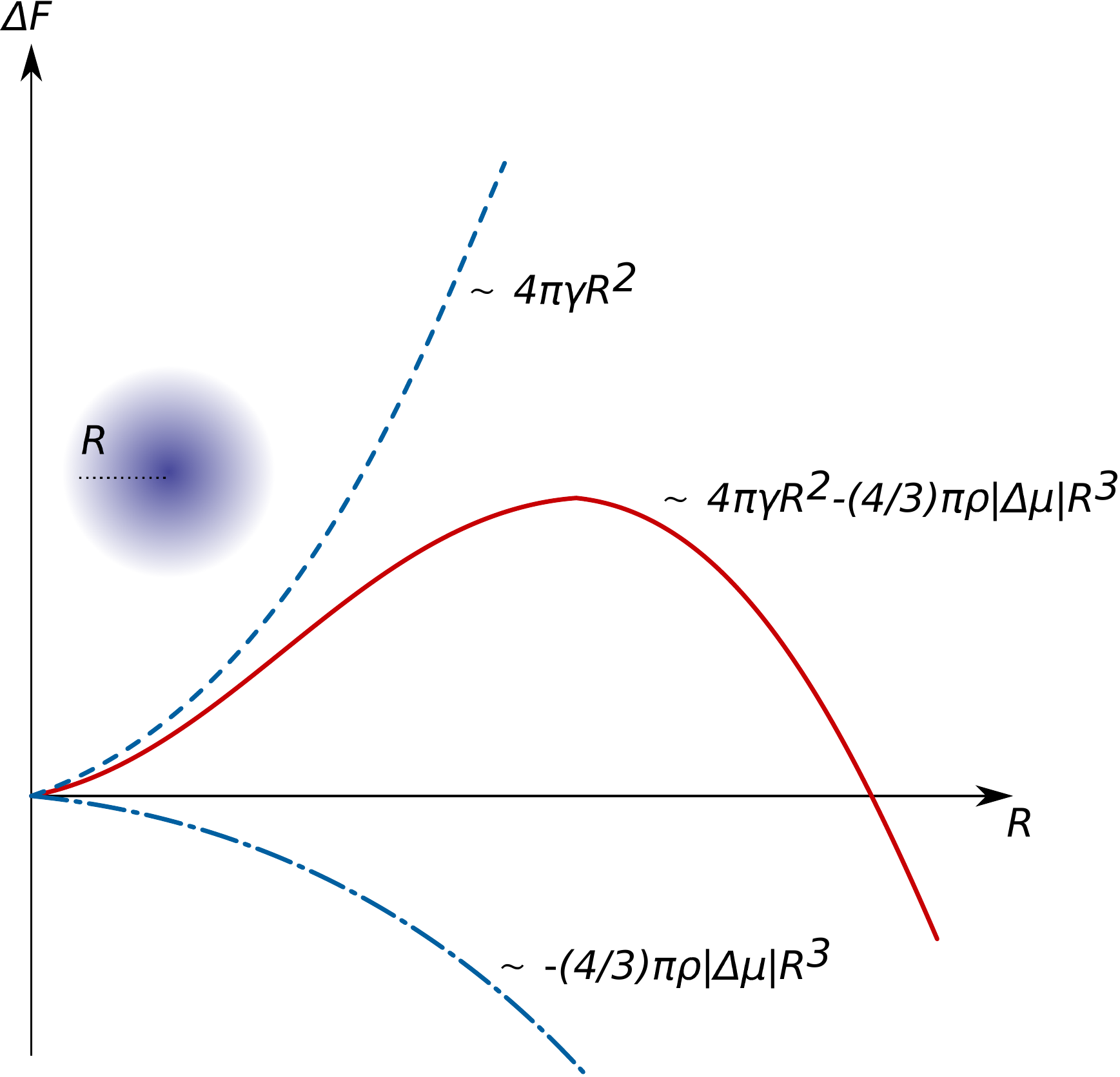}     
\caption{\label{fig:freeenergyCNT} Sketch of the nucleation free energy $\Delta F$ as a function of the nucleus radius $R$ as envisaged in CNT (cf. eq.~(\ref{eq:cnt1})). The total free energy (solid) line is the sum of two separate terms of opposite sign (dashed and dot-dashed lines). The positive term is the cost of forming the interface between the two phases and is proportional to the surface area of this interface. For a spherical nucleus the surface area is simply given by $4\pi R^2$. Multiplication with the surface tension $\gamma$ then yields the free energy of the interface. The negative term, on the other hand, is associated with the lower free energy of the stable phase with respect to the metastable phase. This term is negative and is obtained by multiplying the number of particles in the nucleus, $(4\pi/3)R^3\rho$, (remember that $\rho$ is the number density of the stable phase) with the difference in chemical potential $\Delta \mu$ between the two phases. (To make clear that this volume term is negative we have written it as $-(4\pi/3)R^3\rho|\Delta \mu|$, where $|\Delta \mu |$ is the absolute value of the difference in chemical potential.) Due to the different scaling of the two terms with $R$, the total free energy is dominated by the surface term for small values of $R$, leading to a free energy that is initially uphill. For growing $R$, however, the negative volume term gains importance relative to the surface term and eventually outweighs it, leading to a free energy decreasing with size. In between, the free energy reaches a maximum at the so-called critical radius $R^*$ producing a barrier of height $\Delta F(R^*)$. }
\end{center}
\end{figure}

\subsection{General picture}
\label{subsec:cntGeneral}

Classical nucleation theory \cite{debenedetti, kashchiev:2000}, based on the Gibbs approach \cite{gibbs:1875,gibbs:1877} 
(formulated in $19^{\rm th}$ century), is a combination a series of works by Volmer and Weber \cite{Volmer:1926}, 
Farkas \cite{farkas:1927}, Kaischew and Stranski \cite{stranski:1934, kaischew:1934,kaischew:1934a}, 
Becker and D{\"o}ring \cite{Becker:1935}, Frenkel \cite{frenkel:1939,frenkel:1939a}, and Zeldovich \cite{zeldovich:1942}, which appeared in the first half of the last century.  The central ideas of CNT and its application to various first-order phase transitions ranging from condensation and cavitation to phase decomposition and crystallization had been described in detail in several review articles  \cite{Oxtoby:1998,Binder_RepProgPhys_1987,sear:2007,sear:2012} and books \cite{kelton:2010,debenedetti,kashchiev:2000}.

As mentioned above, CNT views the nucleation process as resulting from an interplay between two terms in the free energy: 
 \begin{equation}
 \label{eq:cnt}
 \Delta F  =  \Delta F_V + \Delta F_I,
\end{equation}
where $\Delta F$ is the total free energy change associated 
with the creation of the nucleus of the new phase. $\Delta F_V$ is the volume-related 
gain in the free energy due to the difference in the chemical potentials of the liquid and the crystal and 
$\Delta F_I$ is the cost of the interface created between them. 
(Here, we refer to the thermodynamic free energy $F$, which, depending on the 
considered ensemble, stands for the Gibbs free energy, Helmholtz free energy, 
or the grand potential.)   
CNT attributes the presence of the free energy barrier stabilizing the 
metastable state to the competition between the two terms appearing in 
eq.~(\ref{eq:cnt}), the first of which drives the system towards the thermodynamically stable state, 
while the second opposes the formation of the new phase. The interplay of the 
free energy contributions allows to illustrate the concept of the critical 
clusters and to connect the reaction rate to the height of the free energy 
barrier. For nuclei smaller than a certain size, the cost for the interface 
creation outweighs the free energy gain due to the formation of the new 
phase and the system is thermodynamically driven back to the initial metastable 
state. When, however, thermal fluctuations manage to form a nucleus of a size larger than critical, 
the thermodynamic force acts in the opposite direction and the new phase 
spontaneously grows to macroscopic dimensions. 

Standard CNT postulates that the transition proceeds through 
the formation of a spherical nucleus of the new phase. If the pressure is kept constant, 
eq.~(\ref{eq:cnt}) can be written as a function of only one variable, $R$, which is the radius of the 
nucleus (see also the sketch in fig.~\ref{fig:freeenergyCNT}):   
 \begin{equation}
 \label{eq:cnt1}
 \Delta F (R) =  -\frac{4\pi R^3}{3}  \rho |\Delta \mu| + 4\pi R^2 \gamma.
\end{equation}
The difference in the chemical potential between the bulk phases, $\Delta \mu$, 
the number density of the nucleating phase, $\rho$, and the surface free energy 
density, $\gamma$, are assumed to be constant. The first term on the right hand side of the above equation is thus nothing else than the number of particles in the nucleus, given as product of volume and density, times the difference in chemical potential (remember that for a one component system the chemical potential is also the free energy per particle). The second term is the surface area of the nucleus, which is assumed to be spherical, times the surface tension $\gamma$. By taking the derivative of $\Delta F(R)$ with respect to the radius $R$ and setting the derivative to zero in order to find the free energy maximum one finds the critical radius of the nucleus,
 \begin{equation}
 \label{eq:cnt2}
 R^* =\frac{2\gamma}{\rho |\Delta \mu|}.
\end{equation}
Classical nucleation theory also yields the nucleation rate $J$ ({\it i.e.}, the average number of nucleation events per unit time and unit volume),
 \begin{equation}
 \label{eq:cnt3}
 J = K \exp\left[-\beta \Delta F(R^*)\right],
\end{equation}
which depends exponentially on the barrier height $\Delta F(R^*)$. Here, $\beta=1/k_BT$, $k_B$ is the Boltzmann constant, $T$ is the temperature, and $K$ is the so-called kinetic prefactor. In the following, we will derive this formula for the nucleation rate and provide an explicit expression for the kinetic prefactor in the framework of classical nucleation theory.

\subsection{Kramers problem and mean first-passage times}
\label{subsec:cntKramers}

\begin{figure}[tb]
\begin{center}
\includegraphics[clip=,width=0.95\columnwidth]{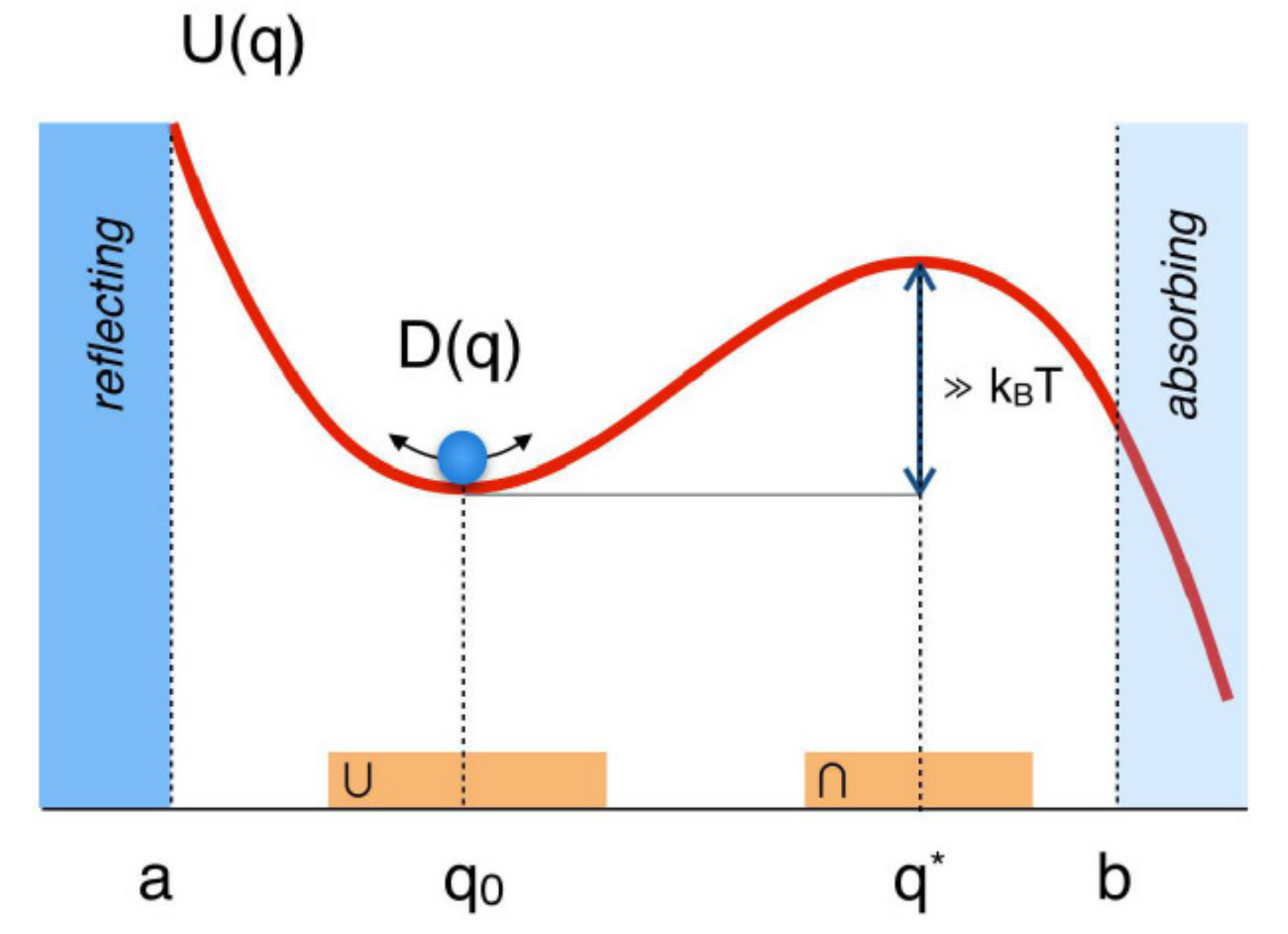}     
\caption{\label{fig:Kramers} The Kramers problem: Escape of a Brownian particle from a deep potential energy well \protect\cite{kramers:1940,haenggi:1990,zwanzig:2001}. The solid line denotes the potential energy landscape $U(q)$ on which the particle, shown by a sphere, evolves diffusively with a diffusion coefficient $D(q)$ along the coordinate $q$. To compute the escape rate from the well a reflective boundary is placed at $a$ and an absorbing boundary at $b$. The horizontal bars on the $q$-axis indicate the regions over which the integrals appearing in the expression for the mean first-passage time are carried out. The barrier height, $\Delta U=U(q^*)-U(q_0)$, is supposed to be much larger than the thermal energy $k_{\rm B}T$.}
\end{center}
\end{figure}

Our starting point is the so-called Kramers problem \cite{kramers:1940}, in which one considers the escape of a one-dimensional particle from a deep potential energy well, as depicted schematically in fig.~\ref{fig:Kramers}. The particle, whose position is specified by the variable $q$, is supposed to move diffusively on the potential energy surface $U(q)$ under the influence of the force $-\partial U / \partial q$ with a position dependent diffusion coefficient $D(q)$. Accordingly, the motion of the particle is described by the overdamped Langevin equation \cite{zwanzig:2001}, 
\begin{equation}
\dot q = -D(q) \frac{\partial \beta U(q)}{\partial q} +\sqrt{2D(q)} \xi(t),
\end{equation}
where $\xi(t)$ is Gaussian white noise mimicking the stochastic effect of a solvent that is thought to surround the particle, but is not represented explicitly. The overdamped Langevin equation is obtained from the full Langevin equation in the limit of large friction. The Kramers problem now consists in calculating the rate at which a particle initially located in the well escapes over the barrier. If the barrier is high compared to the thermal energy $k_{\rm B}T$, the particle will typically fluctuate in the well for a long time before a rare fluctuation drives it over the barrier, leading to a small escape rate. 

Progress can be made by considering the probability density $P(q, t)$ which describes the probability to find the particle at position $q$ at time $t$. It can be shown \cite{zwanzig:2001} that, if the particle position evolves according to the above Langevin equation, the time evolution of the probability density obeys the Smoluchowski equation, a special case of the Fokker-Planck equation valid in the overdamped regime, 
\begin{equation}
\frac{\partial P(q, t)}{\partial t}=\frac{\partial }{\partial q}\left[D(q) \frac{\partial \beta U(q)}{\partial q}P(q, t)+D(q)\frac{\partial P(q, t)}{\partial q}   \right].
\label{equ:Smoluchowski}
\end{equation}
Based on this equation, one can then obtain an expression for the mean first-passage time (MFPT), which is defined as follows. Imagine that the particle is initially placed at a certain position $q_0$ and that there is a reflective boundary located at the position $a$ as well as an absorbing boundary located at $b$. The positions of these boundaries are selected such that $a$ is left of the potential energy well and $b$ is on the other side of the well, beyond the free energy barrier. The initial position $q_0$ is somewhere between $a$ and $b$, not necessarily at the bottom of the well but also not very close to the top of the barrier. The particle let loose at $q_0$ will move under the influence of the random forces $\xi(t)$ and when it happens to hit the wall at $a$ it is simply reflected. When it eventually reaches the wall located at $b$, however, it is absorbed. The time between the start of the particle at $q_0$ and its absorption is called the first-passage time, because it is the time when the particle first reaches $b$ (since we imagine that the particle is absorbed, there are no further passages at $b$). If we now imagine that we repeat this experiment many times, due to the noise the first-passage time will be different even if the particle starts at the same position $q_0$ each time. Therefore, it makes sense to consider a distribution of first-passage times, the average of which is the mean first-passage time, $\tau(q_0)$. Using the Smoluchowski equation (\ref{equ:Smoluchowski}), one can show that $\tau$ for our particle is given by \cite{zwanzig:2001}
\begin{equation}
\tau(q_0)=\int_{q_0}^b dy\,\frac{e^{\beta U(y)}}{D(y)}\int_a^y dz\, e^{-\beta U(z)},
\label{equ:mfpt}
\end{equation}
which in general depends on $q_0$. For a detailed derivation of this equation we refer to the article by Pontryagin, Andronov and Vitt \cite{pontryagin:1933}, which translation is reproduced in the book of Zwanzig \cite{zwanzig:2001}. Note that the above expression for the MFPT is valid for any potential energy surface $U(q)$ and any placement of the reflecting and absorbing barriers and is not restricted to the well escape problem considered here. 

For the Kramers problem the expression for the mean first-passage time of eq.~(\ref{equ:mfpt}) can be simplified significantly. Let us ask ourselves for which values of $y$ the integrand of the outer integral on the right hand side of eq.~(\ref{equ:mfpt}) has the largest contribution. To answer this question, first assume that the diffusion coefficient $D(q)$ does not vary strongly between $a$ and $b$, thus, the change in $\exp[\beta U(y)]/D(y)$ is mainly determined by the variation of the exponential part, which is largest in the region where the potential energy $U(y)$ is large, {\it i.e.}, left and right of the well. For values of $y$ far on the left side of the well, however, the inner integral $\int_a^y dz\, \exp[-\beta U(z)]$ has not picked up any significant contributions yet, because $\exp[-\beta U(z)]$ has large values only near the bottom of the well. So the only region where the integrand of the outer integral has significant values is near the top of the barrier. Hence, for values of $q_0$ in the potential energy well, the integration from $q_0$ to $b$ can be replaced by an integration over the barrier region, denoted by the symbol $\cap$ in fig.~\ref{fig:Kramers}. Also the integration range for the second integral can be simplified similarly. For values of the upper integral limit in the barrier region (these are the only $y$-values we need to consider), the integral over $z$ already encompasses all significant contributions of the integral that come only from the well region, where $\exp[-\beta U(z)]$ has the largest values. Hence, it is sufficient to carry out the integration over $z$ in the well region only, as indicated by the symbol $\cup$ in fig.~\ref{fig:Kramers}. As a result, we can write the MFPT as a product of two independent integrals, one over the barrier region and one over the well region,
\begin{equation}
\tau=\int_\cap dy\,\frac{e^{\beta U(y)}}{D(y)}\int_\cup dz\, e^{-\beta U(z)}.
\label{equ:mfpt2}
\end{equation}
Note that the integration range of the second integral does not depend on $y$ any more. Interestingly, the mean first-passage time obtained here does not depend on the initial point $q_0$, implying that all the points in the well have on average an identical escape time. This is indeed to be expected if equilibration within the well occurs on time scales that are much shorter than the time in which a particle remains trapped by the well. This is the case if the barrier is sufficiently high compared to $k_{\rm B}T$ and the diffusion coefficient $D(q)$ in the well is not significantly lower than near the barrier top. If this second condition is not met, {\it i.e.}, if the diffusion coefficient in the well is very small, the argument used above to justify the separation of the two integrals breaks down and the MFPT cannot be written in the simple form of eq.~(\ref{equ:mfpt2}).

Under certain assumptions, eq.~(\ref{equ:mfpt2}) can be simplified even further. Namely, if the shape of the barrier is close to parabolic near its top ({\it i.e.}, in the region within a few $k_{\rm B}T$ from the top), the potential energy $U(q)$ can be expanded around $q^*$ and truncated after the quadratic term,
\begin{equation}
U(q)\approx U(q^*)- \frac{1}{2}\omega^2(q-q^*)^2,
\end{equation}
where the linear term vanishes because the derivative of $U(q)$ with respect to $q$ is zero at $q^*$. The constant $\omega^2$ is the absolute value of the curvature of $U(q)$ at the barrier top, $\omega^2=|U{''}(q^*)|$. Further assuming that the diffusion coefficient in the barrier region is essentially constant, the first integral in eq.~(\ref{equ:mfpt2}) turns into a Gaussian integral that can be solved analytically, 
\begin{eqnarray}\nonumber
\int_\cap dy\,\frac{e^{\beta U(y)}}{D(y)}&=&\frac{e^{\beta U(q^*)}}{D(q^*)}\int_{-\infty}^{\infty} dy\, e^{-\beta \omega^2(y-q^*)^2/2} \\ 
&=&\frac{\sqrt{2\pi k_{\rm B}T}}{\omega} \frac{e^{\beta U(q^*)}}{D(q^*)},
\end{eqnarray}
where we have extended the integration from $-\infty$ and $\infty$, because the Gaussian function yields significant values only near $q^*$.  Since the escape rate $k$ (number of escapes per unit time) is just the inverse of the mean first-passage time, $k=1/\tau$, we can finally write the escape rate as
\begin{equation}
k=\frac{\omega D(q^*)} {\sqrt{2\pi k_{\rm B}T}}\frac{e^{-\beta U(q^*)}}{\int_\cup dq\, e^{-\beta U(q)}}.
\label{equ:escape_rate}
\end{equation}
Here, the first fraction on the right hand side is the kinetic prefactor $K$, 
\begin{equation}
K=\frac{\omega D(q^*)} {\sqrt{2\pi k_{\rm B}T}},
\end{equation}
which depends on the curvature of the barrier, the diffusion coefficient on the barrier as well as on the temperature.
The second fraction in eq.~(\ref{equ:escape_rate}), on the other hand, is nothing else than the equilibrium probability density $P(q^*)$ to find the particle at $q^*$, properly normalized by the integral 
 \begin{equation}
 P(q^*)=\frac{e^{-\beta U(q^*)}}{\int_\cup dq\, e^{-\beta U(q)}}.
 \end{equation}
 In the following, we will use this result to derive the nucleation rate $J$ within classical nucleation theory. 
 
\subsection{Nucleation rate of CNT}
\label{subsec:cntrate}

\begin{figure*}
\begin{center}
\includegraphics[width=1.65\columnwidth]{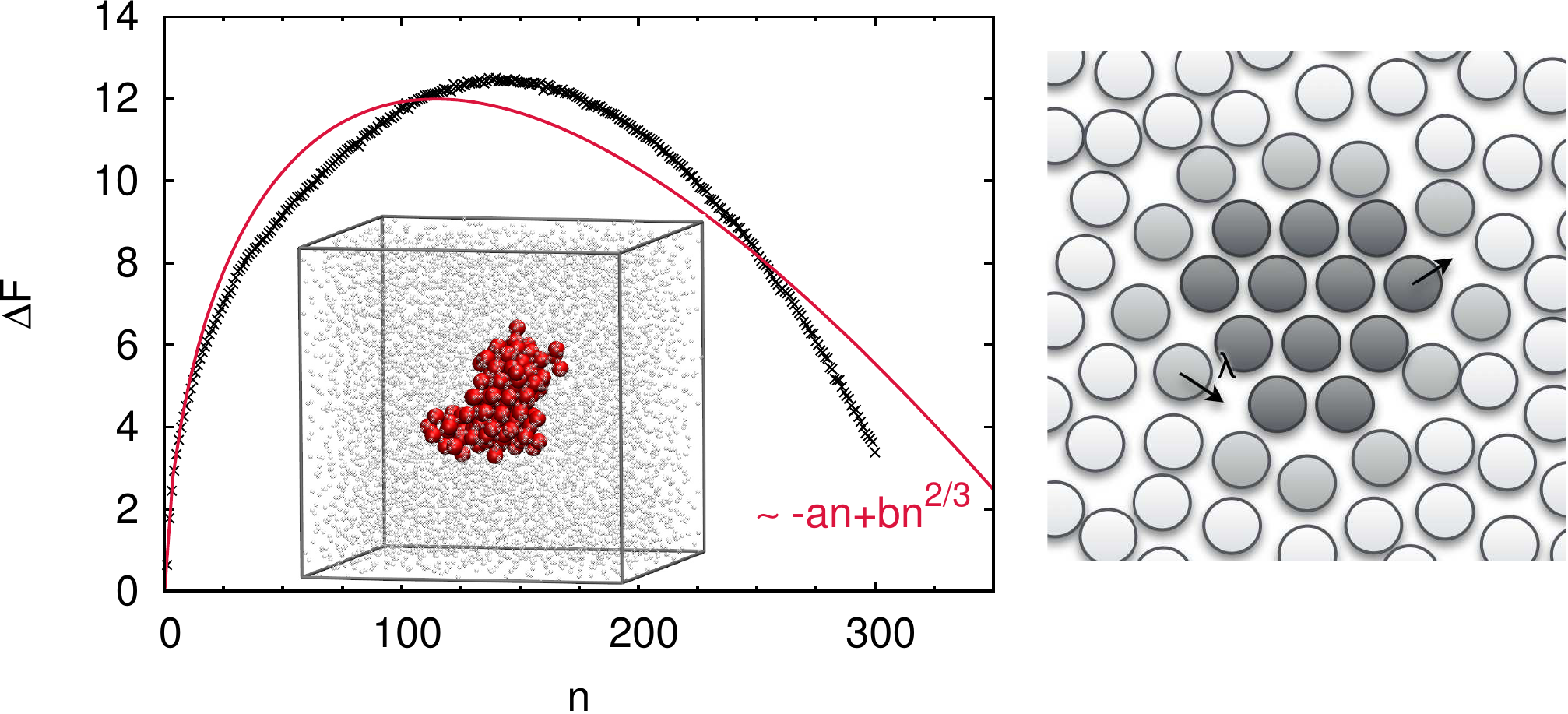}  
\caption{\label{fig:freeenergyLJ} Left: Free energy of a supercooled Lennard-Jones fluid as a function of the size of the largest crystalline cluster \protect\cite{jungblut:2011}. The solid line represents the functional form predicted by CNT (eq.~(\ref{eq:cnt6})) fitted to the simulation data (crosses) with variational parameters $a$ and $b$. The inset depicts a configuration with a critical cluster consisting of large opaque spheres; small transparent spheres indicate positions of other particles in the box. Right: A crystallite (dark grey spheres) in the supercooled liquid (white and light grey spheres) grows or shrinks by the attachment or detachment of particles, respectively, as indicated by the arrows. When one of the liquid particles near the crystallite (light grey spheres) attaches to the cluster, it needs to move by a distance $\lambda$ of molecular dimensions.  }
\end{center}
\end{figure*}

To apply Kramers theory to the crystallization problem we need to introduce a reaction coordinate, a variable that describes the progress of the crystallization process and plays the role of the coordinate $q$ in the derivation laid out above. For this purpose, we consider a certain volume $V$ of the supercooled liquid and imagine that we can identify all little crystallites present in this volume. At any given time there might be several crystallites in the system and, in general, they will have different sizes. Let us now define $n$ as the size of the {\em largest} crystallite found in the given volume. In this way we can assign a well-defined and unique number to any given configuration of the system and in the following we will use $n$ as our reaction coordinate. To establish a connection to the Kramers problem, we view the time evolution $n(t)$ of the size of the largest cluster as a diffusion process with a diffusion coefficient $D(n)$. The diffusive nature of the dynamics of $n$ arises from the stochastic growth and decay of a small crystallite in the liquid occurring through the random attachment and detachment of particles to and from the crystallite. The diffusion process is driven by a thermodynamic force given by the free energy $F(n)=-k_{\rm B}T \ln P(n)$, where $P(n)$ is the probability that the largest crystalline cluster in the system has the size $n$. Since, as discussed earlier, the formation of supercritical crystals is rare, the free energy $F(n)$ typically has a well at small sizes separated by a high barrier from the completely crystalline system. A typical nucleation free energy landscape $F(n)$ is shown in fig.~\ref{fig:freeenergyLJ}.

Having defined the quantities $n$, $D(n)$, and $P(n)$, we can now establish the mapping to the Kramers problem. Accordingly, the nucleation rate $J$, {\it i.e.}, the number of nucleation events per unit time and unit volume, is obtained by dividing the Kramers rate constant $k$ through the sample volume $V$, 
\begin{equation}
J=\frac{k}{V}= {\rm Ze} \frac{D(n^*)}{V}P(n^*),
\label{equ:CNT_nucleation_rate}
\end{equation}
where we introduced the commonly used Zeldovich factor  \cite{frenkel:1946,kelton:2010}, 
\begin{equation}
{\rm Ze}=\sqrt{ \frac{ \left |\Delta F''(n^*)\right|}{2\pi  k_{\rm B}T}}, 
\label{equ:zeldovich}
\end{equation}
which is inversely proportional to the width of the free energy barrier (the Zeldovich factor is given by ${\rm Ze}=1/\sqrt{\pi}\Delta n$, where $\Delta n$ is the width of the free energy barrier 1 $k_{\rm B}T$ below the top).
Here, $n^*$ is the size of the largest crystalline cluster at the top of the free energy barrier, and $P(n^*)$ is the probability that the largest crystalline cluster has this size. $\Delta F'' (n^*)$ and $D(n^*)$ are the corresponding free energy curvature and the diffusion coefficient, respectively.  

In order to compute the nucleation rate $J$, we have to determine the quantities appearing in eq.~(\ref{equ:CNT_nucleation_rate}) and in the following we will do that in the framework of CNT. (In later sections we will discuss how to compute $F(n)$ and $D(n)$ with the help of computer simulations without recurring to the assumptions of classical nucleation theory.) Since we monitor the nucleation process using the size $n$ of the crystallite, we write the nucleation free energy as a function of $n$,
\begin{equation}
 \label{eq:cnt6}
 \Delta F (n) =  -n |\Delta \mu| + (36\pi)^{1/3}(nv)^{2/3} \gamma,
\end{equation}
where $v=1/\rho$ is the volume per particle in the solid and $\rho$ is the corresponding number density.
From this free energy, the size $n^*$ of the critical crystallite is easily obtained by taking the derivative of $\Delta F(n)$ with respect to $n$ and setting it to zero. Solving the resulting equation for $n^*$ then yields
\begin{equation}
n^*=\frac{32\pi v^2\gamma^3}{3 \left |\Delta \mu \right|^3}.
\end{equation}
Inserting this critical size into the nucleation free energy yields a barrier height of
\begin{equation}
 \label{eq:barrier}
 \Delta F (n^*) =  \frac{16\pi v^2\gamma^3}{3 \left |\Delta \mu \right|^2} = \frac{1}{2}\left |\Delta \mu \right| n^*.
\end{equation}
As one can deduce from this equation, the size of the critical nucleus grows with decreasing difference in chemical potential. As one approaches coexistence, where the difference in chemical potential vanishes, the size of the critical nucleus diverges. 

If the nucleation barrier is high, the probability that the largest cluster in the system has size $n^*$ is equal to the probability to find a cluster of size $n^*$. (As clusters of size $n^*$ are very rare, the probability that there is an even larger cluster in the system is negligible.) As a consequence, the probability $P(n^*)$ that the largest cluster in the system has critical size can be written as
\begin{equation}
 \label{eq:P}
P(n^*)=N \exp[-\beta \Delta F(n^*)].
\end{equation}
This probability is the product of the number $N$ of particles in the system, {\it i.e.}, the number of sites where nucleation can start, times the probability $\exp[-\beta \Delta F(n^*)]$ that a crystal of size $n^*$ forms at that site. Using the nucleation free energy from eq.~(\ref{eq:cnt6}), we can also determine the curvature of the free energy at the barrier top,

\begin{equation}
 \label{eq:curvature}
\left|\Delta F''(n^*)\right|=\frac{ \left |\Delta \mu \right|^4}{32\pi \gamma^3 v^2}=\frac{ \left |\Delta \mu \right|}{3n^*}.
\end{equation}
Putting the results we have obtained so far together, the nucleation rate can be written as
\begin{equation}
J= \sqrt{ \frac{\beta \left |\Delta \mu \right|}{6\pi n^*}} \frac{D(n^*)N}{V} \exp \left(-\frac{\beta\left |\Delta \mu \right| n^*}{2}\right).
\end{equation}
All quantities appearing in this expression, except the diffusion coefficient $D(n^*)$, can be determined from $\gamma$, $\Delta \mu$, and $\beta$. (Note that at the beginning of this derivation we explicitly assumed to know the volume $V$ of the supercooled sample, otherwise one computes the Kramers rate constant $k=JV$, which provides the number of nucleation events per unit time in a given volume.) So the only thing left to do for a complete theory of the nucleation rate (constant) is to estimate the diffusion coefficient $D(n)$. We will do that in the following.

First we note that in order to compute the nucleation rate we need to know the diffusion coefficient $D(n)$ only for values of $n$ in the barrier region. Due to the rarity of the nucleation events, only one cluster in this size range is present in the system at any given time (two or more cluster would be extremely unlikely). Therefore, the change in $n$, the size of the largest cluster, happens only through growth or shrinkage of this single large cluster. At smaller cluster sizes, on the other hand, it is possible that the role of being the largest cluster switches from one cluster to another such that it is not sufficient to consider a single cluster in this size regime. In order to determine the diffusion coefficient $D(n)$ near the barrier top we now imagine that the crystalline cluster grows by the attachment of single particles from the liquid. The attachment occurs with an attachment rate $f^+(n)$, which, in general, is a function of the size of the crystallite. Similarly, the crystallite shrinks when particles detach from it and become liquid particles. We imagine that this detachment process also proceeds particle by particle with a detachment rate of $f^-(n)$. We now express the diffusion coefficient $D(n)$ in terms of the attachment and detachment rates. For short times $t$, the mean square displacement of the cluster size $n(t)$ from the cluster size $n(0)$ at time $t$ earlier grows linearly in time, thus defining the diffusion coefficient $D(n)$,
\begin{equation}
\langle [n(t)-n(0)]^2\rangle = 2 D(n) t.
\label{equ:msdn}
\end{equation}
Now, imagine that at time $0$ the cluster has size $n$. Then, the probability that a short time $t$ later the cluster has grown from $n$ to $n+1$ particles is $f^+t$ and the probability that it has lost a particle going from $n$ to $n-1$ is $f^-t$. In both cases the cluster size has changed exactly by one unit. The probability that the cluster size remains the same is $1-f^+t-f^-t$. Since we assume that the time $t$ is so short that $f^+t \ll 1$ and $f^-t\ll 1$, we do not need to consider the cases where the particle number changes by more than one unit because such events would have the probabilities of $(f^\pm t)^2$, which can be neglected. Accordingly, for short times, the mean square displacement of the particle number is 
\begin{eqnarray} \nonumber
\langle [n(t)-n(0)]^2\rangle &=& f^+t +f^-t \\ &=&\left [f^+(n) +f^-(n)\right ] t.
\end{eqnarray}
Comparison with eq.~(\ref{equ:msdn}) then implies
\begin{equation}
D(n) =  \frac{f^+(n) +f^-(n)}{2}.
\end{equation}
Thus, knowing the attachment and detachment rates we know also the diffusion coefficient. 

We can further simplify the expression for the diffusion coefficient by noting that in equilibrium the attachment and detachment rates are related by detailed balance,
\begin{equation}
f^+(n)e^{-\beta \Delta F(n)}=f^-(n+1)e^{-\beta \Delta F(n+1)}.
\end{equation}
Since we are interested in the values of the diffusion coefficient only at the barrier top, we can exploit the fact that the free energy in this region is flat, {\it i.e.},  $\Delta F(n)\approx \Delta F(n+1)$. Therefore, at the critical sizes $n^*$ the attachment and detachment rates are equal,
\begin{equation}
f^+(n^*)=f^-(n^*).
\end{equation}
As a consequence, the diffusion coefficient at the critical sizes is simply equal to the attachment rate,
\begin{equation}
D(n^*)=f^+(n^*).
\end{equation}
The attachment rate is not something we can determine from the quantities we have considered so far such as the surface tension $\gamma$ or the difference in chemical potential $\Delta \mu$. In the following, we will derive an approximation of the attachment rate based on some simple physical considerations. 

To determine the attachment rate $f^+(n)$ of particles to a crystallite of size $n$ we imagine that particles that already are near the crystal need to move by a certain small distance $\lambda$ in order to lock into the right position and become crystalline. These liquid particles that can become crystalline and attach to the crystalline cluster are shown in light grey in the right panel of fig.~\ref{fig:freeenergyLJ}. In this figure, the dark grey particles form the crystalline cluster. The white particles belong to the supercooled liquid just as the light grey particles, but they are too far away from the crystal to attach to it in a short time. How many liquid light grey particles are there now next to the crystal? Assuming that the crystal is spherical with a radius $R$, these liquid particles occupy a layer of thickness $l$ around the crystal with surface area $4\pi R^2$. Thus, the total volume occupied by these particles is $4\pi R^2 l$ and their number is $n_l=4\pi R^2 l/v$, where $v$ is the volume per particle. Identifying the thickness $l$ of the layer with the particle diameter, $l$ is related to the volume per particle by $l = (6v/\pi)^{1/3}$. Inserting this relation into the expression for $n_l$ we obtain
\begin{equation}
n_l=4R^2 (6 \pi^2)^{1/3}v^{-2/3}.
\end{equation}
Using $R=(3nv/4\pi)^{1/3}$, the number $n_l$ of liquid particles in the first layer around the crystalline cluster is finally estimated as
\begin{equation}
n_l=6{n^*}^{2/3},
\end{equation}
where we have replaced $n$ by the the number $n^*$ of particles in the crystal of critical size, because that is the size for which we want to know the attachment rate $f^+$.

We now assume that particles in the layer around the crystal move into their attaching position diffusively with a self-diffusion constant $D_S$. Hence, the time $\tau$ needed on the average for a particle to move by a distance $\lambda$ (remember that $\lambda$ is the distance a particle has to travel in order to become part of the crystal) is given by 
\begin{equation}
\tau=\frac{\lambda^2}{4 D_S}.
\end{equation}
Here, the factor of 4 appears because the diffusion occurs on the two-dimensional surface of the crystal. Also note that by writing this expression we have assumed that a particle on the surface of the crystal diffuses with the same diffusion constant as in the bulk liquid. Since we have a total of $n_l$ particles around the crystal and each of them attaches at a rate of $1/\tau=4 D_S/\lambda^2$, the total attachment rate $f^+(n^*)$ is given by
\begin{equation}
f^+(n^*)=\frac{n_l}{\tau}=\frac{24 D_S}{\lambda^2}{n^*}^{2/3}.
\end{equation}
Putting everything together, we finally find the following expression for the nucleation rate 
\begin{equation}
J= \sqrt{ \frac{\beta \left |\Delta \mu \right|}{6\pi n^*}} \frac{24 D_SN}{\lambda^2 V}{n^*}^{2/3}\exp \left(-\frac{\beta\left |\Delta \mu \right| n^*}{2} \right),
\end{equation}
where, as noted before, the critical nucleus size is given by $n^*=32\pi v^2\gamma^3/(3 \left |\Delta \mu \right|^3)$. This equation is the central result of CNT for the freezing transition. Thus, within the framework of CNT, one can estimate 
the nucleation rate and critical cluster size on the basis of the properties of the reacting bulk phases alone. The expression for the nucleation rate is often used to analyze experimental results. Also, the predictions of CNT can be used to extract quantities from computer simulations that are not so easily calculated directly. For instance, recently critical cluster sizes obtained in simulations using the seeding technique together with chemical potential differences calculated in separate simulations were used to determine the surface tension between liquid water and ice crystals, which was subsequently used to estimate nucleation barriers and rates for the freezing of supercooled water \cite{Vega_ice_2015}. (These simulations are discussed in greater detail in section~\ref{subsec:water}.) Special computer simulation methods are also available for the calculation of the kinetic prefactor \cite{bennett:1977,chandler:1978,tenwolde:1996,ruizmontero:1997,auer:2004}. These methods are based on the concepts of transition state theory and will be discussed in section~\ref{sec:nuclRates}. 

CNT is based on a purely phenomenological approach and gives a very suggestive qualitative picture of the nucleation mechanism.  
Unfortunately, it usually fails to provide a quantitatively accurate description of crystallization in real systems, even  for rather simple examples like crystallization of hard \cite{auer:2001} or Lennard-Jones (LJ) \cite{tenwolde:1996} spheres. To illustrate this point, we have plotted in the left panel of fig.~\ref{fig:freeenergyLJ} the results of computer simulations  for the free energy landscape of crystal nucleation in a supercooled monodisperse  LJ fluid. A fit to the functional form of CNT (eq.~\ref{eq:cnt6}) with $|\Delta \mu|$  and $\gamma$ as variational parameters lies close to the simulation results but does not reproduce them exactly, indicating to the presence of some factors not accounted for by CNT.  Similarly, for the case where the chemical potential difference and the interfacial free energy are derived from other sources, the difference between CNT prediction and simulation amounts to about $20\%$ in the height of the barrier and the critical cluster size is much smaller than the one found in simulations \cite{tenwolde:1996}.  
  
\subsection{When does CNT fail?}
\label{subsec:cntFailure}

Classical nucleation theory relies on several simplifying assumptions, all of which may be violated under certain circumstances \cite{sear:2012}. 

One central assumption is that nucleation is a one-step process in which only one free energy barrier is relevant and the nucleus forming in the metastable phase consists of a small piece of the thermodynamically stable phase. It has been known for a long time that many systems fail to comply with these conditions and usually the transition does not occur immediately into the most stable phase. For LJ freezing, which we will use here as a routine example, it has been shown 
\cite{rahmanII:1977,alexander:1978,tenwolde:1996,moroni:2005,wang:2007,beckham:2011,jungblut:2013a} 
that the crystallization process follows the Ostwald's step rule \cite{ostwald:1897, stranski:1933}. 
This empirical rule says that a metastable system may transform into its final state through 
formation of an intermediate phase if the free energy barrier between the initial and the 
intermediate phases is lower than the one between the initial and the thermodynamically 
most stable states. In the particular case of the LJ crystallization, the intermediate phase is 
a body-centered cubic (bcc) crystal, which forms more easily from a supercooled liquid than the thermodynamically more stable face-centered cubic (fcc) structure. During the crystallization, the structural composition of crystalline clusters changes as they become larger. Thus, the chemical potential difference, assumed constant in CNT, is a function of the 
crystal size (the exact form of this dependence is, however, unknown). The formation of intermediate (in part liquid) phases during crystallization has been observed also for many other systems including small molecules, proteins and ice \cite{sear:2012}. However, even for the simplest gas-liquid transition, which does not involve structural rearrangements, the density in the center of the critical droplet is distinctly below the bulk 
liquid density \cite{oxtoby:1988}. Furthermore, the distribution of particles inside a droplet is also not homogeneous but 
decays smoothly from the center outwards into the gas phase. Thus, the concept of a perfect piece of the stable crystalline structure sharply separated from the surrounding liquid is only a rough approximation that fails in the initial stages of the crystallization.

Another important assumption of classical nucleation theory is the capillary approximation, which states that the surface free energy associated with the interface between the crystallite and the liquid is constant and 
does not depend on the curvature of the interface between two phases. This approximation 
also breaks down on the microscopic level of small droplets. It is possible to 
introduce a curvature dependent correcture to the expression for the free energy in 
terms of a constant Tolman length \cite{tolman:1949}, but, as has been shown recently for liquids, 
this length might  also be a function of the droplet size 
\cite{tenwolde:1998, iwamatsu:1994,granasy:1998,anisimov:2008,troester:2012,troester:2012a}.
 
For small crystallites, one also has to keep in mind that the interfacial free energy of 
a crystal (which is quite small for all faces of an fcc structure \cite{rull:1983,broughton:1986,davidchack:2003} 
and even lower for the bcc phase \cite{davidchack:2005}) depends on which crystal face 
is in contact with the fluid. For comparison of the 
simulation results on the LJ crystallization with theoretical predictions \cite{tenwolde:1996}, the 
differences were assumed marginal and an average over three different (fcc) 
values \cite{broughton:1986} was used. 
In the most cases, however, the determination of the surface free energy density is not 
straightforward, and it is desirable to estimate its value from comparison of the shapes 
of the free energy barriers computed in simulations and predicted by CNT.    
In doing so, the standard CNT is extended to include a correction assuming a constant 
Tolman length \cite{tolman:1949} and the simulation results are fitted to the new functional 
form \cite{lundrigan:2009,jungblut:2011}, yielding a slightly better agreement between CNT and 
simulations.  

Further investigations of LJ crystallization in simulations \cite{trudu:2006} revealed that the 
shape of the crystallites formed in the course of transition is ellipsoidal rather than spherical 
as assumed by CNT, which might be connected to the different values of the interfacial free energies of crystal faces.  
The correction of CNT in this respect does not change the functional form of the free 
energy barrier projected on the number of particles in the largest cluster, but 
shifts the analytic values of the size of the critical cluster and the height 
of the barrier closer to those obtained in simulations.  

In order to address all these issues, CNT has recently been modified to account for the 
non-spherical crystalline droplets, the fuzziness of the liquid-solid interface and its 
thermal fluctuations connected to the change of the interface area 
\cite{prestipino:2012,prestipino:2013,prestipino:2014}.  We would also like to note that recently non-classical nucleation pathways, in which nucleation occurs through the aggregation of so-called pre-nucleation clusters rather than monomer by monomer, have received a lot of attention \cite{baumgartner:2013,gebauer:2014}. While it is clear that CNT does not apply in this case, the ideas and concepts underlying CNT remain useful as they can help to understand the intermediate steps of more complex processes. Accordingly, CNT should not by viewed as quantitative theory for the prediction of nucleation rates, but rather as a helpful framework to think about nucleation, and, in particular, crystallization processes. 

\section{Setting the stage}
\label{sec:freeenergy}

\subsection{Phase diagrams and free energy calculations}
\label{subsec:phaseDiagrams}

A study of a particular transition starts with the identification of its initial and final states. Better yet is to place the process in the 
frame of a phase diagram, which specifies the equilibrium phase as a function of the external conditions such as temperature and pressure. There are many computer simulation methods for the calculation of phase diagrams and in this section we will discuss a few of them. 

The most straightforward way to determine a phase diagram is just to prepare a system under given conditions, let it evolve towards equilibrium and repeat for all conditions of interest. 
Some studies, particularly of complex systems, still use this approach despite its inaccuracy in the identification of truly equilibrium states. 
For example, if the relaxation towards the equilibrium state is very slow, like in a glass, the stable state might not be reached 
on a realistic time scale. 
Some illustrative examples of delayed appearances of new phases can be found in the review by Sear \cite{sear:2012}. 
In some cases, specific Monte Carlo simulation schemes can be constructed that incorporate nonphysical moves designed to accelerate the equilibration. The exact location of the boundary lines between different phases 
is, however, only possible with more sophisticated techniques. A thorough account of them is presented in many books on methods of 
computer simulations \cite{LandauBinderBook:2000,Frenkel_Smit_book,Allen_Tildesley_book}, and numerous reviews 
\cite{panagiotopoulos:2000,rickman:2002,vega:2008,mobley:2012,sweatman:2015} assess their 
applicability in particular cases. Here, we briefly survey only a few techniques particularly suited to the studies of crystallization and omit 
those which applicability is restricted to less dense systems, such as the Gibbs ensemble method \cite{panagitopoulos:1987,panagitopoulos:1988}. 

Among the schemes that are employed for the determination of phase diagrams involving crystalline phases, one can differentiate 
between techniques that rely on the computation of the free energies and those which do not. 
In the latter \cite{kerrache:2008,pedersen:2013}, two phases are simulated simultaneously, adjusting the conditions in a way to bring 
them in equilibrium. The analysis then has to include the consideration of an interface in between and its effects on the phases in 
contact. A possible drawback of this approach consists in the finite width of the interface, which may be large enough to influence the apparently bulk phases, leading to considerable finite size effects. It is possible to leave out the interface in the Gibbs ensemble, but, 
as mentioned above, the simulation of a very dense crystalline phase becomes then quite inefficient.    
On the other hand, it is also possible to simulate the phases separately and relate them to each other in terms of the 
free energy, which, on the microscopic level, can be written as  
 \begin{equation}
 \label{eq:partitionfuncitonFE}
 F  =  - k_{\rm B}T \ln Z,
\end{equation}
where $Z$ is the partition function calculated as a sum over all microstates of the system weighted according to 
their total energy,
 \begin{equation}
 \label{eq:partitionfuncitonZ}
 Z  = \sum_m e^{-\beta E_m}.
\end{equation}
Hence, the probability to find the system in a particular configuration $m$ with energy $E_m$ 
is given by 
\begin{equation}
\label{eq:partitionfuncitonP}
P(m)=\frac{e^{-\beta E_m}}{Z}=e^{-\beta (E_m-F)}.
\end{equation}
Generally, it is impossible to account for all microstates in one simulation, but the relative frequency of 
the occurrence of macroscopic phases is sufficient to compute free energy differences. There are, however, 
some relatively simple systems for which the free energy can be calculated analytically. 
Hence, the computation of the free energy difference between these systems and others, for which there is no analytic solution, 
leads to the absolute value of the free energy of the sample. 
Accordingly, methods for the calculation of free energies can be split into techniques calculating absolute free energies of individual 
phases or free energy differences between them. Phase coexistence then is characterized by either equal absolute free energies 
of two phases or, equivalently, by a vanishing free energy difference.  

The thermodynamic integration method \cite{LandauBinderBook:2000,Frenkel_Smit_book,schilling:2009} is widely used for calculation of 
the absolute free energies of fluids and solids. In this technique, one defines a coupling parameter $\alpha$ which allows 
to gradually transform the system from the examined state into the reference state the free energy of which is supposed to be known. It is important that the transformation is possible without hysteresis. Thus, the state in question and the reference state should not be separated by a first-order phase transition. The absolute free energy is then computed by the integration along a suitable thermodynamic path connecting the states 
 \begin{equation}
 \label{eq:feTI}
 F  =  F_{\rm ref}+\int_0^{1} d\alpha \left \langle \frac {dH(\alpha)}{d\alpha} \right \rangle _{\alpha} ,
\end{equation}
where $F_{\rm ref}$ is the free energy of the reference system. The Hamiltonian $H(\alpha)$ specifies 
the pathway between the two states and is defined such that $H(0)$ corresponds to the reference state and $H(1)$ to the state we are interested in. The angular brackets $\langle \cdots \rangle_\alpha$ indicate an average calculated for the value $\alpha $ of the switching parameter. To calculate the integral in the above equation, several simulations for different values of $\alpha$ are carried out, in which the average $ \langle dH(\alpha) / d\alpha \rangle _{\alpha}$ is determined. From these values, the integral is then computed numerically. The exact form of the $H(\alpha)$ depends on the considered phases. 
For instance, the absolute free energy of solids is usually calculated within the Frenkel-Ladd method \cite{frenkel:1984,polson:2000}, 
where the particles are coupled to their respective positions in an Einstein crystal ($r_i^0$) by harmonic springs with 
strength $\alpha$, such that the modified Hamiltonian reads 
 \begin{equation}
 \label{eq:FLcrystal}
H({\alpha})  =  H_0+\alpha\sum_{i=1}^{N}(r_i-r_i^0)^2 ,
\end{equation}
where $N$ is the number of particles. $H_0$ and, thus, ${F(\alpha=0)}$ refer to the real solid, while $F(\alpha \gg 1)$ is 
the free energy of the Einstein crystal, which is known analytically.
For the determination of coexisting phases, however, a direct computation of the free energy differences between the phases turns out 
to be more precise from the statistical point of view, since absolute free energies are rather large numbers and the 
deviations accumulate \cite{wilms:2012}. 

The phase- and lattice-switch Monte Carlo techniques \cite{bruce:1997,bruce:2000,wilding:2000,jackson:2002} allow to compute 
free energy differences directly, omitting thereby the simulation of the intermediate states between the coexisting bulk phases. 
As indicated by the name, these methods make use of MC moves that allow a direct switch between the phases. Hence, the 
system can freely transform between the states and the corresponding free energy difference is computed from their relative occurrences: 
 \begin{equation}
 \label{eq:feDiff}
\Delta F  =  - k_BT\ln \frac{Z_1}{Z_2} =  - k_BT \ln \frac{P_1}{P_2},
\end{equation}
where $Z_i$ and $P_i$ are the partition function and the probability to find the system in phase $i=\{1, 2\}$, respectively. 

At coexistence, two phases are stable, occur with the same frequency, and thus possess the same free energy. 
If the average probability of occurrence for one of the states is smaller than for the other, the first phase  
is metastable. The likelihood to find the system in this state at the end of an unbiased simulation run will depend on the free energy 
difference but also on the height of the barrier between the phases, and, thus, on the rate with which the system transforms from one to the other. 
In other words, starting a simulation from a metastable state, we may stay there through the whole run if the free energy barrier that has 
to be overcome on the way to equilibrium is too high. Similarly, if the barrier is low, thermal fluctuations will facilitate 
frequent visits of the metastable state on the time scale of the simulation. The same applies to coexisting stable states, where furthermore 
the position of the critical point characterized by a vanishing free energy barrier between the phases, should the examined transition possess 
one, can be determined by a finite size analysis \cite{binder:2008}.       
Thus, the shape of the free energy landscape in the region between the reacting states is of particular importance 
for the studies of transition kinetics. In order to assign a set of physical microstates visited in the course of transition to one of the 
reacting phases or to the intermediate region between them, we need to define a collective variable on which the free energy is projected. 
In doing so, we distinguish between an order parameter and a reaction coordinate, which may, in some applications, be interchangeable but 
serve as such different purposes. 
Thus, a general distinction between two phases is usually drawn by means of an order parameter, which uniquely characterizes 
the initial and the final states of a transition. The more specific reaction coordinate, on the other hand, is continuous and 
describes the progress of the reaction at all stages. The order parameter can be recovered from the 
reaction coordinate by assigning all states that are located left and right of the top of the barrier 
to the respective reacting and product states. The barrier itself is constructed by the projection of the free 
energy landscape on a reaction coordinate $q$, which can be computed from the probability to find the system at a given 
stage of transition, defined according to eq.~(\ref{eq:partitionfuncitonP}) as 
\begin{equation}
\label{eq:pofq}
 P (q)  =  \sum_{m_q} e^{-\beta (E_{m_q}-F)} = Z(q)e^{\beta F},
\end{equation}
where $m_q$ is the ensemble of microstates contributing to $q$ ({\it i.e.}, the microstates in which the reaction coordinate has the particular value $q$). The free energy profile is then given by 
 \begin{equation}
 \label{eq:feBarrier} 
F (q)  =  - k_BT\ln P(q) + {\rm const},
\end{equation}
where the constant, which equals the free energy $F$, indicates that the free energy landscape is measured on a restricted set of microstates, which 
belong to some global ensemble. Thus, the computation of the microscopic 
distribution of states in the reacting states and along the transition path yields the shape of the free energy barrier. 
Unfortunately, in a straightforward computer simulation, the probability to find the system outside of the (meta)stable states is proportional to the height of the barrier and usually rather low. Hence, computational methods have been constructed to estimate the free energy barrier by enhancing the sampling of these rarely occurring states. 

One of the most popular techniques for the calculation of free energies as a function of a given reaction coordinate $q$ is the umbrella sampling \cite{torrie:1974}. In this method, the energy landscape is biased 
by a predefined function $w(q)$ to enhance sampling of less probable states in the transition region and the real distribution 
is obtained by reweighing the simulated distribution via 
 \begin{equation}
 \label{eq:reweighing}
P (q)  =  P^{\rm sim}(q) e^{-\beta w(q)}.
\end{equation}
Furthermore, the transition region is divided into overlapping windows and each of them is simulated separately with an appropriate 
weight function. The complete probability distribution of states along the barrier is then obtained by matching the computed probability distributions in the overlapping areas. The success of this approach depends on the reasonable choice of the 
weight function. 

In contrast, multicanonical sampling \cite{berg:1992} and the metadynamics scheme \cite{laio:2002,barducci:2008} 
use an effective Hamiltonian, which is self-consistently adjusted during a simulation 
run in order to uniformly sample all possible states along $q$. The resulting weight function is then simply related to the 
free energy landscape. The convergence of the technique, however, might need some effort.   
A similar approach is followed in the Wang-Landau sampling method \cite{wang:2001, wang:2001a, landau:2004}, but the weights 
are adjusted to obtain a flat distribution of the energy. The method allows for a very precise calculation of the density of states, from which other thermodynamic quantities of interest can be calculated. 
In single and multiple histograms methods \cite{ferrenberg:1988,ferrenberg:1989}, the distributions computed at conditions, at which the height of the barrier is comparatively low and the system can easily move between 
the reacting states, are reweighed in order to extrapolate the shape of the barrier. 
Similarly, parallel tempering \cite{swendsen:1986,earl:2005}, also known as replica exchange Markov Chain Monte Carlo 
sampling, allows the exchange of configurations between a number of systems simulated simultaneously but at 
different temperatures, such that all states in the low-temperature regime can be accessed.    
Successive umbrella sampling \cite{virnau:2004,chopra:2006}, which resembles the canonical umbrella 
sampling in the use of the windows, computes the relative probabilities of very narrow regions 
along a reaction coordinate. In this case, the free energy differences between the adjacent windows are very 
small and the sampling can be performed either without any weights or with the weights extrapolated into the 
next window. The resulting probability distribution is constructed by successive sampling of all 
segments of the path connecting (and including) the reacting states.

\subsection{Order parameter and reaction coordinate(s) for crystallization}
\label{subsec:orderParameter}

In his book on statistical mechanics \cite{SethnaBook:2006}, James P. Sethna wrote in 
2006: ``Choosing an order parameter is an art.'' We would like to add that choosing a reaction 
coordinate turns out to be the next stage of complexity. In simulations of crystallization it important to be able to 
distinguish between crystalline and liquid regions, ideally with single particle resolution. Over the years, many algorithms to detect 
crystalline structures in computer simulations have been put forward. Quite recently, Santiso and Trout \cite{santiso:2011} 
presented a framework to construct order parameters for molecular crystals. In parallel, 
Keys, Iacovella and Glotzer \cite{keys:2011} proposed to use a set of shape matching functions to identify various 
structures. Still, for simple monoatomic system, Steinhardt bond order parameters \cite{steinhardt:1983} provide a rather 
convenient and widely established method for identification of the bulk crystalline phases.   
Some years ago, ten Wolde, Ruiz-Montero and Frenkel \cite{tenwolde:1995} modified the scheme 
in order to account for local crystallinity, such that the size of the largest set of connected 
crystalline particles can be used as reaction coordinate for crystallization.  
In the original paper, however, the local crystallinity parameters were only employed in the examination of the cluster structure, 
while the progress of reaction was monitored by means of the global crystallinity $Q_6$.    
Some other earlier works \cite{rahmanI:1977,rahmanII:1977} related the order parameter and reaction 
coordinate to the global and local structure factors for the purpose of establishing a connection to experiments. 
In more recent crystallization experiments carried out using colloids \cite{gasser:2001}, 
which behave similarly to atomistic systems \cite{herlach:2010}, particle positions are determined 
with high accuracy in real space such that the comparison between 
simulations and experiments can be performed by means of Steinhardt bond order parameters.

In this scheme, structures are analyzed in terms of spherical harmonics, which provide characteristic structural finger prints. 
Each particle is evaluated on the basis of its neighborhood, which includes all particles which are closer than a certain distance $d_{\rm th}$. 
This value is usually chosen to correspond to the first minimum of the pair correlation function in an fcc crystal at coexistence.
The environment of each particle, numbered by the index $i$, is then represented by the complex vector
\begin{equation}
\label{eq:qvector}
q_{6m}(i) = \frac{1}{n_b} \sum_{j=1}^{n_b} Y_{6m}\left(\theta \left(r_{ij}\right), \phi \left(r_{ij}\right)\right), 
\end{equation}
where $Y_{6m}(\theta, \phi)$ are spherical harmonics, $\theta (r_{ij})$ and 
$\phi (r_{ij})$ are the angular spherical coordinates of a bond $r_{ij}$ in a fixed reference 
frame. The sum runs over all  $n_b$ neighbors of particle $i$.
Then, for every pair of particles $i$ and $j$, one computes the normalized scalar product of their complex vectors 
$q_{6m}$, 
\begin{equation}
\label{eq:sij}
s_{ij}= \frac {\sum_{m=-6}^6 q_{6m}(i) q_{6m}^{\ast}(j)}{\left (\sum_{m=-6}^6 |q_{6m}(i)|^2 \right )^{1/2} \left (\sum_{m=-6}^6 |q_{6m}(j)|^2 \right )^{1/2}},
\end{equation}  
which quantifies the degree of correlation between the structures surrounding the two particles.    
If the value of this product is larger than a certain value $s_{\rm th}$, the particles are considered to be connected to 
each other. A particle is identified as crystalline, if its number of connections 
is larger than a certain threshold $n_{\rm th}$. 
Ten Wolde, Ruiz-Montero and Frenkel \cite{tenwolde:1995} chose the threshold values for the bond strength, $s_{\rm th}$, 
and the number of connections, $n_{\rm th}$, on the basis of corresponding probability distributions in the liquid 
and solid states at coexistence. Once one knows for each particle whether it is crystalline or not, one can determine clusters
of crystalline particles. Two crystalline particles that are neighbors are defined to belong to the same cluster. In a given  
configuration $x$ (here $x$ includes the information about the positions of all particles) there may be several crystalline
clusters of different sizes. The number of particles in the largest one of these crystalline clusters is then
defined as the reaction coordinate to monitor to progress of the crystallization. Note that this reaction coordinate 
is a function of the threshold values used 
to define the crystallinity of individual particles, $n(x; d_{\rm th}, s_{\rm th}, n_{\rm th})$. 
The global crystallinity value is recovered by
\begin{equation}
\label{eq:globalq6}
Q_6=\left(\frac{4\pi}{13}\sum_{m=-6}^{6}\left | \frac{\sum_{i=1}^N n_b(i)q_{6m}(i)}{\sum_{i=1}^N n_b(i)} \right|^2 \right)^{1/2}. 
\end{equation}  
This parameter measures the average crystallinity of the entire sample without giving information about size and location of local crystalline regions.

The choice to use the complex vector $q_{6m}$ in the above analysis is a consequence of the six-fold symmetry of the crystalline states we 
intend to detect while studying freezing into fcc and bcc structures. In general, however, Steinhardt bond order 
parameters \cite{steinhardt:1983} were defined to 
address all possible types of crystals. In this respect, if the final solid state possesses a different symmetry, {\it e.g.}, is a cubic lattice, another 
complex vector like $q_{4m}$ or $q_{8m}$ (with $m$ running from $-4$ to $4$ and from $-8$ to $8$, respectively) 
would be more appropriate. Furthermore, if one is interested not only in detecting crystals but also in differentiating between various 
symmetries, a combination of Steinhardt bond order parameters can be assigned to different lattice types. Recently, Lechner and Dellago \cite{lechner:2008} 
proposed to distinguish locally between fcc, hcp, and bcc structures by means of a combination of the averaged parameters $\bar{q}_{4}(i)$ and $\bar{q}_{6}(i)$ 
defined similarly to the global crystallinity above as    
\begin{subequations}
\begin{eqnarray}
\label{eq:averageqs}
\bar{q}_4(i)&=&\left(\frac{4\pi}{9}\sum_{m=-4}^{4}\left | \frac{\sum_{j=1}^{\tilde{n}_b} q_{4m}(j)}{\tilde{n}_b} \right|^2 \right )^{1/2}\\
\bar{q}_6(i)&=&\left(\frac{4\pi}{13}\sum_{m=-6}^{6}\left | \frac{\sum_{j=1}^{\tilde{n}_b} q_{6m}(j)}{\tilde{n}_b} \right|^2\right )^{1/2},
\end{eqnarray}  
\end{subequations}
where $\tilde{n}_b=n_b+1$ and the sum runs over the particle $i$ and its nearest neighbors.  
In fact, this combination also detects the liquid state in a unique range of values, such that one can simultaneously find crystalline 
clusters and identify their structure \cite{lechner:2011}. 

The reaction coordinate for crystallization defined as the size of the largest crystalline cluster identified on the basis of Steinhardt 
bond order parameters is relatively robust, 
handy and has been  used widely  \cite{chushak:2000, leyssale:2003, auer:2004, valeriani:2005, moroni:2005, vanmeel:2008, bokeloh:2011, lechner:2011, russo:2012}.  
Still, it has been suggested that this size alone is not
sufficient to serve as a reaction coordinate for the crystallization transition \cite{tenwolde:1999, moroni:2005, lechner:2011}, 
and a few variations have been proposed in the meantime \cite{klumov:2010, schilling:2010, kawasaki:2011}. 
In section~\ref{subsec:reactionCoordinate}, we return to the issue of the definition and performance of reaction coordinates for 
crystallization, but would like to note here already 
that, currently, the size of the largest cluster still scores best in the set of proposed alternatives \cite{beckham:2011}.   

\subsection{Free energy landscapes are not unique}
\label{subsec:feLandscapes}

While the analysis of free energy landscapes (for instance the free energy $F(q)$ defined above) may yield useful insights, it is important to keep in mind that they are not unique.  In contrast to the energy, which is uniquely defined, free energies  always depend on the specific collective variable (reaction coordinate or order parameter) one considers \cite{Frenkel:2013}. As a consequence, there is no such thing as {\em the} free energy profile. Rather, every collective variable (or a set of collective variables) has its own free energy landscape and these landscapes may differ drastically from each other. In fact, one can make free energy barriers appear or disappear at will by a simple transformation of variables. To be more explicit, consider a collective variable $q(x)$ in a system in which the microscopic degrees of freedom are represented by $x$. The probability density function of $q$ is then given by
\begin{equation}
\label{eq:fe1}
P_Q(q)=\int \,dx \rho(x) \delta [q - q(x)],
\end{equation}
where $\rho(x)$ is the equilibrium distribution and $\delta (q)$ is the Dirac delta function. Note that here we consider a continuous configuration space, while above we have looked at a discrete set of microscopic states. The free energy (up to a constant, which we drop in the following since the change of coordinates does not modify the global ensemble of microstates) as a function of the collective variable is defined as
\begin{equation}
\label{eq:fe2}
F_Q(q)=-k_BT\ln P_Q(q). 
\end{equation}
The choice of the particular collective coordinate is arbitrary, so we may as well consider the free energy as a function of another collective variable $z(x)$ that is obtained from $q(x)$ by a simple transformation of variables, $z=\varphi[q(x)]$. If the transformation is unique ({\it i.e.}, if the function $\varphi(q)$ is invertible), no information is lost due to the transformation. The probability density functions of $q$ and $z$ are then related by 
\begin{equation}
\label{eq:fe3}
P_Z(z)=P_Q[\varphi^{-1}(z)]\left|\frac{d\varphi^{-1}}{dz}\right|.
\end{equation}
Accordingly, the free energy $F_Z(z)=- k_BT\ln P_Z(z)$ can be written as
\begin{equation}
\label{eq:F_Jacobian}
F_Z(z)=F_Q[\varphi^{-1}(z)]- k_BT\ln \left|\frac{d\varphi^{-1}}{dz}\right|. 
\end{equation}
The last term on the right hand side results from the Jacobian of the variable transformation from $q$ to $z$ and can be viewed as an entropic contribution that takes into account expansions and contractions arising from the transformation. If the transformation of variables is linear, the Jacobian is constant. It then leads only to an irrelevant shift but no other change of the free energy profile. If the transformation is non-linear, however, the form of the free energy profile is changed and, in particular, barrier heights change. In fact, by choosing the transformation $\varphi(q)$ appropriately, one can obtain any free energy profile in the new variable $z$ that one wants. Imagine that we would like to obtain a certain free energy profile $F_Z(z)$. We can get this particular free energy by choosing the transformation $z=\varphi(q)$ such that it satisfies
\begin{equation}
\label{eq:fe4}
\left|\frac{d\varphi^{-1}}{dz}\right|=\exp\left\{-\beta \left[F_Z(z)-F_Q[\varphi^{-1}(z)]\right]\right\}.
\end{equation}
For instance, we can get a completely flat free energy profile $F_Z(z)$ by applying the transformation
\begin{equation}
\label{eq:fe5}
\varphi (q)=\int_{q_0}^q dq'\, \exp[-\beta F_Q(q')].
\end{equation}
In this case,
\begin{equation}
\label{eq:fe6}
\left|\frac{d\varphi^{-1}}{dz}\right|=\left|\frac{d\varphi(q)}{dq}\right|^{-1}=\exp[\beta F_Q(\varphi^{-1}(z))]
\end{equation}
such that eq.~(\ref{eq:F_Jacobian}) implies 
\begin{equation}
\label{eq:fe7}
F_Z(z)= F_Q[\varphi^{-1}(z)]-k_BT\ln \exp[\beta F_Q(\varphi^{-1}(z))] = 0.
\end{equation}
Based on this flat free energy profile, we can now go a step further and transform the uniform probability density $P_Z(z)$ into a distribution $P_W(w)$ of an arbitrary shape by carrying out another change of variables to the new variable $w=\psi(z)$,
\begin{equation}
\label{eq:fe8}
P_W(w)=C\left|\frac{d \psi^{-1}}{dw} \right|=C\left|\frac{d \psi}{dz} \right|^{-1},
\end{equation}
where $C$ is the constant that normalizes the probability density of $z$. Choosing
\begin{equation}
\label{eq:fe9}
\psi^{-1}(w)=\int_{w_0}^{w} dw' \, P_W(w')
\end{equation}
and then inverting $\psi^{-1}(w)$ to obtain $z=\psi(w)$ yields the desired distribution. Concatenation of the two transformations of variables, $w=(\psi \circ \varphi)(q)=\psi(\varphi(q))$, produces a collective variable $w$ that is distributed according to a given arbitrary probability density function $P_W(w)$. The particular features of $P_W(w)$, and hence of the corresponding free energy $F_W(w)=-k_BT\ln P_W(w)$, may be completely unrelated to the features of the original distribution $P_Q(q)$ and its free energy $F_Q(q)$. 

\begin{figure}[ht]
\begin{center}
\includegraphics[clip=,width=0.95\columnwidth]{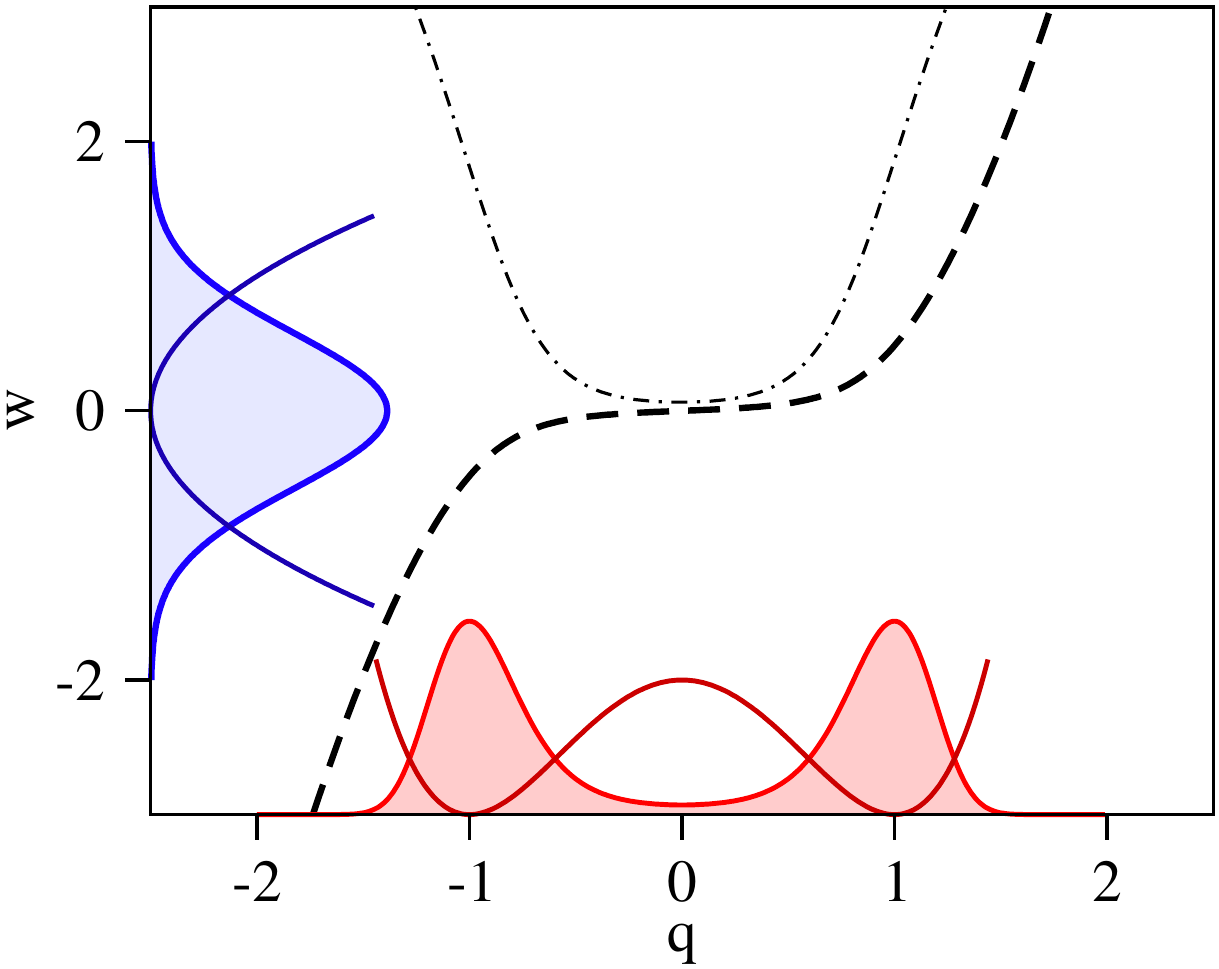}
\end{center}
\caption {\label{fig:FE_trafo} Effect of a transformation of variables on the free energy profile for the example discussed in the main text. The original probability density function $P_Q(q)$ is shown as shaded area on the horizontal axis together with the corresponding bistable free energy $F_Q(q)$. Applying the transformation $w=(\psi\circ\phi)(q)$, shown as dashed line, yields the probability density function $P_W(w)$ displayed as shaded area on the vertical axis alongside the resulting free energy $F_W(w)$, which displays a single minimum and no barrier. The dot-dashed line is the Jacobian of the transformation, $\left| dw/dq\right|$. For this example calculation, the parameters were set to $\kappa=1$, $k=1$, $q_0=1$, and $\beta=3$.}
\end{figure}

To illustrate this important point further, we consider a variable $q$ that is distributed according to 
\begin{equation}
\label{eq:fe10}
P_Q(q)=\frac{1}{\mathcal Q}\exp[-\beta \kappa (q_0^2-q^2)^2],
\end{equation}
where $\mathcal Q=\int dq\, \exp[-\beta \kappa (q_0^2-q^2)^2]$.
The free energy, given by 
\begin{equation}
\label{eq:fe11}
F_Q(q)= \kappa (q_0^2-q^2)^2+k_{\rm B}T \ln \mathcal Q,
\end{equation}
has two minima located at $\pm q_0$ separated by a barrier of height $\Delta F = \kappa q_0^4$ located at $q=0$. We first carry out the transformation 
\begin{equation}
\label{eq:phi}
z=\varphi(q)=\int_{q_0}^q dq'\,\frac{1}{\mathcal Q}\exp[-\beta \kappa (q_0^2-q'^2)^2].
\end{equation}
This transformation yields a uniformly distributed new variable $z$. We next carry out the second transformation $\psi(z)$ defined by 
\begin{eqnarray}\nonumber
\label{eq:fe12}
\psi^{-1}(w)&=&\int_{w_0}^{w} dw' \, \sqrt{\frac{\beta k}{2\pi}}\exp(-\frac{\beta kw'^2}{2})\\&=&\frac{1}{2}{\rm erf} \left(\sqrt{\frac{\beta k}{2}} w\right)+\frac{1}{2},
\end{eqnarray}
where $k$ is a constant and we have chosen $w_0$ so small that ${\rm erf} (\sqrt{\beta k/2}w_0) = -1$. Hence the function $\psi (z)$ is given by 
\begin{equation}
\label{eq:psi}
\psi(z)=\sqrt{\frac{2}{\beta k}}{\rm erf}^{-1}(2z-1).
\end{equation}
The result of the two transformations, carried out one after the other according to $w=(\psi\circ\varphi)(q)$ is a free energy 
\begin{equation}
\label{eq:fe13}
F_W(w)=\frac{\beta kw^2}{2}-\frac{1}{2 \beta} \ln  \frac{\beta k}{2\pi}
\end{equation}
that has only one minimum located at $w=0$ and no barrier at all. The effect of this transformation of variables is depicted in fig.~\ref{fig:FE_trafo}. 

As another example for how the choice of reaction coordinates affects the shape of the free energy consider a nucleation process and imagine that the free energy as a function of the nucleus diameter $d$ follows classical nucleation theory (similar to eq.~(\ref{eq:cnt1}) but with reference free energy set to zero),
\begin{equation}
\label{eq:fe14}
F_D(d)=\pi \gamma d^2 - \frac{\pi}{6}d^3 \rho \left| \Delta \mu \right |.
\end{equation}
By taking the derivative of $F_D(d)$ with respect to $d$ and setting it to zero, one finds the size of the critical nucleus
\begin{equation}
\label{eq:fe15}
d^*=\frac{4\gamma}{\rho \left| \Delta \mu \right |}
\end{equation}
and the height of the nucleation barrier
\begin{equation}
\label{eq:fe16}
\Delta F_D^*=\frac{16 \pi \gamma^3}{3(\rho\Delta \mu)^2}.
\end{equation}
If the droplet diameter $d$ is a good reaction coordinate for the nucleation process, the volume $v=(\pi/6)d^3$ of the nucleus should work as well. According to eq.~(\ref{eq:F_Jacobian}), the free energy $F_V(v)$ as a function of the nucleus volume $v$ is given by 
\begin{equation}
\label{eq:fe17}
F_V(v)=\pi \gamma\left(\frac{6v}{\pi} \right)^{2/3} -v\rho \left |\Delta \mu \right |+\frac{k_{\rm B}T}{3} \ln \frac{9\pi v^2}{2} .
\end{equation}
The last term is the entropic contribution originating from the Jacobian of the transformation. While this logarithmic term also affects the free energy in the barrier region, the most pronounced changes occur in the limit of small $v$. For $v\rightarrow 0$, the logarithm goes to $-\infty$ such that the barrier height diverges, $\Delta F_V^*=F_V(v^*)-F_V(0)=\infty$. Thus, although $d$ and $v$ are uniquely related to each other and they are expected to serve equally well as reaction coordinate, they lead to completely different barrier heights underlining the arbitrariness of free energy profiles.  Despite this arbitrariness, free energies computed as functions of particular variables may be very useful in the analysis of molecular simulations. In such an analysis, however, it is important to bear in mind the physical significance of the variables as a function of which the free energy is computed. 
As discussed in the next section, where we return to this example, for the computation of nucleation rates the arbitrariness inherent in the choice of reaction coordinate is less relevant. In this case, transforming from one reaction coordinate to another leads to simultaneous changes in the height of the free energy barrier and the dynamical correction factor, which cancel each other such that the same nucleation rate is obtained in both cases.

\section{Computing nucleation rates}
\label{sec:nuclRates}

\subsection{Macroscopic view}
\label{subsec:macroscopic}

Nucleation is a stochastic process: if a system is repeatedly prepared in exactly the same 
macroscopic conditions, the times needed by the system to convert into the new phase will be different.
When we talk about nucleation rates, 
we quantify the average time we have to wait to see this 
new phase.   
This time is related to the nucleation rate $J$ via
 \begin{equation}
 \label{eq:meantime}
 \langle t \rangle = \frac{1}{JV},
\end{equation}
where $V$ is the volume of the sample. Thus, the nucleation rate $J$ is the average number of nucleation events occurring 
per unit time and unit volume. We will also consider the rate constant for nucleation, $k$, defined as the average
number of nucleation events occurring in a system, but not normalized by the volume. $J$ and $k$ are related by 
 $k=JV$. 

As discussed earlier, the crucial event during the nucleation process is the crossing of the free energy barrier 
associated with the formation of the critical nucleus. For high barriers, this is a rare event similar to a chemical reaction, where 
some activation energy is needed to transform reactants ($A$) into products ($B$). On the macroscopic 
level, the kinetics of such reactions is well captured by so-called kinetic equations, which describe the time evolution 
of the populations in the reactant and product states. Since we will use these concepts later to introduce methods for 
an efficient simulation of rare events, we will first write down the kinetic equations for a general two-state system.

Consider many copies of a system that can exists in two different states, $A$ and $B$, for instance a molecule in solution with two different 
conformations between which rare transitions occur. The concentrations of the two species, $c_A$ and $c_B$, then evolve in time according to 
\begin{subequations}
 \begin{eqnarray}
 \label{eq:concentration}
\frac{dc_A(t)}{dt}&=&-k_{A\rightarrow B}c_A(t)+k_{A\leftarrow B}c_B(t) \label{eq:concentrationA}\\ 
\frac{dc_B(t)}{dt}&=&-k_{A\leftarrow B}c_B(t)+k_{A\rightarrow B}c_A(t), \label{eq:concentrationB}
\end{eqnarray}
\end{subequations}
where $k_{A\rightarrow B}$ and $k_{A\leftarrow B}$  are the forward and backward rate constants for 
the reaction $A\rightleftharpoons B$, respectively. By constructing these equations, one just assumes that the number 
of transitions per unit time from $A$ to $B$ is proportional to the population in $A$ and the proportionality constant is the
rate constant. Similarly, the number of transitions per unit time from $B$ to $A$ is proportional to the population in $B$. 
Thus, the gain and loss terms in the above equations arise from the transitions between the two states. 

In equilibrium, the concentrations of the two substances do not change with time and their 
ratio is equal to the ratio of backward and forward rate constants: 
 \begin{equation}
 \label{eq:ratioC}
 \frac{\langle c_A \rangle}{\langle c_B \rangle} = \frac{k_{A\leftarrow B}}{k_{A\rightarrow B}}.
\end{equation}
Here, the angular brackets $\langle \cdots \rangle$ denote the equilibrium value of the respective concentration. 

To determine the rate constants one usually prepares the system in one of the reaction states, say $A$, and monitors
 its decay into the second state, $B$. Starting from such an initial state the relaxation of the concentrations towards the 
 equilibrium state is given by 
\begin{subequations}
 \begin{eqnarray}
 \label{eq:concentrationTT}
c_A(t)&=&\langle c_A \rangle + \left[c_A(0) - \langle c_A \rangle \right ] \exp \left ( -\left[k_{A\rightarrow B}+k_{A\leftarrow B}\right ] t\right ) \nonumber \\ 
&\ & \label{eq:concentrationTTA}\\
c_B(t)&=&\langle c_B \rangle \left [1 - \exp \left ( -\left [k_{A\rightarrow B}+k_{A\leftarrow B} \right ] t\right )\right ].  \label{eq:concentrationTTB}
\end{eqnarray}
\end{subequations}
Hence, the approach to equilibrium is determined by the reaction time 
 \begin{equation}
 \label{eq:sumrates0}
 \tau_{\rm rxn}^{-1}=k_{A\rightarrow B}+k_{A\leftarrow B},
\end{equation}
which depends on both the forward and the backward reaction rate constants. Using eq.~(\ref{eq:ratioC}), the reaction time can be expressed in terms of the equilibrium concentrations, 
 \begin{equation}
 \label{eq:sumrates}
 \tau_{\rm rxn}^{-1}=k_{A\rightarrow B}+k_{A\leftarrow B}=k_{A\rightarrow B}\frac{\langle c_A \rangle +\langle c_B \rangle}{\langle c_B \rangle}.
\end{equation}

The kinetic equations written down above apply to the general case, where the system constantly fluctuates between $A$ and $B$. In the case of crystallization, however, once the nucleation has occurred and the crystal has formed it never returns to the metastable liquid (unless the conditions are changed). In such irreversible cases, the 
backward rate constant ($k_{A\leftarrow B}$) vanishes effectively and the evolution of 
concentrations is given by 
\begin{subequations}
 \begin{eqnarray}
 \label{eq:concentrationT}
c_A(t)&=&c_A(0)\exp \left ( -k_{A\rightarrow B} t\right ) \label{eq:concentrationTA}\\ 
c_B(t)&=&c_A(0)\left [ 1 - \exp \left ( -k_{A\rightarrow B} t\right ) \right ]. \label{eq:concentrationTB}
\end{eqnarray}
\end{subequations}
The average waiting time of eq.~(\ref{eq:meantime}) for the decay of the initial state then corresponds to the reaction time and is given by the inverse of the 
rate constant $k_{A\rightarrow B}$,
 \begin{equation}
 \label{eq:taureaction}
 \tau_{\rm rxn}=k_{A\rightarrow B}^{-1}.
 \end{equation}

In the following, we present a number of methods used in computer simulations to 
calculate stationary reaction rate constants. We restrict the 
analysis to the forward rate constant $k_{A\rightarrow B}$, since the backward rate 
$k_{A\leftarrow B}$ can be obtained in the same way. 

\subsection{A pedestrian approach}
\label{subsec:pedestrians}

The most straightforward way to determine rate constants is, of course, to allow the 
system to fluctuate between the reactant and product states in a long MD 
simulation. One then simply counts the number of jumps out of the initial state and determines 
the time spent there. The reaction rate constant is then calculated according to 
its definition as  
 \begin{equation}
 \label{eq:rateMD}
 k_{A\rightarrow B}^{\rm MD}=\frac{n_{A\rightarrow B}}{t_A},
\end{equation}
where $n_{A\rightarrow B}$ is the number of jumps out of the reactant state and 
$t_{A}$ is the total time spent in the reactant state.

The catch of this approach is in the time scale. The transition under study usually is not only 
a stochastic but also a rare event and the frequency of its occurrence depends on 
the height of the free energy barrier between the states. To transform from 
one state into the other, the system has to reach the top of the free energy 
barrier either by a single rather large statistical fluctuation or by a series 
of smaller fluctuations, which consequently drive the system towards the critical 
state on top of the free energy barrier. The probability of both scenarios decreases 
exponentially with the free energy 
difference between the initial and the critical states. 
Therefore, if the free energy barrier between the reacting states is relatively high, 
the system will spend a substantial amount of simulation time in one of the basins of 
attraction.  In this case, the collection of reasonable statistics for a reliable 
estimation of reaction rate constants requires an 
unrealistically long simulation run, a large part of which is spent in the stable states and 
does not contribute to the analysis. Hence, for a successful calculation of the rate constants in a long MD simulation run, 
the free energy barrier should be relatively low and the forward and backward reaction rate 
constants comparable. These requirements make the technique specifically unsuitable to study 
the decay of metastable states, in which the system virtually never returns to $A$ after reaching $B$.

\subsection{Mean first-passage time} 
\label{subsec:mfpt}

As discussed in section~\ref{subsec:cntKramers}, in 1940 Kramers \cite{kramers:1940, haenggi:1990} proposed to 
model chemical reactions as the process in which a Brownian particle escapes from a potential well
over a barrier. Therefore, the probability density of the particle position $q$ moving on the potential energy $U(q)$ 
evolves according to the Smoluchowski equation. By applying this formalism to the calculation of 
nucleation rates, the potential energy $U(q)$ is identified with the free energy determined as a function of the 
reaction coordinate $q$. 

In section~\ref{subsec:cntKramers}, we considered the mean time to reach an absorbing boundary located beyond 
the barrier for the first time. Instead, one can also consider the mean first-passage time $\tau(q^*)$ through a boundary located 
exactly at the top of the barrier, $q^*$. Then the reaction rate constant for an escape from $A$ is given by 
 \begin{equation}
 \label{eq:rateKramers}
 k_{A\rightarrow B}^{\rm MFPT}=\frac{1}{2\tau(q^*)}.
\end{equation}
The factor of $1/2$ arises from the fact that the probability to proceed to the final state from the transition state equals the probability 
to relax back into the initial state.

In general, the mean first-passage time $\tau(q)$ to reach a certain point $q$ starting from point $q_0$ is given by \cite{pontryagin:1933}:
 \begin{equation}
 \label{eq:mfpt}
\tau(q, q_0)=\int_{q_0}^{q} dy\frac{1}{D(y)} \exp \left[ \beta U \left(y\right) \right ] \int_{a}^y dz \exp \left[ -\beta U \left(z\right) \right ],
\end{equation}
where we have again assumed that there is a reflecting wall left of the potential well. Provided that the relaxation in the potential 
well is fast compared to the escape time, the mean first-passage time to reach a point $q$ near the barrier 
is the same for all initial points $q_0$ in the well region. Using the same approximations employed in section~\ref{subsec:cntKramers}
one can show that for values of $q$ near the top of the barrier the MFPT as a function of $q$ is given by the function 
\begin{equation}
\label{eq:mfptErf}
\tau(q)=\frac{\tau_J}{2}\left [ 1 + {\rm erf} \left (\left[q-q^*\right]c\right)\right ], 
\end{equation} 
where ${\rm erf}(z)=(2/\sqrt{\pi})\int_0^z\exp(-y^2)dy$ is the error function and $c=\sqrt{\beta \left| U^{''}(q^*)\right|/2}$ 
is the respective local curvature, which also determines the 
Zeldovich factor introduced in CNT via ${\rm Ze}=c/\sqrt{\pi}$ (see eq.~(\ref{equ:zeldovich})).

\begin{figure}[ht]
\begin{center}
\includegraphics[clip=,width=0.95\columnwidth]{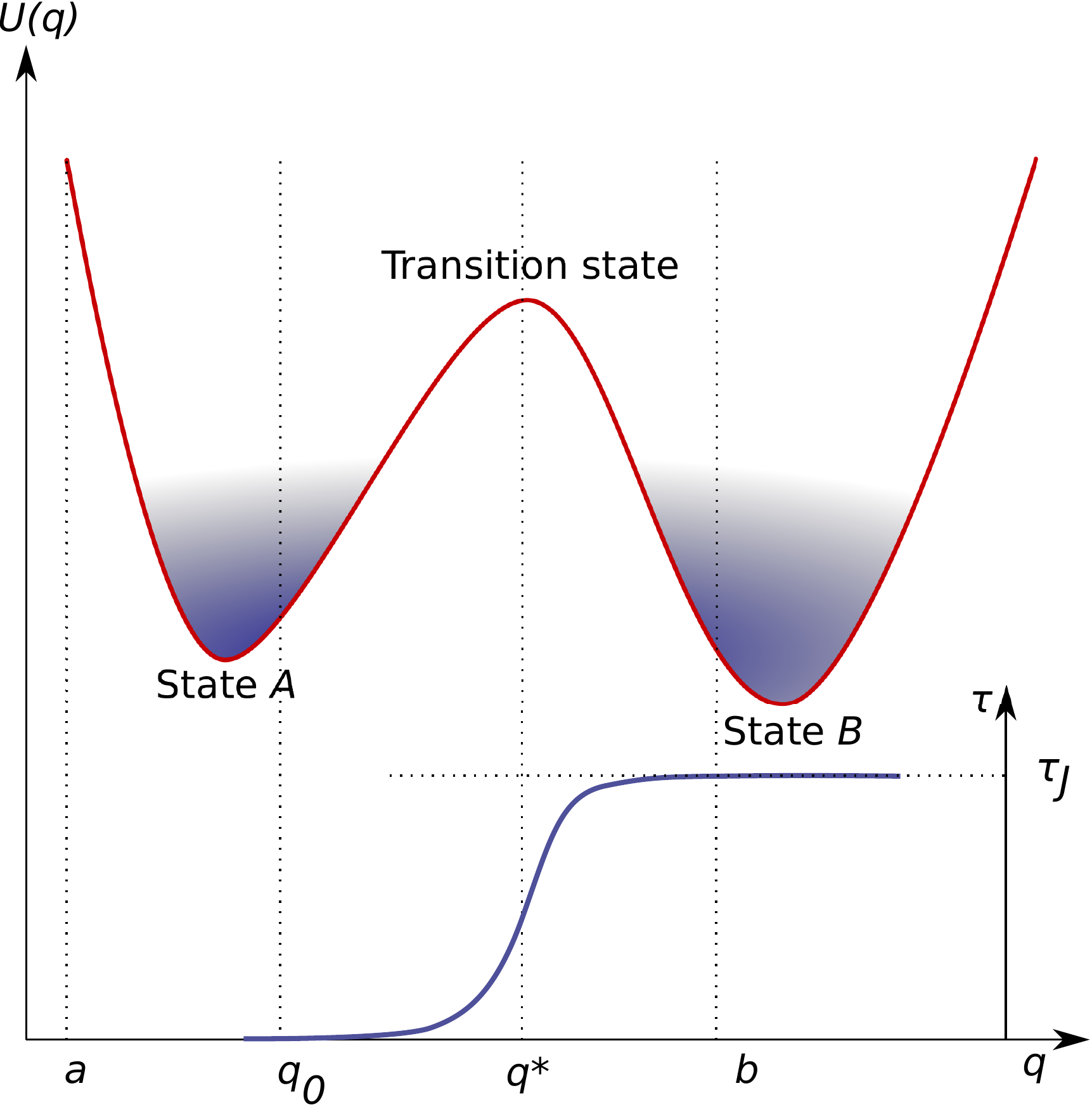}     
\caption{\label{fig:freeenergyKramers} Sketch of the Kramers approach \protect\cite{kramers:1940, haenggi:1990}. 
Diffusion of the system along a potential energy landscape $U(q)$ is described by the Fokker-Planck 
equation with a reflecting boundary fixed at $a$ and a moving absorbing boundary $b$. 
The reaction rate is inversely proportional to twice the mean time required to escape through $b=q^*$ when 
starting at $q_0$ (eq.~(\ref{eq:rateKramers})). For a sufficiently high barrier, the MFPT $\tau (q)$ 
can be approximated by an error function (eq.~(\ref{eq:mfptErf})).
}
\end{center}
\end{figure}

The above equation for the mean first-passage time can be used for the calculation of reaction 
rates provided that the free energy barrier is not too high, as suggested  by Wedekind, Strey and Reguera \cite{wedekind:2007}.
In this approach, one prepares a number of representative configurations in $A$ and starts MD 
simulations from each of them, following the systems until they cross the barrier and reach $B$. 
From these trajectories the mean first-passage time is then calculated as a function of the reaction coordinate $q$ 
and subsequent fitting of eq.~(\ref{eq:mfptErf}) to the simulation data with $\tau_J$, $q^*$, and $c$ as variational 
parameters yields the reaction rate constant as  
\begin{equation}
 \label{eq:nuclrate}
k_{A\rightarrow B}^{\rm MFPT}=\frac{1}{\tau_J}.
\end{equation} 
The situation considered in this approach is illustrated in fig.~\ref{fig:freeenergyKramers}.

This MFPT technique can be extended to reconstruct the free energy landscape underlying the 
reaction from the collection of nucleating trajectories \cite{wedekind:2008}. 
Unfortunately, the application of the analysis is quite limited by the prescription of the 
functional form of the MFPTs, which is valid only for substantially high and symmetrical 
nucleation barriers, where the simulation data are well described by eq.~(\ref{eq:mfptErf}). 
It has, however, been shown that the analysis is feasible even though more involved for transitions in which 
the time scales of nucleation and growth are not well separated \cite{mokshin:2014,shneidman:2014}. 
The occurrence of this case can be clearly seen in the MFPT computed from the simulation, which then does not 
display a plateau after passing the transition region but increases continuously with the rate of growth. 
Aside from that, in the illustrative example below, we demonstrate that also a poor choice of the reaction 
coordinate makes the application of this technique unreliable \cite{jungblut:2015}.      

Furthermore, there is the mean lifetime (MLT) \cite{baidakov:2011a} or direct observation 
method \cite{chkonia:2009}, which is based on the same formalism as the original MFPT analysis 
and employs the times that are measured for states well beyond the transition region. 
Similar to the approach introduced by Wedekind, Strey and Reguera, this method does not rely on the exact 
definition of the transition state \cite{talkner:1987} and allows a distinction between 
nucleation and growth regimes \cite{shneidman:2014}.
Assuming that the times to reach a particular state are Poisson distributed, simulated values 
for the probability to observe the transition to $B$ in a given time interval are 
fitted to the distribution 
\begin{equation}
\label{eq:distribution}
H(t|q\in B)=g t \exp (-h t), 
\end{equation}
using $g$ and $h$ as fitting parameters. The reaction rate constant $k_{A\rightarrow B}$ is 
then equal to $h$: 
\begin{equation}
\label{eq:rateMLT}
k_{A\rightarrow B}^{\rm MLT}=h. 
\end{equation}

The calculation of reaction rates with mean first-passage times and mean lifetimes 
is suitable for comparatively small nucleation barriers, such that at least one 
transition event can be observed on the time scale of reasonably long 
straightforward MD simulations. In comparison to a single simulation with 
multiple transitions, this approach does not require an equilibrium between the 
reactant and the product states and can be applied for calculation of the decay rates of 
metastable states. For most applications, however, the free energy barrier 
between the states will be rather high, and the transition events truly rare. One of 
the possibilities to deal with such cases is the Bennett-Chandler method, 
based on transition state theory.

\subsection{Transition state theory (TST)}
\label{subsec:tst}

\begin{figure}[ht]
\begin{center}
\includegraphics[clip=,width=0.95\columnwidth]{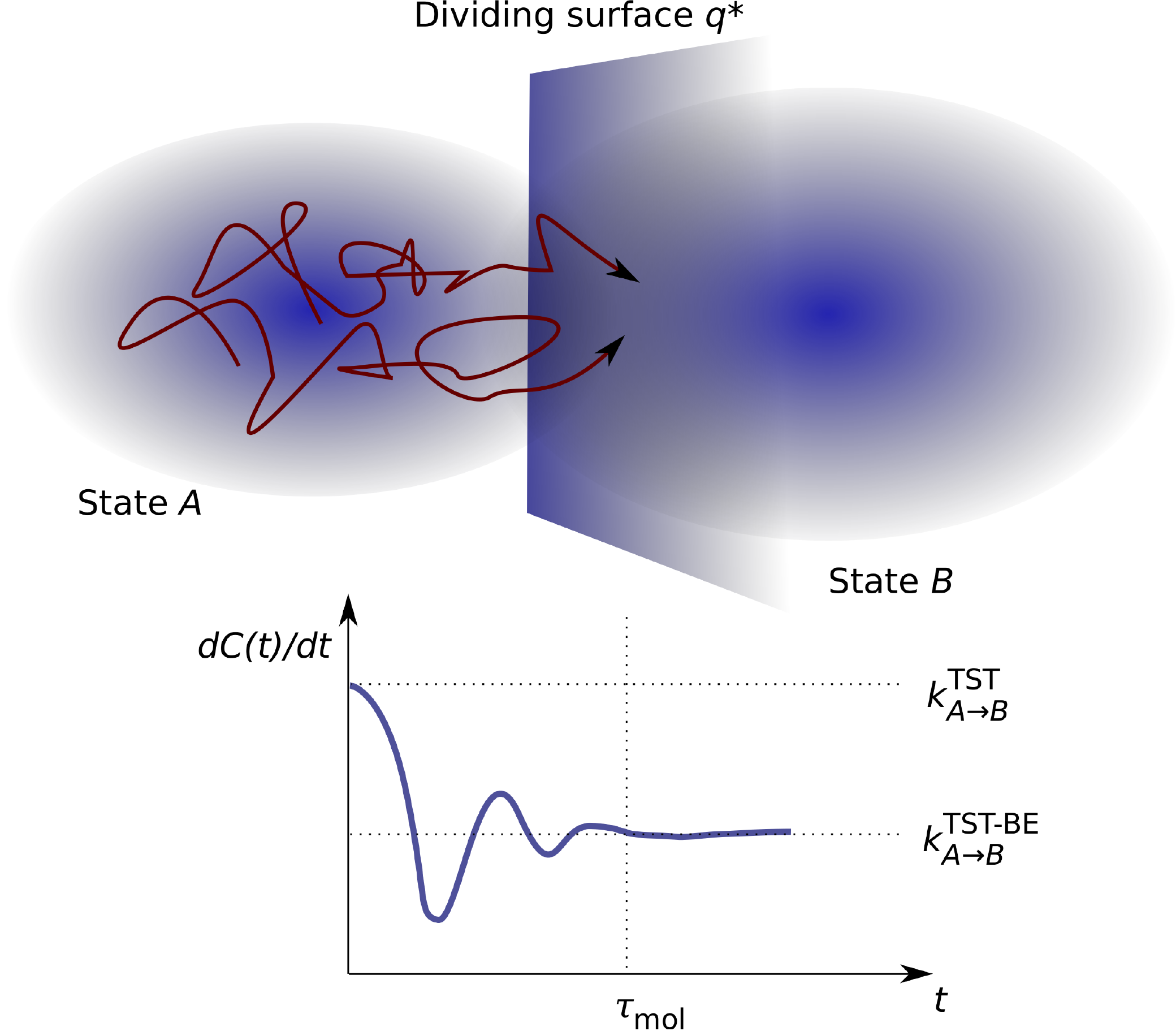}    
\caption{\label{fig:freeenergyTST} Schematic illustration of a reaction considered in TST. The 
initial state $A$ is separated from the product state $B$ by a dividing surface, 
which is placed perpendicularly to the reaction coordinate at the saddle point of the 
free energy landscape. The reaction rate constant is the product of the probability to 
find the system in the activated state, 
{\it i.e.}, at the dividing surface, and the mean velocity of all trajectories 
crossing $q^*$ in the direction of $B$. On short time scales, correlated recrossings of the 
dividing surface may occur, effectively reducing the flux out of $A$. 
The Bennett-Chandler approach corrects for such recrossings.     
}
\end{center}
\end{figure}

At about the same time as Kramers introduced the idea of viewing chemical 
reactions as diffusion processes, the concepts of the transition state theory 
\cite{eyring:1935, eyring:1938, wigner:1938,chandler:1978} (for a review on its evolution, see ref.~\cite{pollak:2005})   
were developed on the basis of thermodynamic rather than kinetic (in contrast to MFPT) considerations. 
Still, both theories are closely connected and in fact, 
as has been shown later, Kramers rate expression can be derived in the framework of TST \cite{pollak:1986}.
Central to the idea of TST is the concept of a transition state, a point on top of the barrier that 
has to be crossed during the transition. To be more precise, in TST the transition state, or activated state, is 
characterized as a saddle point in the 
free energy landscape, such that the curvature is negative only for one degree of 
freedom, which corresponds to the reaction coordinate. 
One can then place an imaginary surface perpendicular to the direction of the reaction 
coordinate (fig.~\ref{fig:freeenergyTST}) and identify all states that lie on this surface as belonging to the 
transition state ensemble. 
Considering only trajectories passing through this surface that started in the initial state, 
the reaction rate constant 
is then deduced from the ratio of the partition functions of the activated ($Z^*$) and 
initial ($Z_A$) states, multiplied by the average velocity, $\langle v^* \rangle$, with which the 
activated complex slides across the dividing surface: 
\begin{equation}
\label{eq:rateTST0}
k_{A\rightarrow B}^{\rm TST}=\kappa \langle v^* \rangle \frac{Z^*}{Z_A}. 
\end{equation}
The prefactor $\kappa$ was initially introduced by Eyring \cite{eyring:1935} to account for the possibility of 
recrossings, but was subsequently set to unity. The derivation of its precise form followed only decades 
later \cite{chandler:1978} and resulted in the Bennett-Chandler routine \cite{bennett:1977,chandler:1978} to 
correct the values of reaction rate constants computed with TST, which we discuss in detail below. 

In computer simulations \cite{Frenkel_Smit_book}, the ratio of the partition functions and 
the flux through the dividing surface are computed separately.    
In doing so, the average velocity of the activated complex is determined by 
preparing an ensemble of configurations at the top of the free energy barrier and considering the 
velocities only of those configurations that relax to the product state. 
The calculation of the ratio of the partition functions corresponds to the computation of the 
probability to find the system on the top of the free energy energy barrier sampled in the 
initial state ensemble. Using the free energy $F(q)$ computed as a function of the reaction coordinate $q$ (for instance,
with one of the methods mentioned in section~\ref{subsec:phaseDiagrams}), the TST rate can be written as 
\begin{equation}
\label{eq:rateTST1}
k_{A\rightarrow B}^{\rm TST}=\frac{1}{2} \langle |\dot{q}|\rangle_{q=q^*} \frac{e^{-\beta F(q^*)}}{\int_{-\infty}^{q^*}e^{-\beta F(q)}dq}, 
\end{equation}
where the conditional average $\langle \cdots\rangle_{q=q^*}$ is calculated separately in a system restricted to the 
top of the free energy barrier. 
The absolute value of $|\dot{q}|$ (with a factor of $1/2$) utilizes all configurations 
at the dividing surface instead of only those with a positive velocity. This replacement is 
justified if the reaction coordinate does not depend on 
the momenta of particles and the dynamics of the system is time reversible. The form of the 
equation closely resembles the CNT rate in eq.~(\ref{eq:cnt3}), giving another interpretation of the 
kinetic prefactor $K$.    

The fundamental problem of this TST approach is that it is based on the assumption that 
any trajectory crossing the dividing surface  coming from the reactant side will relax into the 
product state. In practice, however, fluctuations on the molecular time scale will induce some correlated 
recrossings of the dividing surface, causing TST to overestimate the reaction rate constant. 
Adjusting the dividing surface to minimize the number of recrossings, as proposed by the variational TST \cite{keck:1960}, 
reduces the deviation, but an accurate rate can be determined in this way only if a dividing 
surface exists that excludes recrossings completely. In most cases, such an optimum dividing surface 
does not exist \cite{mullen:2014}. It is, however, possible to account for 
the recrossings directly, as it is done in the Bennett-Chandler approach \cite{chandler:1978} explained 
in the next section, which combines the concepts of TST with information obtained from dynamical trajectories. 

\subsection{Bennett-Chandler method (TST-BC)}
\label{subsec:tstBE}

The Bennett-Chandler approach is a computational method that permits the exact (up to 
statistical errors) calculation of rate constants. In this method, one imagines to have 
a reaction coordinate $q$, which changes continuously as the transition proceeds from reactant 
to the product state. Based on this reaction coordinate, one defines characteristic functions for 
the reactant state, $A$, and the product state $B$,  
\begin{subequations}
 \begin{eqnarray}
 \label{eq:htheta}
h_A\left(q\right)&=&\theta\left(q^*-q\right)=\left \{ \begin{matrix} 1 & {\rm \qquad if\ } q < q^* \\ 0 & {\rm else} \end{matrix} \right . \label{eq:hthetaA}\\ 
h_B\left(q\right)&=&\theta\left(q-q^*\right)=1-h_A\left(q\right), \label{eq:hthetaB}
\end{eqnarray}
\end{subequations}
where $\theta(q)$ is the Heaviside theta function. The functions $h_A$ and $h_B$ indicate 
whether a configuration with reaction coordinate $q$ lies on the reactant or product side of the 
dividing surface located at $q=q*$. 
Using these characteristic functions, the time correlation function $C(t)$, {\it i.e.}, the conditional probability to find the 
system in state $B$ at time $t$ provided 
that it was in $A$ at time $0$, can be expressed as 
\begin{equation}
\label{eq:coft0}
C(t)=\frac{\langle h_A\left[ q(0) \right ] h_B\left[ q(t) \right ] \rangle}{\langle h_A \rangle}.  
\end{equation}
Here, $\langle h_A \rangle = \langle c_A \rangle /(\langle c_A \rangle +\langle c_B \rangle)$ is 
the fraction of equilibrium concentration in $A$ as introduced in eq.~(\ref{eq:ratioC}). 
If a system is prepared such that $c_B=0$ initially, the time evolution of the product concentration
can be expressed in terms of the correlation function,
\begin{equation}
\label{eq:concentrationBcoft0}
c_B(t)=(\langle c_A \rangle +\langle c_B \rangle)C(t). 
\end{equation}
As mentioned above, original TST assumes that all trajectories connecting $A$ and $B$ cross the 
dividing surface only once. For time scales longer than the typical time scale $\tau_{\rm mol}$
of molecular fluctuations, one expects the populations to evolve according to the 
rate equations~(\ref{eq:concentrationTT}). Thus, in the time interval $\tau_{\rm mol} < t \ll \tau_{\rm rxn}$, 
the comparison of eqs.~(\ref{eq:concentrationTTB}) and~(\ref{eq:concentrationBcoft0}) yields   
\begin{equation}
\label{eq:coft00}
C(t)=\langle h_B \rangle \left [ 1-\exp \left ( -\left [k_{A\rightarrow B}+k_{A\leftarrow B} \right ] t\right )\right ]. 
\end{equation}
For times $t$ that are larger than $\tau_{\rm mol}$ but still much smaller than the reaction time 
$\tau_{\rm rxn}=(k_{A\rightarrow B}+k_{A\leftarrow B})^{-1}$, this expression reduces to 
\begin{equation}
\label{eq:coft000}
C(t)\approx k_{A\rightarrow B} t.  
\end{equation}
Thus, for times larger than the time scale of molecular fluctuations, when all correlated recrossings have occurred, 
the time derivative $\dot{C}(t)$ is equal to the reaction rate constant, 
\begin{equation}
\label{eq:rateTSTBC}
k_{A\rightarrow B}^{\rm TST-BC}= \left . \frac{dC(t)}{dt} \right |_{t>\tau_{\rm mol}} .
\end{equation}
Using the definition given in eq.~(\ref{eq:coft0}) and exploiting the time translation invariance of the correlation function, the time 
derivative of the correlation function $C(t)$ can be expressed as \cite{chandler:1978} 
\begin{equation}
\label{eq:dtcoft000}
\frac{dC(t)}{dt}=\frac{\langle \dot{q}(0)\delta \left [q(0)-q^*\right ] h_B\left [q(t)\right ] \rangle}{\langle h_A \rangle}. 
\end{equation}

If recrossings are neglected, the passages through the dividing surface become uncorrelated in 
time and we can take the limit $t \rightarrow 0$, 
in which the difference in $h_B\left [q(t)\right ]=\theta \left [ q(t)-q^* \right ]$ is replaced by the 
derivative $\theta \left [ \dot{q}(0)\right ]$, thus, eliminating the time dependence:     
\begin{equation}
\label{eq:rateTST2}
k_{A\rightarrow B}^{\rm TST}=\frac{\langle \dot{q}(0)\delta \left [q(0)-q^*\right ] \theta \left [\dot{q}(0)\right ] \rangle}{\langle \theta(q^*-q) \rangle}.
\end{equation}
In this approximation, the reaction rate constant can be determined using only equilibrium averages without the need to 
calculate dynamical trajectories. 
Dividing and multiplying the above formula with the factor $\langle \delta (q^*-q) \rangle$ yields
\begin{equation}
\label{eq:rateTST3}
k_{A\rightarrow B}^{\rm TST}=\frac{\langle \dot{q}(0)\delta \left [q(0)-q^*\right ] \theta\left [\dot{q}(0)\right ] \rangle}{\langle \delta (q^*-q) \rangle} \times \frac{\langle \delta (q^*-q) \rangle}{\langle \theta(q^*-q) \rangle}. 
\end{equation}
Thus, we recover the TST expression of eq.~(\ref{eq:rateTST1}), since the first factor on the right hand side corresponds to the 
expectation value of the positive velocity and the second term is the conditional probability to find 
the system at $q^*$. The deviation of the TST estimate from the exact value of the rate constant, for instance computed using the Bennett-Chandler 
approach, is described by the transmission coefficient 
\begin{equation}
\label{eq:kappa}
\kappa=\frac{k_{A\rightarrow B}^{\rm TST-BC}}{k_{A\rightarrow B}^{\rm TST}}.
\end{equation}
Since TST always overestimates the reaction rate constant, the transmission coefficient is a number between 0 and 1. 
The more correlated recordings occur, the smaller is the transmission coefficient $\kappa$, which roughly measures 
the fraction of crossings of the dividing surface that lead to successful transitions from reactants to products. 

Transition state theory and its more sophisticated variants such as the Bennett-Chandler approach rely on 
the knowledge of a good reaction coordinate capable of describing the progress of the reaction in consideration. 
(Poor choices of the reaction coordinate lead to a transmission coefficient that is statistically
indistinguishable from zero.)
For complex systems, however, such a reaction coordinate often cannot be identified beforehand. 
Then, the mechanism and kinetics of the reactions can be studied with transition path sampling, a 
computational methodology based on the statistical description of ensembles of trajectories which does 
not require a priori knowledge of the reaction mechanism and the definition of a valid reaction coordinate. 
The transition path sampling method is the subject of the next section. 

\subsection{Transition path sampling}
\label{subsec:tps}

The central idea of transition path sampling (TPS)
\cite{dellago:1998,bolhuis:1998,dellago:2000,dellago:2002,bolhuis:2002,dellago:2009}  
is to define the transition path ensemble (TPE), the set all trajectories connecting the initial and the final states.
To every trajectory in this ensemble one assigns a statistical weight, which depends 
on the underlying dynamics as well as on the distribution of states from which the trajectories 
can start. The trajectories of the transition path ensemble are then sampled using 
a Monte Carlo procedure that generates a Markov chain of pathways, occurring according to the 
statistical weight they have in the transition path ensemble. The harvested trajectories can then 
be analyzed to reveal the microscopic mechanism of the transition and identify a good 
reaction coordinate that captures the essential mechanistic features. The transition path 
sampling framework allows also the calculation of reaction rate constants without the 
need to define a valid reaction coordinate. A detailed account on the basic principles of 
transition path sampling as well as practical issues arising in its implementation can be found, {\it e. g.}, 
in ref.~\cite{dellago:2002}. In the following, we will focus on the use of transition path 
sampling for the calculation of reaction rate constants and, in particular, nucleation rates
for crystallization. 

In TPS, every trajectory $x(t)$ is composed of a sequence of ${\mathcal N}+1$ microscopic states, 
or time slices,  $x_{\tau}$, and has a time length $t = {\mathcal N}\Delta t$. Here, $x_\tau$ indicates the 
state of the system at time $\tau$ along the path, specified either by the position of all particles in the 
system, $x_{\tau}=\{r\}$, or by their positions and momenta, $x_{\tau}=\{r,p\}$, and $\Delta t$ is the time interval 
between subsequent states. The dynamics underlying the 
time evolution of the pathways is assumed to be Markovian, such that the probability of a given 
trajectory is  
\begin{equation}
\label{eq:pathprobabilityTPS}
{\mathcal P}\left[x(t)\right]= \rho(x_0)\prod_{i=0}^{t/\Delta t-1}p(x_{i\Delta t}\rightarrow x_{(i+1)\Delta t}), 
\end{equation}
where $\rho(x_0)$ is the distribution of the initial conditions and 
$p(x_{\tau}\rightarrow x_{\tau+\Delta t})$ are the single time step transition 
probabilities. Both of these probabilities are normalized and depend on the specific details of the system 
that is considered. Note that most types of dynamics usually considered in computer simulations, including Newtonian, 
Langevin or Monte Carlo dynamics, are Markovian such that they can easily be treated 
in the TPS framework. 
 
\begin{figure}[h]
\begin{center}
\includegraphics[clip=,width=0.95\columnwidth]{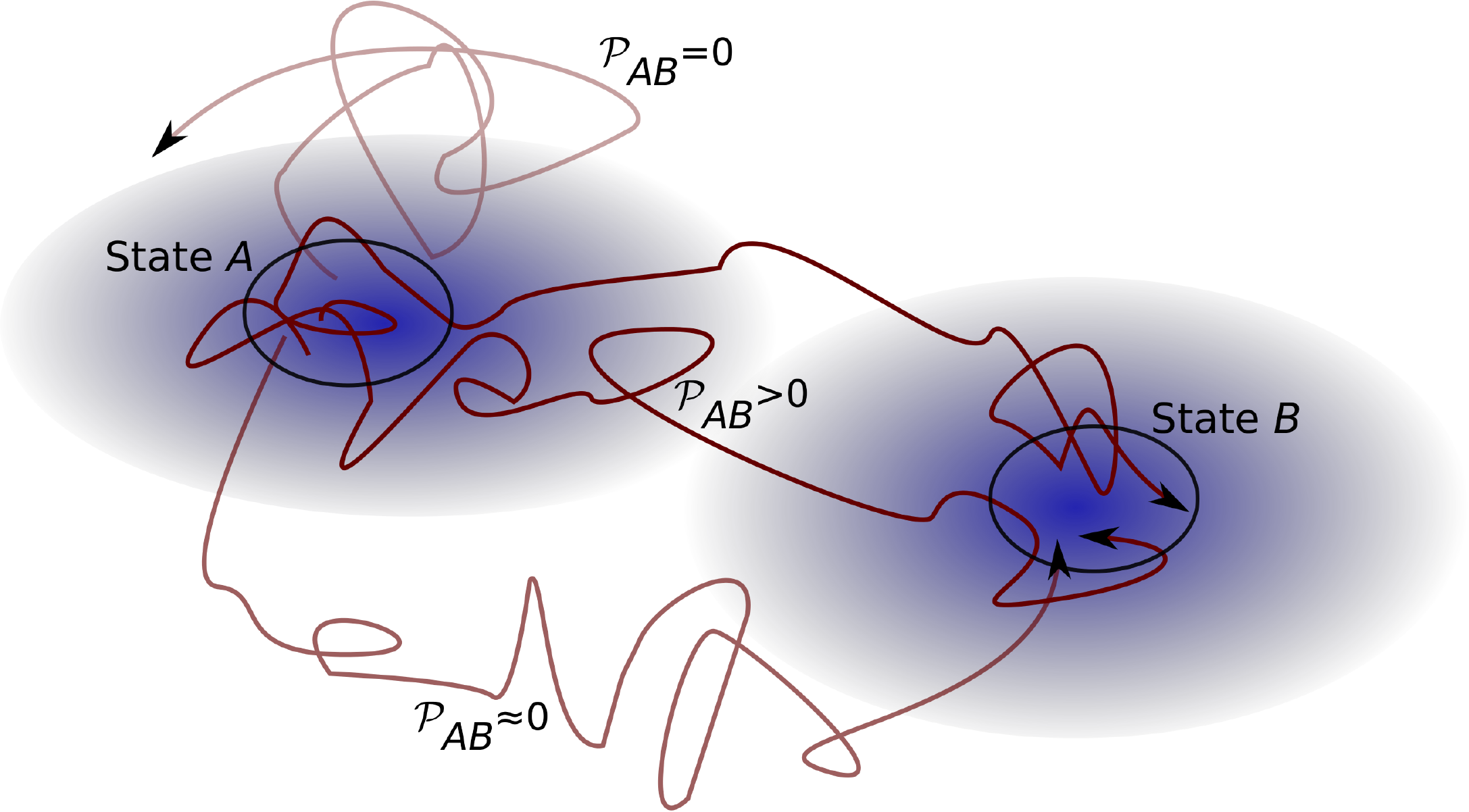}     
\caption{\label{fig:pathsTPS} Stable states $A$ and $B$ and various trajectories considered in TPS. 
The borders of the initial and the final states of the reaction are defined by means of a suitable order parameter which differentiates 
between them, but no further assumptions are made about the reaction coordinate. The probability ${\mathcal P}_{AB}$ to 
observe a given pathway  connecting $A$ and $B$ is the product of all single time step 
transition probabilities (eq.~(\ref{eq:pathprobabilityABTPS})) and the probability of the initial state. 
Trajectories connecting $A$ and $B$ may have a finite statistical weight while all other trajectories
are assigned a weight of zero in the TPE.}
\end{center}
\end{figure}

In contrast to the methods for calculation of reaction rates discussed previously, TPS does not follow the 
progress of a transition along a predefined reaction coordinate 
but rather samples trajectories connecting spatially separated regions $A$ and $B$, which are identified by means 
of a suitable order parameter. These regions do not need to be adjacent like the regions considered in TST and related 
approaches. Accordingly, we redefine the auxiliary functions (see eq.~(\ref{eq:htheta})), which 
indicate whether a state $x$ belongs to regions $A$ and $B$, respectively:
\begin{subequations}
 \begin{eqnarray}
 \label{eq:hthetaTPS}
h_A(x)=\left \{ \begin{matrix} 1 & {\rm \qquad if\ } x\in A \\ 0 & {\rm else} \end{matrix} \right .\\ 
h_B(x)=\left \{ \begin{matrix} 1 & {\rm \qquad if\ } x\in B \\ 0 & {\rm else} \end{matrix} \right .
\end{eqnarray}
\end{subequations}
Using these functions, the probability of a trajectory $x(t)$ in the TPE is given by (see also fig.~\ref{fig:pathsTPS})  
\begin{equation}
\label{eq:pathprobabilityABTPS}
{\mathcal P}_{AB}\left[x(t)\right]= \frac{h_A(x_0){\mathcal P}\left[x(t)\right]h_B(x_t)}{{\mathcal Z}_{AB}}, 
\end{equation}
where the partition function ${\mathcal Z}_{AB}$ of the paths connecting the initial and 
the final states of the reaction is defined as 
\begin{equation}
\label{eq:partitionfunctionTPS}
{\mathcal Z}_{AB}(t)= \int {\mathcal D}x(t) h_A(x_0){\mathcal P}\left[x(t)\right]h_B(x_t)  
\end{equation}
and the integration
\begin{equation}
\label{eq:pathintegralTPS}
 \int {\mathcal D}x(t)= \int \dots \int dx_0dx_{\Delta t}dx_{2\Delta t} \dots dx_{t}
\end{equation}
is performed over all states along a trajectory. Note that all pathways not starting in $A$ and ending in $B$ are
assigned a statistical weight of zero in the TPE. Accordingly, the sampling of the TPE yields only reactive trajectories,
{\it i.e.}, trajectories connecting the two regions. 

The time correlation function defined in eq.~(\ref{eq:coft0}), which contains all the information 
needed to determine the reaction rate constant, can be rewritten 
in the TPS framework as the ratio of two path ensemble averages: 
\begin{equation}
\label{eq:coftTPS0}
C(t)=\frac{\int {\mathcal D}x(t){\mathcal P}\left[x(t)\right]h_A(x_0)h_B(x_t)}{\int {\mathcal D}x(t){\mathcal P}\left[x(t)\right]h_A(x_0)}.
\end{equation}
Hence, the time correlation function $C(t)$ can be viewed as the fraction of all paths of length $t$
starting in the initial state that make it to the final state of the reaction. Note that the denominator 
in the above equation does not depend on time since, due to the 
normalization of the single time step probabilities used in the dynamical weight 
${\mathcal P}\left[x(t)\right]$, it effectively reads $\int dx_0 h_A(x_0) \rho(x_0)$, 
which is the equilibrium population in $A$, $\langle h_A \rangle$. In principle, $C(t)$ can  
be calculated by starting trajectories of length $t$ in $A$ and counting how many of them reach $B$. 
In most interesting cases, however, the transition from $A$ to $B$ is a rare event and 
this straightforward approach is impractical. Instead, we first transform the 
expression for the time correlation function $C(t)$ into a form that is more convenient for 
computations. In order to do so, we introduce an order parameter $\lambda(x_{\tau})$ 
that can describe the product state $B=\{x_{\tau}: \lambda_{\rm min}^B<\lambda(x_{\tau})<\lambda_{\rm max}^B\}$ 
or the entire configuration space, $-\infty <\lambda(x_{\tau})< \infty$, including $A$. 
Substitution of the indicator function $h_B(x_{\tau})$ into eq.~(\ref{eq:coftTPS0}) and change of the 
integration order lead to 
\begin{eqnarray}
\label{eq:coftTPS1}
C(t)&=&\frac{1}{\langle h_A \rangle} \int {\mathcal D} x(t){\mathcal P}\left[ x(t)\right] h_A(x_0) \int_{\lambda_{\rm min}}^{\lambda_{\rm max}}d\lambda \delta \left [\lambda - \lambda(x_t) \right ] \nonumber\\
 &=& \int_{\lambda_{\rm min}}^{\lambda_{\rm max}} d\lambda \langle \delta \left[ \lambda - \lambda (x_t) \right] \rangle_A = \int_{\lambda_{\rm min}}^{\lambda_{\rm max}}d\lambda P_A(\lambda, t),
\end{eqnarray}
where $\langle \cdots \rangle_A$ denotes the path average over all trajectories originating in $A$. 
The function $P_A(\lambda, t)$ is the probability that at time $t$ a path 
has reached $\lambda$, provided it started in $A$ at time 0. The transition in question is a rare 
event and only a few trajectories of the whole ensemble will eventually end in $B$. Thus, in 
practice, the calculation of the probability distribution is performed by umbrella sampling 
method \cite{torrie:1974}, appropriately reformulated for transition pathways. The 
entire range of order parameter $\lambda(x_{\tau})$ is divided into overlapping windows $W_i$ 
with $\lambda_i^{\rm min}<\lambda(x_{\tau})<\lambda_i^{\rm max}$ and the weighted probability to reach 
a particular window is calculated as 
\begin{eqnarray}
\label{eq:pawi}
P_{AW_{i}}(\lambda, t)&=&\frac{\int {\mathcal D}x(t){\mathcal P}\left[x(t)\right]h_A(x_0)h_{W_i}(x_t)\delta \left[ \lambda - \lambda (x_t) \right]}{\int {\mathcal D}x(t){\mathcal P}\left[x(t)\right]h_A(x_0)h_{W_i}(x_t)} \nonumber\\
 &=&  \langle \delta \left[ \lambda - \lambda (x_t) \right] \rangle_{AW_{i}} .
\end{eqnarray}
The distributions in adjacent windows are matched and normalized to yield $P_A(\lambda, t)$, 
of which the section belonging to $B$ is used to calculate the time correlation function $C(t)$ via 
eq.~(\ref{eq:coftTPS1}).  

 The reaction rate constant is then determined, in a way similar to the TST-BC method (eq.~(\ref{eq:rateTSTBC})), as 
the time derivative of the correlation function at time scales larger than $\tau_{\rm mol}$. 
Since, however, the technique of evaluation of $C(t)$ for a fixed time $t$ specified above is rather involved to 
be applied to all times in order to numerically derive this function, a 
computationally more convenient modification \cite{dellago:1998,dellago:1999} can be employed 
for the calculation of time derivative. The variation consists in the shift of the time 
dependence into a prefactor, which is computed in a path sampling simulation with a fixed length $t$:  
\begin{equation}
\label{eq:coftTPS2}
C(t)=\frac{\langle h_{B}(t)\rangle _{AB}}{\langle h_{B}(t')\rangle _{AB}}C(t'). 
\end{equation}
In this way, the computationally expensive umbrella path sampling simulation has to be performed only for a 
single time $t'$, which can be much shorter than $t$. The reaction rate constant then reads  
\begin{equation}
\label{eq:rateTPS}
k_{A\rightarrow B}^{\rm TPS}=\left .\frac{\langle \dot{h}_{B}(t)\rangle _{AB}}{\langle h_{B}(t')\rangle _{AB}}C(t')\right |_{t>\tau_{\rm mol}}. 
\end{equation}
This method for the calculation of reaction rate constants in the framework of TPS is computationally
demanding. An improvement in efficiency can be obtained with the transition interface sampling method
explained in the next section.

\subsection{Transition interface sampling (TIS)}
\label{subsec:tis}

\begin{figure}[ht]
\begin{center}
\includegraphics[clip=,width=0.95\columnwidth]{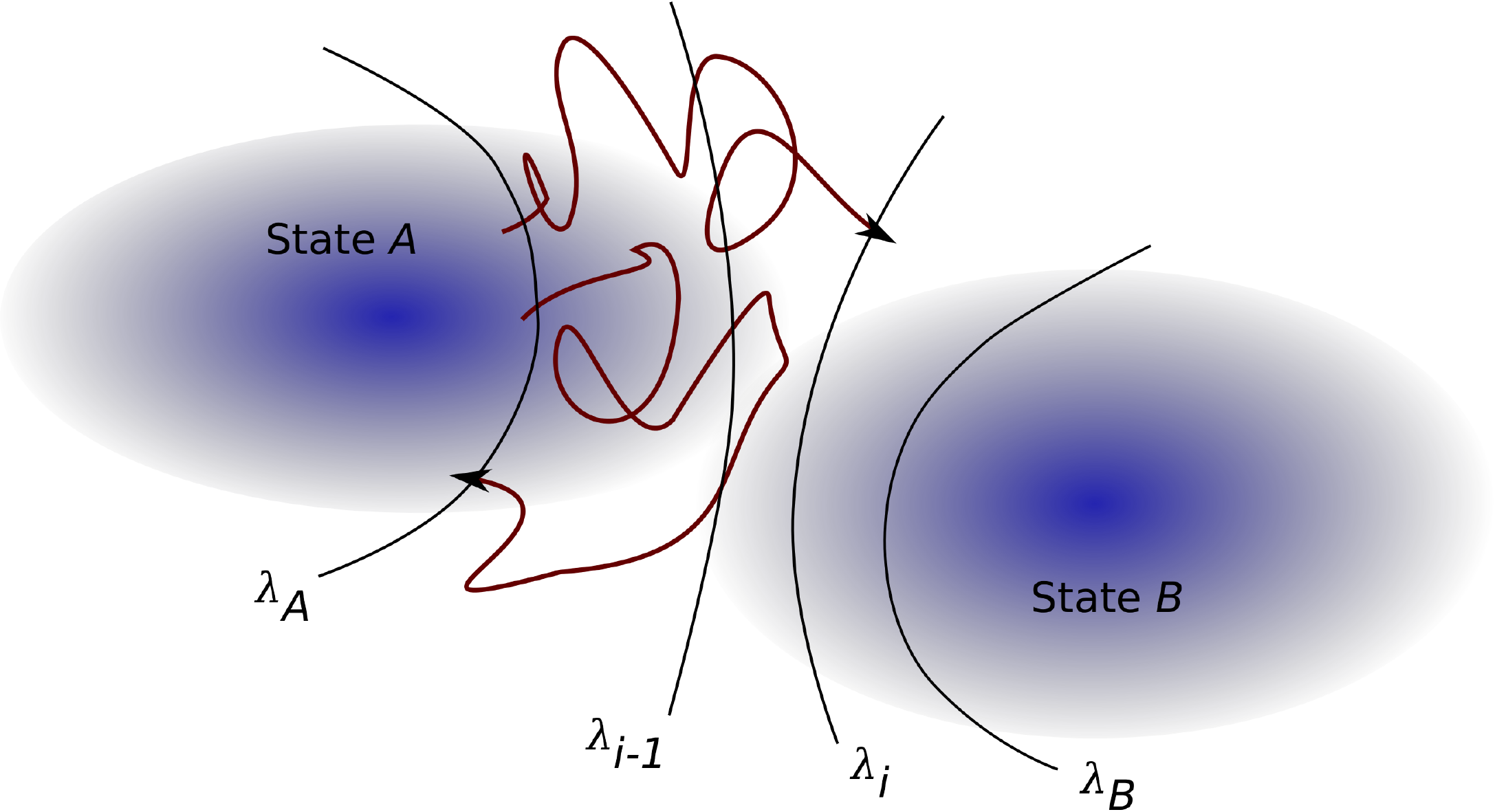}     
\caption{\label{fig:pathsTIS} Sketch of the TIS paths. The region between the initial and the 
final states is subdivided into non-overlapping windows by placing multidimensional interfaces 
at increasing values of a suitable order parameter $\lambda$. The first interface, corresponding to $\lambda_A$, and the 
last interface, corresponding to $\lambda_B$, are the borders of the reactant and the product states, respectively. All  
paths sampled in one window cross its left interface ($\lambda_{i-1}$) coming directly from $A$. The 
integration of a trajectory is stopped when the system reaches either the next interface ($\lambda_{i}$) or the 
border of the initial state.}
\end{center}
\end{figure}

In the transition interface sampling method \cite{vanerp:2003,vanerp:2005,vanerp:2007,bolhuis:2008}, the 
requirement of a fixed length for trajectories is relaxed and the integration of pathways stops 
when the system reaches certain points in phase space. This is achieved by dividing 
the phase space between the initial and the final state of the reaction into
windows, similar to those used in TPS for the calculation of the time correlation function in eq.~(\ref{eq:pawi}) 
but not overlapping. 
The multidimensional interfaces that limit the windows, and give the method its name, are placed at fixed values 
of the order parameter, $\lambda(x)=\lambda_i$, such that the borders of the reactant state $A$ and the product state $B$ 
coincide with the first $\lambda_A=\lambda_0$ and the last $\lambda_B=\lambda_n$ interfaces, respectively.  
In every window, the sampling is restricted to paths coming from the border of the initial state and crossing the 
left interface ({\it i.e.}, the interface closer to the reactant) of the window (the concept is also sketched in fig.~\ref{fig:pathsTIS}).   

In the derivation of the TIS formalism, the reaction rate constant is defined as the positive steady flux 
through the border of the product state $B$, $\lambda_n$. One can then expand the region of 
the final state by adding adjacent windows successively, such that, in the first step, the product state is defined 
by $\lambda(x_{\tau})>\lambda_{n-1}$ and the flux is reduced by a factor that accounts for the number of paths that 
pass the $\lambda_{n-1}$ interface but return to $\lambda_{0}$ before reaching $\lambda_{n}$. 
The procedure is repeated for every window, and finally the interface $\lambda_1$ is placed so close to the 
initial state that the flux through it can be easily computed in a straightforward MD simulation carried out
in the reactant state.     
The reaction rate constant is then given by the product of the flux out of the initial state, $\Phi_{1,0}$, and 
the probability to end in the product state, $P_A(\lambda_n|\lambda_1)$, under the condition that the trajectories reaching $\lambda_n$ originate in the reactant state and cross the first interface $\lambda_1$: 
\begin{equation}
\label{eq:rateTIS}
k_{A\rightarrow B}^{\rm TIS}=\Phi_{1,0}P_A(\lambda_n|\lambda_1).  
\end{equation}
The last term is the product of all probabilities to reach the next interface after crossing the previous one, 
starting with $\lambda_1$, for which the reactive flux is known, and going through the whole phase space 
between the reacting states to the border of the product state, 
\begin{equation}
\label{eq:paTIS}
P_A(\lambda_n|\lambda_1)=\prod_{i=1}^{n-1}P_A(\lambda_{i+1}|\lambda_i). 
\end{equation}
In practice, not all probabilities to reach the next interface have to be computed, since, as soon as the system 
has crossed the transition state region
between the states (wherever it should lie), all paths reaching a window will continue towards the product 
state $B$ yielding a progression probability of unity. At this stage, the conditional probability 
$P_A(\lambda_n|\lambda_1)$ saturates to a plateau and the sampling of the remaining windows can be omitted. 

For the analysis of the transition mechanism, one recovers the ensemble of transition paths, 
similar to the one of TPS, by letting the paths from the last simulated window evolve until they reach $B$. 
  
For reactions that occur diffusively, transition pathways may become very long hampering their sampling within a TIS simulation. 
To overcome this problem, the partial path transition interface sampling (PPTIS) method was developed \cite{moroni:2004,moroni:2005a}.
This method, which is applicable if segments of pathways in different windows are statistically independent from each other, 
also allows to simultaneously sample the free energy landscape underlying the reaction.  
 
The sampling of paths in TPS and TIS requires that the reacting system is in microscopic equilibrium, {\it i.e.}, 
the dynamics of the trajectories is time reversible. Although the framework of TPS has been reformulated to 
address systems out of equilibrium \cite{geissler:2004}, another method, namely the forward flux sampling, 
was purposely designed to investigate such systems.  

\subsection{Forward flux sampling} 
\label{subsec:ffs}

The forward flux sampling (FFS) technique \cite{allen:2005,allen:2006,borrero:2007,allen:2009,escobedo:2009} deals with 
stochastic non-equilibrium systems, without requiring  knowledge of the phase space density. 
The formalism for the calculation of the rate constant is 
identical to TIS (eq.~(\ref{eq:rateTIS})), but the reaction paths are produced in a different way. 

\begin{figure}[ht]
\begin{center}
\includegraphics[clip=,width=0.95\columnwidth]{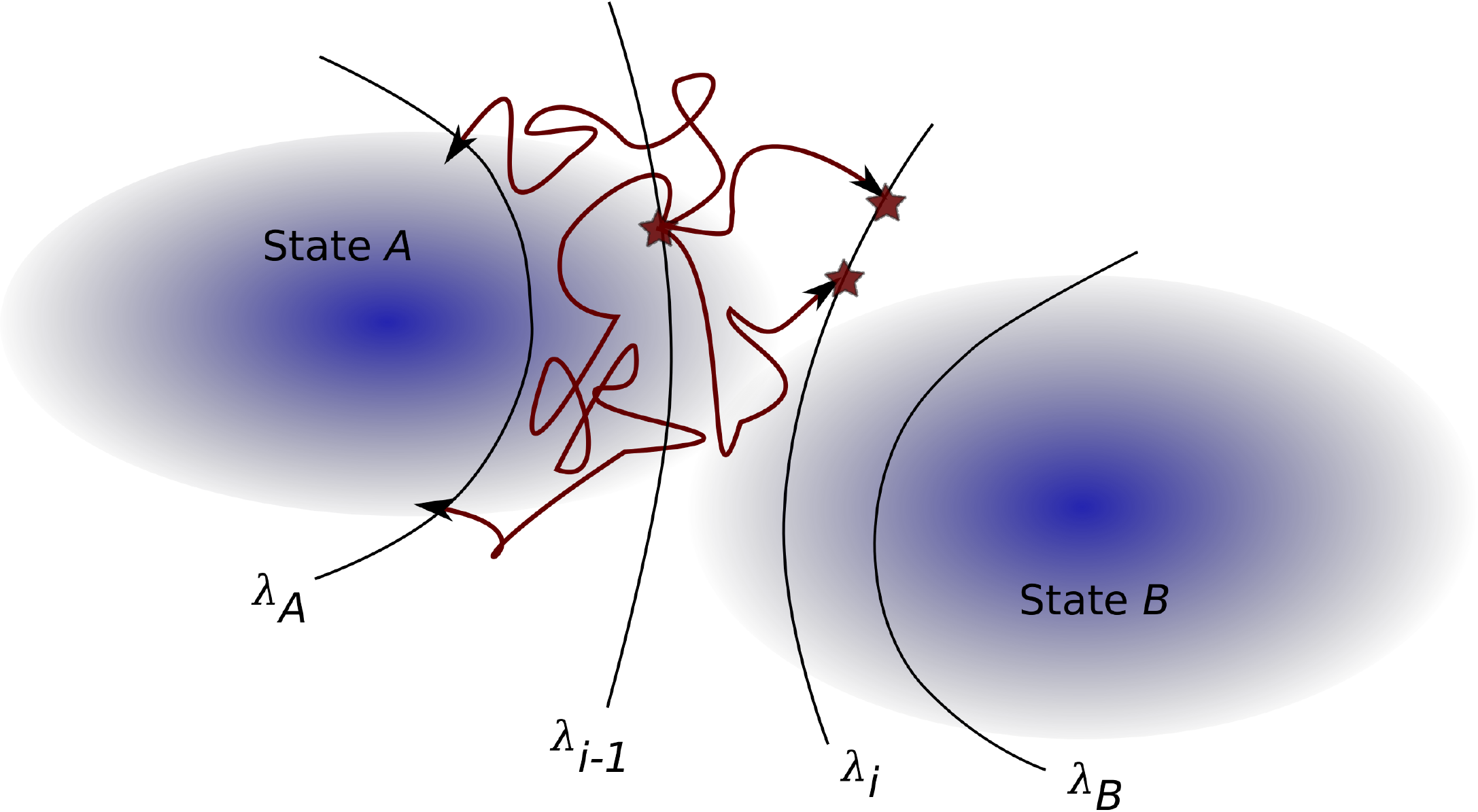}    
\caption{\label{fig:pathsFFS} Sketch of FFS paths. In a window, defined similar to TIS, the 
trajectories start from configurations (indicated by stars) at the left interface ($\lambda_{i-1}$) and end either in 
the initial state or at the next interface ($\lambda_{i}$). Endpoints of the pathways that terminate at the right 
border are stored to be used in the simulations of the next window. }
\end{center}
\end{figure}

Given a set of non-overlapping interfaces, identical to those used in TIS, the flux out of the initial 
state is calculated in a straightforward MD simulation. Along the way, configurations at the 
first interface are stored in order to be used as starting points for trajectories of the first 
window. Then, in the direct FFS \cite{allen:2005,allen:2006} illustrated in fig.~\ref{fig:pathsFFS}, 
a number of trajectories is started 
from randomly selected configurations and the fraction of pathways evolving to the next interface 
instead of returning to the initial state 
contributes the conditional probabilities $P_A(\lambda_{i+1}|\lambda_i)$ used in eq.~(\ref{eq:paTIS}). 
Configurations at the endpoints of trajectories that reached the next interface are again collected as starting points 
for the probability calculation 
in the next window. At the end, continuous transition paths are constructed by joining the pieces 
which belong to the neighboring windows.   

There are also alternative FFS sampling schemes, like branched growth or the Rosenbluth method \cite{allen:2006}, 
that first generate complete (branched) paths by starting trial runs from the configurations at interfaces 
and subsequently sample the conditional probabilities $P_A(\lambda_n|\lambda_1)$ from averages over these paths. 
Unfortunately, the performance of FFS strongly depends on the proper definition of the first-interface 
ensemble and an insufficient sampling at this stage might propagate (without evident indications) 
through the whole analysis even in very simple systems as has been demonstrated in a recent review 
by van Erp \cite{vanerp:2012}. In contrast, TPS and TIS 
methods do not suffer from this shortcoming, since they allow for backward shooting and the trajectories 
are able to relax in both directions.

\subsection{Evolution of methods}
\label{subsec:otherMethods}

The methods discussed above were presented roughly sorted according to their increasing computational costs, 
but at the same time their order mirrors, to some degree, the historical development of the field. 
The more recent techniques are more accurate and reliable but also rather 
involved and only applicable due to the availability of powerful computers.
However, the performance of these methods is still not sufficient to enable the analysis of 
rare transitions in arbitrary systems, since the demand for more realistic representation of the underlying potential energy 
surface evolves 
along with the increase of the complexity of the methods. Thus, even with the vast computing power
available today, one has to choose methods that find a compromise between speed and accuracy,  
and provide as much information on the details of the system as possible. 
Hence, although the application of transition path sampling techniques yields more 
information about the reaction in question, less involved techniques based on transition state 
theory or on the Kramers approach have their virtue in being relatively fast and, in some cases, 
as precise as TPS in the calculation of reaction rate constants. Furthermore, the development and 
improvement of TPS related techniques is still going on \cite{bolhuis:2015}.  
Currently, one can observe two routes followed in the development of new methods. While one 
is concerned with the increase of the sampling efficiency, the other one is directed towards including 
non-standard types of dynamics and aims to release the equilibrium and steady state assumptions.    

Along these lines, TPS has been expanded to include parallel tempering of pathways \cite{vlugt:2001} and permutation shooting  \cite{mullen:2015},
explore reactions on diffusive free energy barriers \cite{bolhuis:2003,miller:2007,gruenwald:2008,chopra:2008}, 
perform aimless shooting \cite{peters:2006,peters:2007}, and sample pathways 
connecting multiple stable states \cite{rogal:2008}. In addition, TPS was combined with the Wang-Landau algorithm 
for simultaneous sampling of different temperatures \cite{borrero:2010} and biased dynamics to steer the 
sampled trajectories towards the transition \cite {guttenberg:2012}. 
The efficiency of TIS has been increased by optimized interface placement \cite{borrero:2011}
and using several path replica exchange algorithms \cite{vanerp:2007,bolhuis:2008,du:2013,swenson:2014}, 
which additionally facilitate sampling of multiple states and reaction channels. 
FFS has been extended to apply to non-stationary \cite{becker:2012,becker:2012a} states, 
simultaneously compute stationary probability distributions and rates in non-equilibrium 
systems \cite{valeriani:2007}, and optimized to reduce the computational cost by adaptive 
choice of the number of trial runs at each interface as well as of the position of interfaces \cite{borrero:2008}. 

Of course, there are many more methods for the estimation of reaction rate constants including
milestoning \cite{faradjian:2004,west:2007}, the weighted ensemble method \cite{huber:1996,zhangB:2007,zhangB:2010}, 
non-equilibrium umbrella sampling \cite{warmflash:2007}, discrete path sampling \cite{wales:2002,wales:2005}, or 
barrier method \cite{adams:2010}, just to mention a few of them. There are also methods which were developed to 
study driven systems \cite{williams:2013} and non-steady states \cite{berryman:2010}.  

While many methods are available to study rare transitions in complex systems, the choice of the right method 
is an important issue. In fact, as we will demonstrate in the following using the crystallization of a Lennard-Jones liquid
as example, selection of inappropriate methods may lead to invalid results that may seem acceptable at first sight. 

\subsection{Crystallization rates of a supercooled LJ fluid} 
\label{subsec:example}

Along with the freezing of hard spheres, the freezing of a supercooled monodisperse LJ fluid is 
the crystallization transition most studied with computer simulations. 
Up to date, crystal nucleation rates for the LJ system have been 
calculated with TST-BC \cite{tenwolde:1996}, MFPT \cite{lundrigan:2009}, 
MLT \cite{baidakov:2011a,baidakov:2011b}, FFS \cite{vanmeel:2008}, 
TPS \cite{trudu:2006}, and (PP)TIS \cite{moroni:2005} methods. Yet, the degree 
of supercooling, the pressure, and other parameters used in these simulations 
varied from work to work, such that a direct comparison of the results is feasible only
in a few cases.  

For example, Lundrigan and Saika-Voivod \cite{lundrigan:2009} found that free energy barriers 
obtained with umbrella sampling are consistent with those extracted from the MFPTs 
according to the scheme proposed by Wedekind and Reguera \cite{wedekind:2008}. 
Likewise, the attachment rates to the crystalline cluster at the top of the free barrier computed 
within the TST-BC and 
MFPT frameworks were found to be similar. There was, however, a disagreement in the position of the 
transition state compared to the results presented by Wang, Gould and Klein \cite{wang:2007}, 
who previously also used umbrella sampling to compute the free energy landscape at the same 
conditions. The authors explained this discrepancy by different criteria used to 
identify crystalline clusters. 

At different conditions, Moroni, ten Wolde and Bolhuis \cite{moroni:2005} computed 
the rates using the TIS and PPTIS techniques and found good agreement between them. 
The free energy barrier computed within PPTIS was, however, about $5$ times 
higher than the one calculated previously with the umbrella sampling method by ten Wolde, 
Ruiz-Montero and Frenkel \cite{tenwolde:1996}. 
As for the comparison of the reaction rate constants obtained with the TST-BC and (PP)TIS 
methods, the former crystallization rate was orders of magnitude higher or lower than the TIS 
rate, depending on whether the prefactor of ref.~\cite{tenwolde:1996} was used with 
the free energy barrier from PPTIS or TST-BC simulations.   
The authors attributed the failure of the TST-BC method to the importance of the structural 
component of the reaction coordinate, which results in a broad distribution of 
critical cluster sizes and hence forbids the definition of a unique dividing 
surface between the reactant and product states in terms of the size of the crystalline cluster. 

Baidakov and collaborators observed a good agreement between the 
crystallization rates computed with the  MLT and TIS \cite{baidakov:2011b} as well as 
the MLT and MFPT \cite{baidakov:2011a} methods.  

\begin{figure}[ht]
\begin{center}
\includegraphics[clip=,width=0.95\columnwidth]{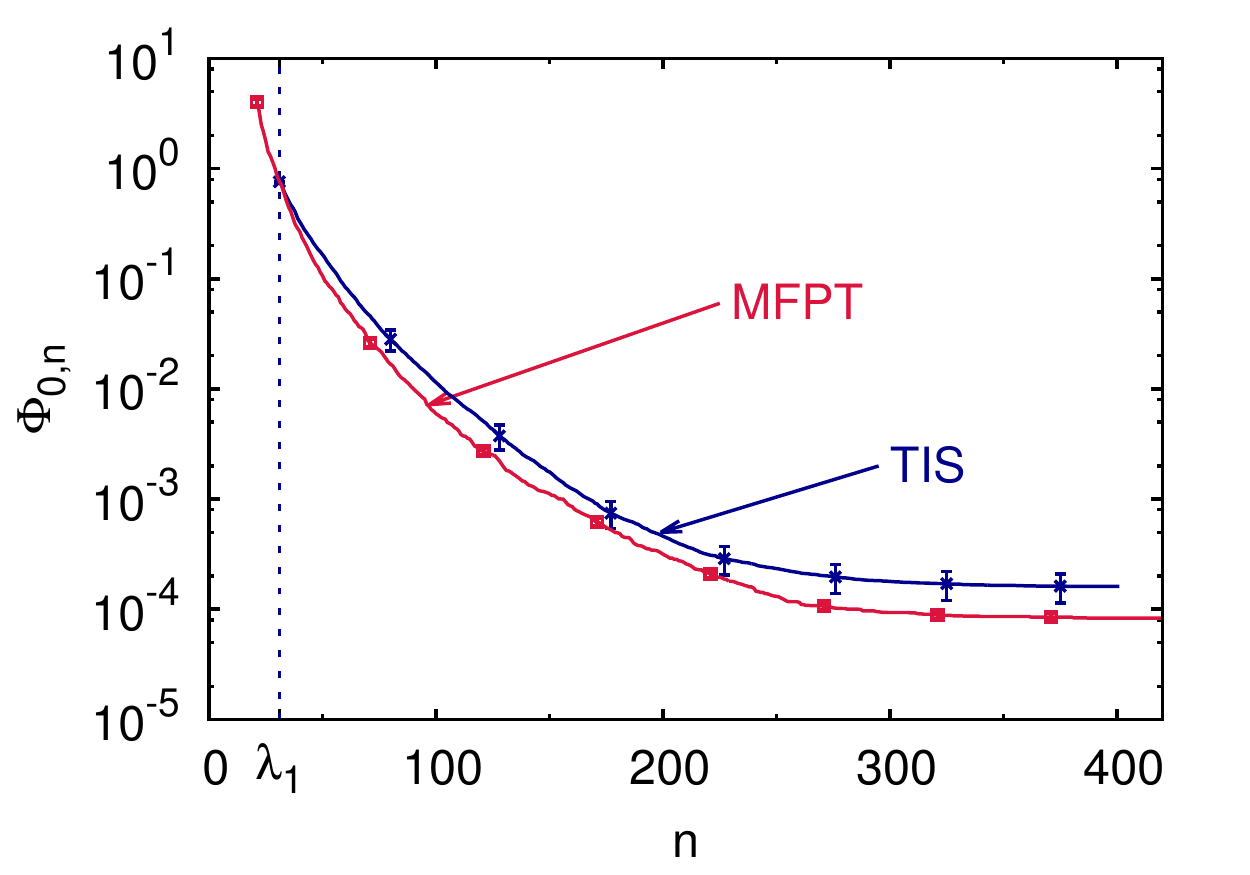}     
\caption{\label{fig:fluxLJ} Positive flux, $\Phi_{n,0}$, through an imaginary interface, 
placed at $n$, calculated with the TIS (eq.~(\ref{eq:fluxTISLJ})) and MFPT (eq.~(\ref{eq:fluxMFPTLJ})) 
techniques. Here, $n$ is the size of the largest crystalline cluster in the system. 
Both curves saturate to the respective 
values of the rate constant, which differ by a factor of two. Deviations between the results 
start to grow right from the first interface, where the fluxes calculated with the two methods are 
equal. Statistical errors are indicated for selected data points. The dashed line 
marks the position of the first TIS interface.}
\end{center}
\end{figure}

Recently, we have compared the MFPT rate with the one calculated in TIS simulations \cite{jungblut:2015}.   
The MFPT method in the formulation by Wedekind, Strey and Reguera \cite{wedekind:2007}
does not require previous knowledge about the position of the top of the 
free energy barrier, but rather allows to identify this position from the time 
evolution of the crystallizing trajectories. Since the formal derivations of the 
rate expressions in TIS and MFPT were similar, we did not expect any significant 
discrepancies. The resulting rates differed, however, by a factor of two, which might appear 
insignificant in comparison with the orders of magnitude variances in the TST-BC 
values seen previously, but is remarkable if one considers that the two methods have a 
similar theoretical foundation as both of them view the transition as a steady state diffusion 
process along an appropriate progress coordinate. As has been done in many other studies, we used the 
size $n$ of the largest crystalline cluster present in the system for this 
purpose. 

To illustratively compare the performance of the two methods, we define the positive 
flux, $\Phi_{n,0}$, of trajectories coming from the initial state and leaving the 
system through the interface placed at $n$. In the TIS framework, $\Phi_{n,0}$ 
is the flux out of the initial phase through the first TIS interface modified by the 
probability to propagate from there to the given cluster size $n$:     
\begin{equation}
\label{eq:fluxTISLJ}
\Phi_{n,0}=\Phi_{1,0}P_A(n|\lambda_1).
\end{equation}
This positive flux saturates to the value of the TIS rate constant defined in eq.~(\ref{eq:rateTIS}).  
Configurations with $n\leq 20 = \lambda_0$ belong to the initial supercooled liquid state 
and the first TIS interface $\lambda_1$ is placed at $n=30$. 

In the MFPT framework, the flux can be written as the inverse of the 
MFPT at a given crystalline cluster size, 
\begin{equation}
\label{eq:fluxMFPTLJ}
\Phi_{n,0}=\tau^{-1}(n).  
\end{equation}
In correspondence with the TIS scheme, the starting point of the MFPT trajectories 
is placed at $q_0=\lambda_0$. 
In the formalism proposed by Wedekind, Strey and Reguera \cite{wedekind:2007}, MFPTs close to 
the transition region are described by an error function (eq.~(\ref{eq:mfptErf})), the magnitude of which 
contributes the reaction rate according to eq.~(\ref{eq:nuclrate}). For a supercooling of 
about $28\%$, this approximation perfectly describes the behavior of the MFPTs, 
particularly beyond the transition region, indicating a clear separation between the nucleation 
and growth regimes. Thus, the MFPT rate constant is given by the value of the 
plateau in fig.~\ref{fig:fluxLJ}, just as in the case of the TIS calculation. 

\begin{figure}[ht]
\begin{center}
\includegraphics[clip=,width=0.95\columnwidth]{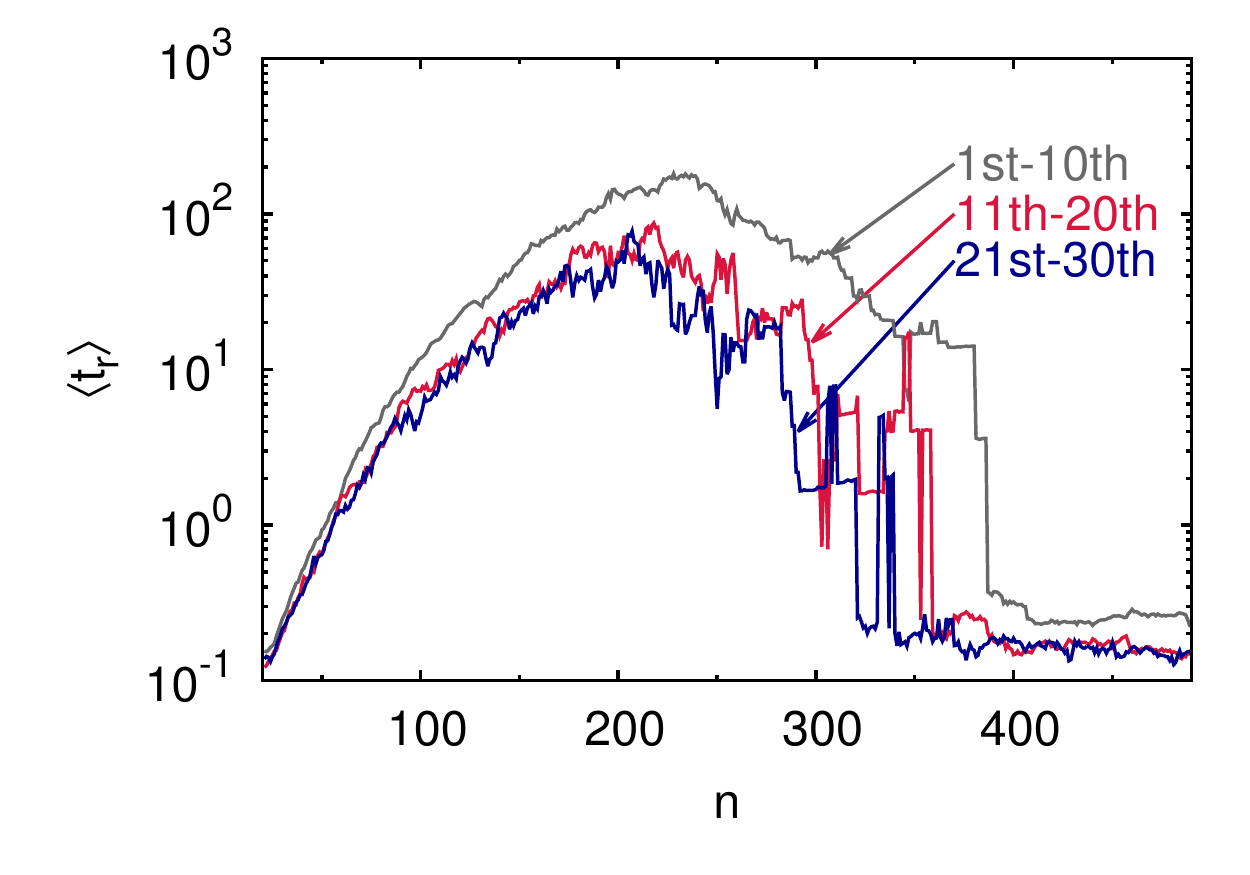}     
\caption{\label{fig:rtimesLJ} Mean recurrence times for configurations with given 
cluster sizes, averaged over ten subsequent passages to improve statistics. The system 
needs distinctly more time to return to a given state in the period of the first ten than in 
the next ten crossings. As the number of passages increases, the recurrence times of every 
cluster size approach a constant value.}
\end{center}
\end{figure}

As can be seen in fig.~\ref{fig:fluxLJ}, the flux values obtained with the two methods coincide only at the 
first TIS interface, after which the difference between them slowly grows until both techniques 
reach a plateau in the same size regime.  
A closer examination of the MFPT trajectories revealed the non-Markovianity of 
the dynamics of the system, if the number of particles in the 
largest crystalline cluster is used as reaction coordinate.       
In Kramers theory \cite{kramers:1940,haenggi:1990}, a central assumption made in the 
calculation of the rate constant is that the dynamics has the Markov property, {\it i.e.}, that 
a system sliding along the free energy barrier is not affected by its past. 
We found \cite{jungblut:2015}, however, that this is not the case if the process of crystallization  
is projected on a poor reaction coordinate, such as the number of particles in the 
largest crystalline cluster. 
To demonstrate the appearance of memory effects, we measured not only the MFPTs at a 
given state $n$, but also times for subsequent passages. The mean recurrence time obtained 
in this way is defined as half of the time between two passages through an imaginary interface.
This definition takes into account that, due to the shape of the underlying free 
energy landscape, the times to reach a given cluster size from above will differ from those 
to reach it from below. For a Markov process, the recurrence time of a state, calculated in this way, should not depend 
on whether the system has previously visited this state or not. This requirement is, however, not fulfilled for the crystallization transition 
as can be seen in fig.~\ref{fig:rtimesLJ}, where the intervals between successive passages through 
an interface clearly decrease with the number of crossings. Since Markovianty of the dynamics is assumed 
in the application of the MFPT approach, the validity of nucleation rates obtained from a MFPT analysis 
is unclear. 

In ref.~\cite{jungblut:2015}, we have shown that the memory effects decay exponentially, 
strongly indicating the presence of a coordinate not taken into account in the analysis. 
This phenomenon is most likely due to the structural relaxation not captured in the 
size of the crystalline clusters used as reaction coordinate, which also affected the 
TST-BC reaction rate \cite{moroni:2005}. Furthermore, we would 
like to note that, although the factor of two in the difference between the MFPT and TIS rate 
constants might seem insignificant, particularly with respect to the comparison between PPTIS and TST-BC, 
an intentional manipulation of the importance of the structural component of the reaction coordinate by introducing 
small pre-structured seeds resulted in disagreement factors of up to $13$ \cite{jungblut:2015a}.  
In the next section, we address the methods for selecting and evaluating the performance 
of reaction coordinates, but would like to emphasize with this example that a good choice 
of the method used to compute the reaction rate is equally important, since, after all, 
TIS requires only an order parameter suitable to differentiate between the reactant and the 
product states, but provides a solid value for the reaction rate constant without relying on the 
Markovianity of the dynamics projected on this order parameter.

\subsection{Transition rates are unique} 
\label{subsec:trates}

As noted earlier, transition rates calculated using a particular collective variable should be independent of this choice (of course, the accuracy of such a calculation will strongly depend on the quality of the chosen reaction coordinate). In section~\ref{subsec:feLandscapes}, we have seen, using some simple examples, that free energy landscapes depend on the collective variable on which they are projected. Here, we return to one of these examples and compare the reaction rates determined for the same simple system in two different representations. 
\begin{figure*}[ht]
\begin{center}
\includegraphics[height=60mm]{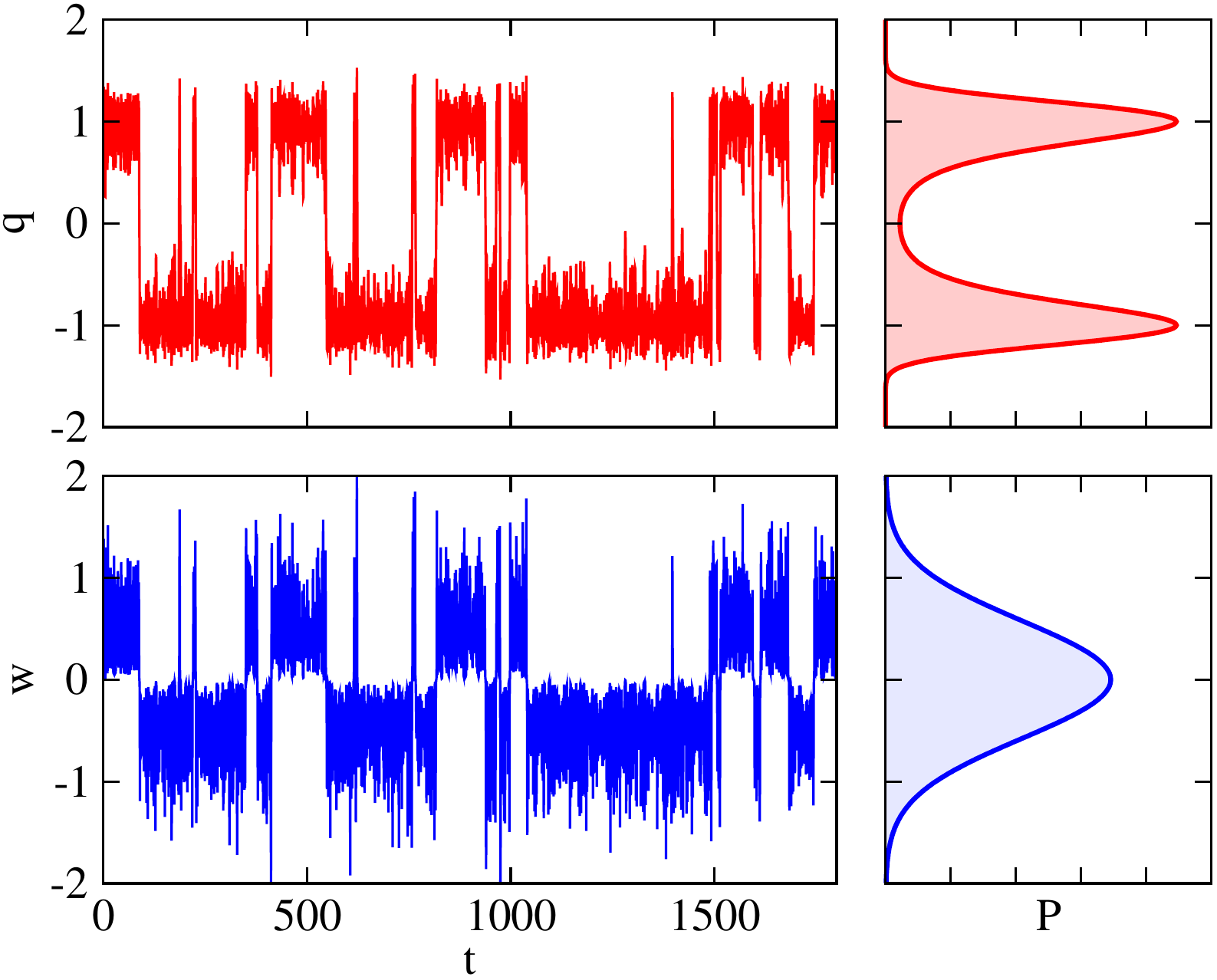}
\hspace{0.2cm}
\includegraphics[height=59mm]{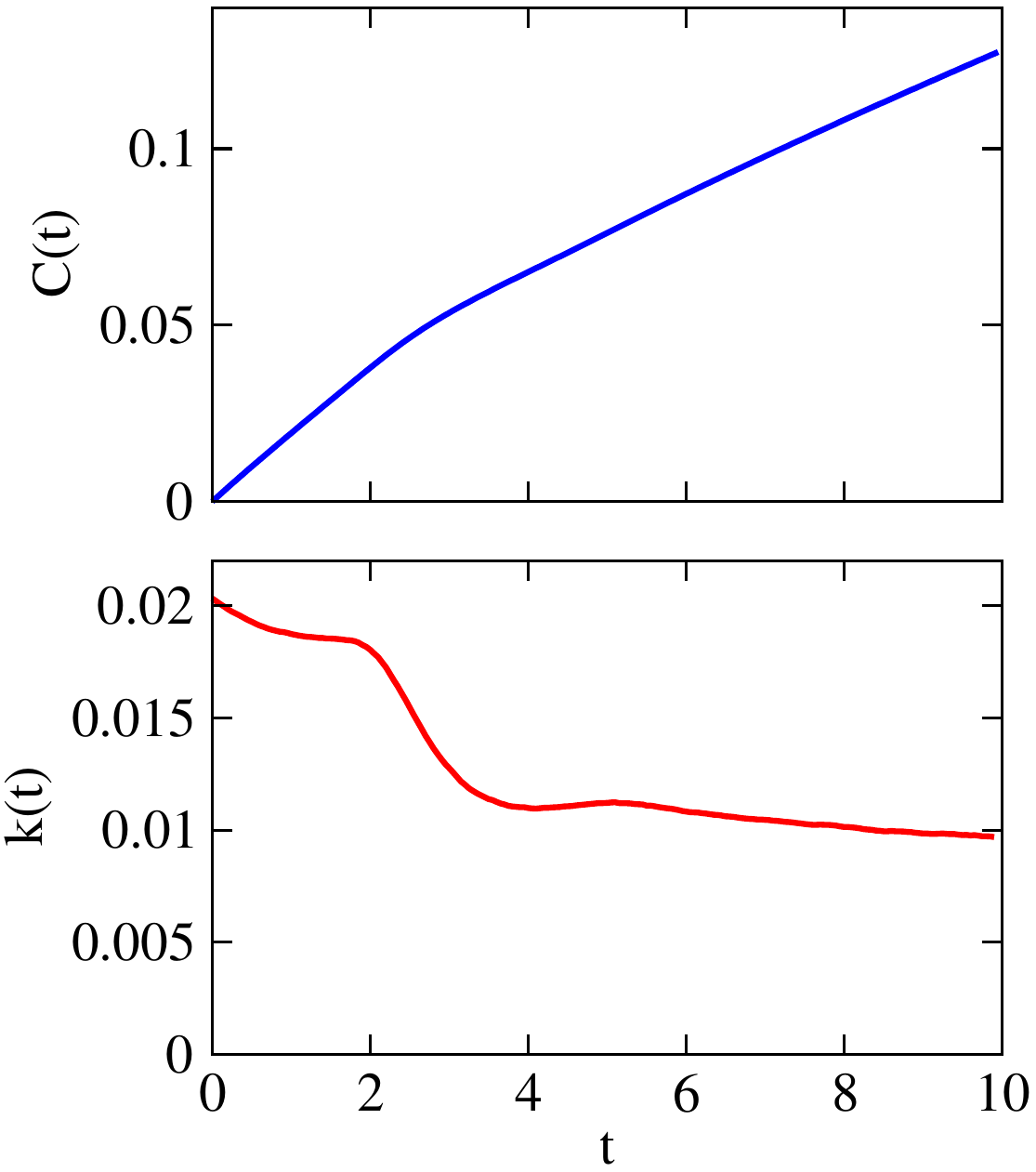}
\end{center}
\caption {\label{fig:FE_trace} Top left: Time trace of the position $q$ of a particle of mass $m$ evolving at reciprocal temperature $\beta$ in the bistable potential $V(q)=\kappa (q_0^2-q^2)^2$ according to the Langevin equation with a friction constant $\gamma$. The parameters used in the simulation were  $\kappa=1$, $q_0=1$, $\beta=3$, $m=1$ and $\gamma=0.4$. The particle spends most of its time in the two potential energy minima located at $q=\pm q_0$, but occasionally transitions between the minima occur. Top middle: Probability density $P_Q(q)$ of $q$. Bottom left: Time trace of the variable $w=(\psi \circ \phi)(q)$ obtained by transforming the time trace shown in the top left panel. The transformation was carried out according to eqs.~(\ref{eq:phi}) and~(\ref{eq:psi}) for $k=1$. Bottom middle: probability density $P_W(w)$ of the transformed variable. Top right: Time correlation function $C(t)$ for transitions between the two wells. Bottom right: Time derivative $k(t)=dC(t)/dt$.}
\end{figure*}
We first consider the position $q$ of one-dimensional particle that evolves in a bistable potential of the form $V(q)=\kappa (q_0^2-q^2)^2$ as a reaction coordinate. If the particle is in contact with a heat bath at low temperature, {\it i.e.}, $k_{\rm B}T$ is much smaller than the barrier height $\kappa$, the particle will be preferentially located near the potential energy minima at $q=\pm q_0$, but once in a while a transition from one of these long-lived states to the other will occur. The rate at which such transitions occur decreases exponentially with increasing barrier sizes (or, equivalently, with decreasing temperature). A typical trace $q(t)$ and the resulting probability density are shown in the top left and middle panels of fig.~\ref{fig:FE_trace}, respectively. The time evolution of the system viewed through the variable $w$, obtained by applying the transformation $w=(\psi \circ \phi)(q)$ to the variable $q$ as described in section~\ref{subsec:feLandscapes}, looks different from the time evolution of $q$. Nevertheless, one can clearly discern transitions between the two wells also in this variable, although neither the stable states nor the barrier between them are revealed in the free energy $F_W(w)$.

The kinetic rate constants obtained from the analysis of the time traces of two variables $q(t)$ and $w(t)$ are exactly the same. To see this, just imagine that we place the dividing surface between the two states somewhere in the barrier region at $q=q^*$. So the system is in region $A$ if $q\le q^*$ and in region $B$ if $q>q^*$. Equivalently, one may define the dividing surface by $w=w^*$ where $w^*=(\psi \circ \phi)(q^*)$ and regions $A$ and $B$ by $w\le w^*$ and $w>w^*$, respectively. Since the mapping from $q$ to $w$ is monotonic, the two definitions of regions $A$ and $B$ are completely equivalent. Accordingly, the same time correlation function $C(t)=\langle h_A(0)h_B(t) \rangle / \langle h_A \rangle$ is obtained from the time traces $q(t)$ and $w(t)$ and, therefore, also the transition rate constants are the same in both cases. 

Another way to look at this issue is by considering the reactive flux $k(t)$, {\it i.e.}, the derivative of the time correlation function, $k(t)=dC(t)/dt$. The time correlation function as well as its time derivative are shown in the right panels of fig.~\ref{fig:FE_trace}. As discussed earlier, the transition rate constant for transitions from $A$ to $B$ is equal to the plateau value of $k(t)$ after transient effects due to correlated recrossing events on the molecular time scale $\tau_{\rm mol}$ have decayed. The reactive flux for a dividing surface placed at $q^*$ is given by (eq.~(\ref{eq:rateTST2})) 
\begin{equation}
k_Q(t)=\frac{\langle \dot q(0) \delta[q(0)-q^*]\theta[q(t)-q^*] \rangle}{\langle \theta[q^*-q(0)]\rangle}.
\label{eq:rflux}
\end{equation}
Transforming to a new variable $w=w(q)$ and using the monotonicity of the mapping $w(q)$ one finds
\begin{eqnarray}
\label{eq:kofq}
k_Q(t) & =& \frac{\langle \dot w(0)\left(\frac{\partial w}{\partial q}\right)^{-1}\left(\frac{\partial w}{\partial q}\right) \delta[w(0)-w^*] 
\theta[w(t)-w^*] \rangle}{\langle \theta[w^*-w(0)]\rangle} \nonumber \\
 & =& \frac{\langle \dot w(0) \delta[w(0)-w^*] 
\theta[w(t)-w^*] \rangle}{\langle \theta[w^*-w(0)]\rangle} = k_W(t),
\end{eqnarray}
because the two Jacobian factors arising from the time derivative and the transformation properties of the delta function cancel each other. Thus, the reactive flux obtained from variables $q(t)$ and $w(t)$ is the same, implying that also the transition rate constants are identical. 

\section{Analyzing the nucleation mechanism}
\label{sec:nuclMechanism}

\subsection{Reaction coordinate}
\label{subsec:reactionCoordinate}

While computer simulations yield very detailed information on the structure and dynamics of complex systems, they do not automatically generate understanding. In fact, vast amounts of data are produced each day by computer simulations running on high performance computers and often the challenge no longer is to create data, but to make sense of them. In the case of crystallization, a collection of crystallization trajectories, obtained by brute force MD or by path sampling methods and stored on a computer in form of long lists of positions and momenta of all particles at different times, do not yield insights into the transition mechanism directly. For instance, even deciding whether the process follows the general picture of nucleation theory and proceeds by a localized formation of a crystalline embryo that then grows subsequently is not a completely trivial issue and early simulation studies of crystallization focused on exactly that \cite{rahmanI:1977,rahmanII:1977}. Some progress may be made by visualizing the crystallization trajectories with a molecular viewing program, perhaps highlighting regions of pronounced local crystallinity with different colors, but only further statistical analysis of these trajectories can give a reliable description of the mechanism with which the system goes from the disordered to the ordered state.  Such mechanistic understanding is expressed in terms of models that capture the essential features of the process under study. These models should have a minimal number of degrees of freedom and are ideally expressed in terms of physically meaningful variables to render them understandable to humans. Classical nucleation theory is an example of such a low-dimensional model, in which a single variable (the size of the crystalline nucleus) is used to describe the progress of the reaction and some simplifying assumptions are made on the dynamics of this variable. Thus, the central goal of the analysis of transition pathways is to identify a reaction coordinate, {\it i.e.}, a physical variable that captures the relevant physical features of the process and quantifies the progress of the transition. 

The reaction coordinate $q(x)$, usually defined as a function of the configuration $x$ containing the positions of all atoms, is a variable that is supposed to provide a measure for how far the process under study has proceeded. It should, for instance, tell us whether a particular configuration is a transition state from which both stable states are equally accessible. In the context of rare event simulations, it is important to distinguish between {\em reaction coordinates} and {\em order parameters}. While from an order parameter we require only the ability to discriminate among stable states between which the transition occurs, a reaction coordinate should tell us how far the reaction has advanced and what is likely to happen next. To make the significance of this terminology clearer, consider the crystallization of a supercooled liquid. In this case, one may use the structure factor $S(k)$ for an appropriate reciprocal vector $k$ or a bond order parameter $Q_l$ (usually $l=6$ for LJ, but the exact choice depends on the lattice type one intends to detect) averaged over the whole sample to tell apart a liquid and a crystal with their typical fluctuations. Thus, $S(k)$ and $Q_l$ may serve as order parameters, but they usually do not work well as reaction coordinate (even if they were used in this role in some earlier studies \cite{rahmanII:1977,tenwolde:1996}). Particularly for strong supercooling, where the critical nucleus is small, the signal produced in the order parameter by a crystalline embryo that is just a little beyond the critical size may be so small that it disappears in the $\sqrt{N}$-fluctuations of the order parameter. Clearly, to work for a nucleation process, a reaction coordinate cannot be global, but must be able to detect local variations of crystallinity. Which collective variable one chooses as reaction coordinate is arbitrary to some degree. Just as in the case of free energy landscapes discussed earlier, there is no such thing as {\em the} reaction coordinate. Rather, there are usually many choices for a reaction coordinate, some better and some worse.

\begin{figure}[ht]
\begin{center}
\includegraphics[clip=,width=0.95\columnwidth]{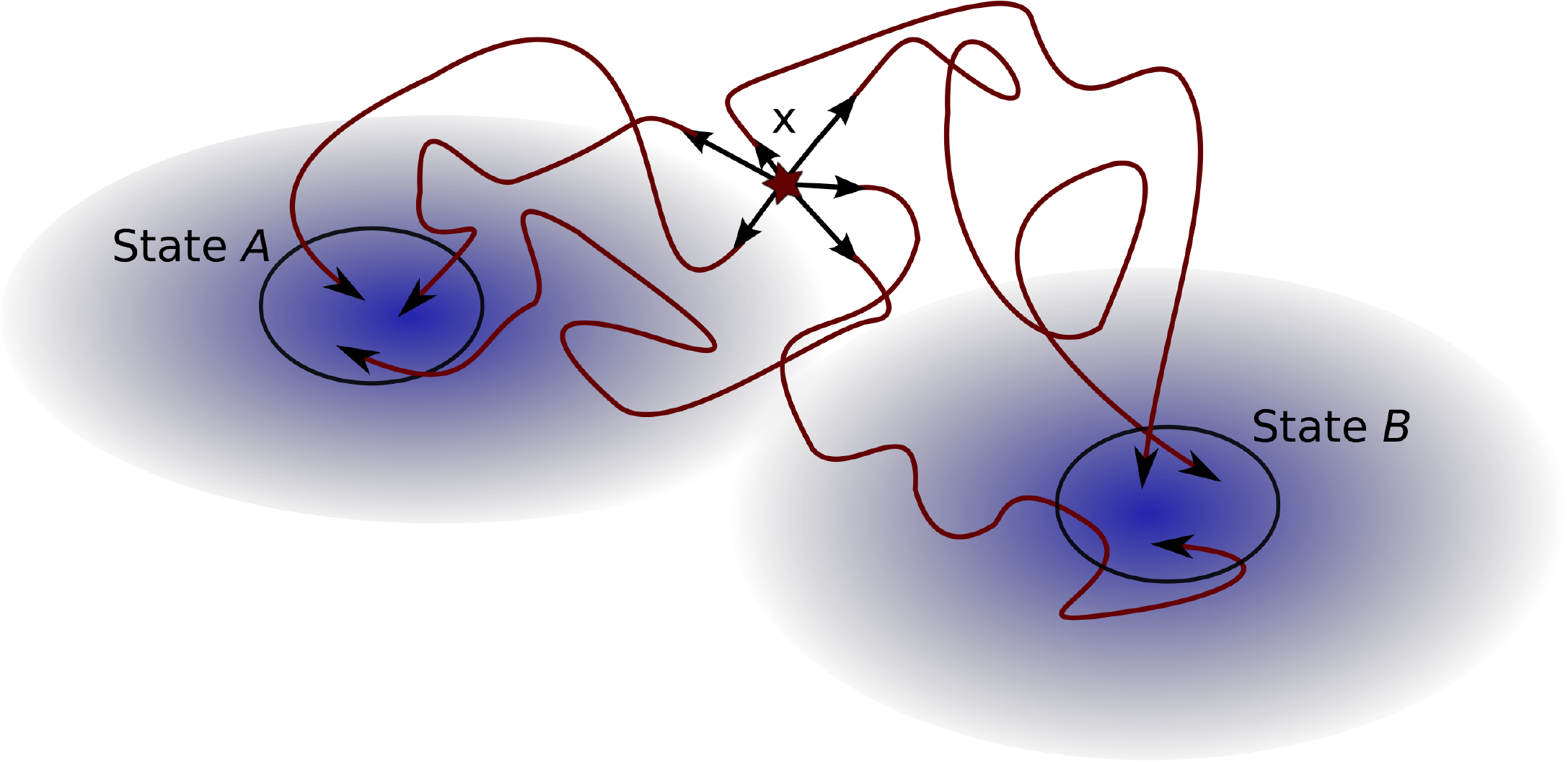}     
\caption{\label{fig:committor} The committor $p_B(x)$ of a configuration $x$ is the probability that a trajectory initiated at $x$ reaches state $B$ before state $A$. The committor can be estimated numerically by starting a certain number of trajectories from $x$ with random momenta drawn from the appropriate Maxwell-Boltzmann distribution and counting how many of them go to state $B$. }
\end{center}
\end{figure}

\subsection{Committor}
\label{subsec:committor}

How well a given collective variable works as a reaction coordinate can be assessed by considering the commitment probability, or committor, $p_B(x)$. For a given configuration $x$, the committor is defined as the probability that trajectories started from $x$ reach region $B$ before region $A$. The committor $p_A(x)$ for region $A$ is defined analogously. The concept of committor, first introduced by Onsager as splitting probability in the analysis of ionic separation \cite{TS_ONSAGER}, is illustrated schematically in fig.~\ref{fig:committor}. In a simulation, the committor for a particular configuration $x$ can be computed by initiating $m$ trajectories from $x$ with random momenta drawn from a Maxwell-Boltzmann distribution. These trajectories are terminated when they reach either $A$ or $B$. From the number $n_B$ of trajectories that reach $B$ the committor is then estimated, $p_B\approx n_B/m$. (Note that for overdamped dynamics momenta are irrelevant and the average is done over noise histories.) The committor is a statistical measure for how ``committed'' configuration $x$ is to state $B$. Configurations with a committor $p_B\approx 1$ lead to trajectories most often relaxing to $B$ and therefore can be viewed as being near $B$. Analogously, configurations with $p_B\approx 0$ most often go to $A$ and can be viewed as being near $A$. Configurations with a committor of $p_B \approx 1/2$, on the other hand, are equally likely to go either side, and it is natural to define these configurations as transition states half way between regions $A$ and $B$. The definition of transition states via the committor goes back at least to Ryter  \cite{RYTER,SZABO2}, and was used by several authors in the theory of activated stochastic processes \cite{TS_SCHUSS,POLLAK,TALKNER}. The committor, which has also proven very useful in the analysis of protein folding \cite{TS_PANDE}, lies at the heart of transition path theory, a probabilistic framework to study the statistical properties of rare event trajectories \cite{Eric_TPT_Erice,TPT_2006}.

In the study of crystallization, the committor can be used as a dynamical criterion for the definition of the critical nucleus. Already in very early molecular dynamics simulations of crystallization of a Lennard-Jones fluid, carried out by Mandell, McTague and Rahman \cite{rahmanI:1977,rahmanII:1977}, this concept was used to determine the approximate time at which a growing crystalline embryo reached critical size. For this purpose, the authors of these studies first generated a molecular dynamics trajectory at very large supercooling for which crystallization happened spontaneously during the relatively short simulation time of about 50 reduced time units. At various configurations taken from this trajectory, new trajectories were initiated with random initial momenta for a time long enough to determine whether the crystallization completes or subsides. In the former case, the corresponding configuration was considered post-critical and in the latter pre-critical. A statistically more meaningful determination of critical nuclei for the crystallization of a Lennard-Jones liquid, with several trajectories started at each configuration, was carried out by Honeycutt and Andersen \cite{Honeycutt:1984, Honeycutt:1986}, who also defined, for the first time, the nucleus to be critical if the configuration containing it lead to trajectories with equal probability to completely crystallize or liquefy.  

What we expect from a good reaction coordinate $q(x)$ is to tell us whether the configuration $x$ is at the beginning or the end of the reaction or if it is a transition state. So, by looking at $q$ we should be able to tell whether the reaction has just started or is almost completed. In a sense, the committor itself is the ideal reaction coordinate, because it tells us how far the reaction has proceeded and what is likely to happen next \cite{E:2005,Best:2005,TS_HUMMER1}. For instance, a committor of $p_B\approx 1/2$ tells us that the corresponding configuration is a transition state half way between states $A$ and $B$. But while the committor serves as a precise measure of the transition, it usually does not convey a physically transparent meaning because of its generality. For instance, imagine we study a crystallization transition that roughly follows the nucleation mechanism envisaged by classical nucleation theory, but that we do not know about this theory. Now, the fact that a particular configuration has a certain committor $p_B$ does not immediately tell us that the transition occurs by nucleation and growth. To find that out we would have to analyze configurations with different committor values and realize that the configurations contain a crystalline nucleus, whose size correlates with the committor. Based on this insight, we could then hypothesize about different contributions to the free energetics of nucleus formation and finally come up with a simple model such as classical nucleation theory. Thus, while the committor is a very useful concept in the analysis of rare transition pathways, it does by no means directly lead to physical insight and should be considered the starting point for further analysis rather than an end by itself. 

One very useful property of the committor is that it can be used as a criterion to gauge the quality of a reaction coordinate $q$. From a good reaction coordinate we expect that it provides a measure for the progress of the reaction. As such, one should be able to predict the committor $p_B(x)$ of a particular configuration based solely on the reaction coordinate $q(x)$ of that configuration. In other words, a good reaction coordinate should parametrize the committor, $p_B(x)=p_B[q(x)]$. For a bad reaction coordinate, on the other hand, there is no such relation with the committor and configurations with the same value of the reaction coordinate may have rather different values of the committor.

\begin{figure}[ht]
\begin{center}
\includegraphics[clip=,width=0.95\columnwidth]{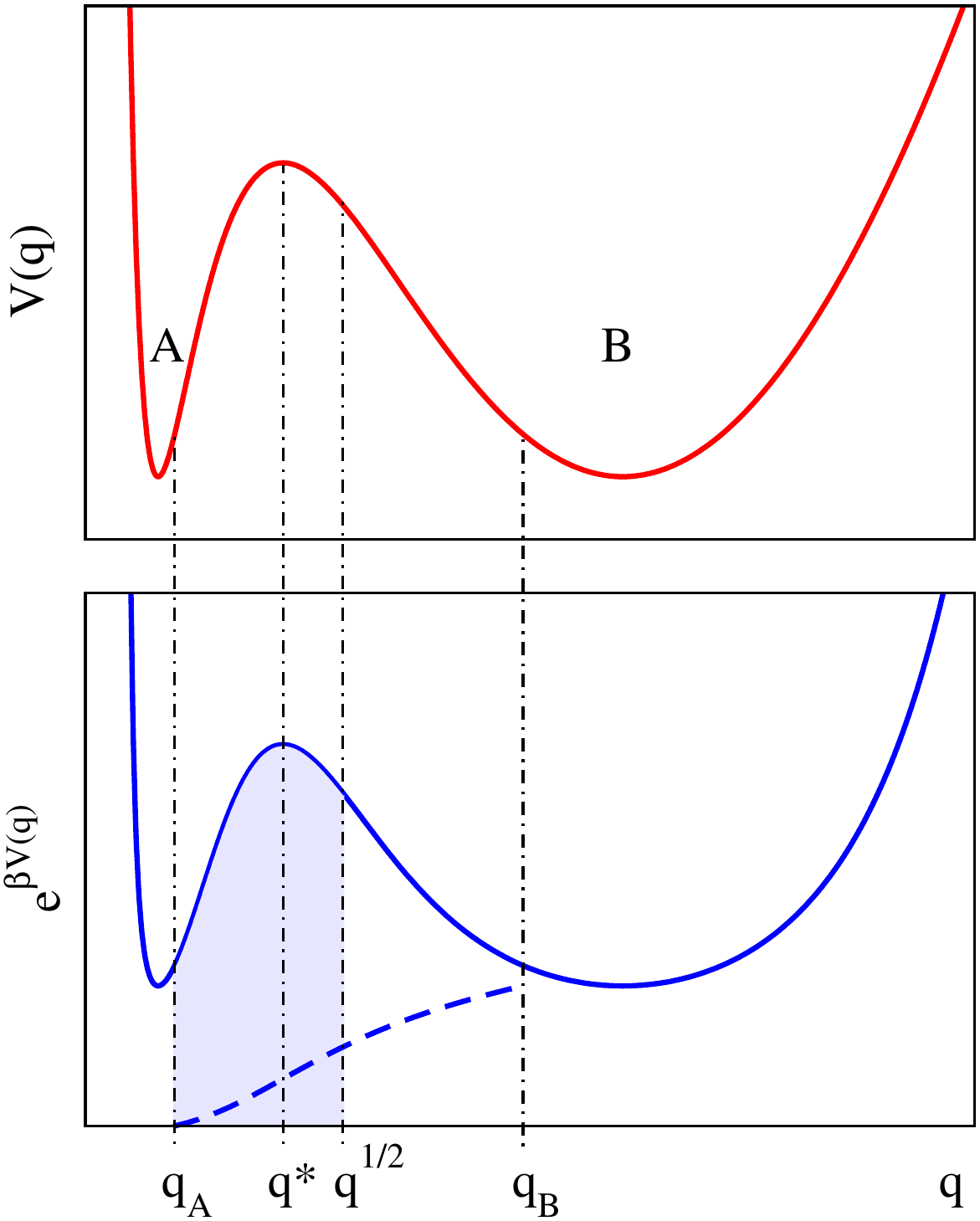}     
\caption{\label{fig:F1d} Asymmetric bistable potential energy $V(q)$ (top) and corresponding exponential $\exp[\beta V(q)]$ (bottom). Here, $q_A$ and $q_B$ denote the boundaries of regions $A$ and $B$, respectively, and $q^*$ is the position of potential energy barrier between the two stable states. The shaded area indicates the integral $\int_{q_A}^{q} \exp[\beta V(q')]dq'$ appearing in the expression for the committor (eq.~(\ref{eq:pBofq})) as discussed in the main text. The committor is $1/2$ at the point $q^{1/2}$, for which the area between $q_A$ and $q^{1/2}$ is equal to that between $q^{1/2}$ and $q_B$. The dashed line in the bottom panel line is the committor, which changes from 0 at $q_A$ to 1 at $q_B$.
}
\end{center}
\end{figure}

One may ask the question, whether for a perfect reaction coordinate $q(x)$, {\it i.e.}, a reaction coordinate that uniquely determines the committor, the top of the free energy barrier separating two stable states always coincides with the isocommittor surface corresponding to $p_B=1/2$. The answer to this question is no. As a simple example, consider a particle moving in one dimension on the potential energy $V(q)$ according to the overdamped Langevin equation with constant diffusion coefficient $D$.  The potential, shown schematically in fig.~\ref{fig:F1d}, is supposed to be asymmetric with two wells $A$ and $B$ of different width. The committor $p_B(q)$ of a point $q$ in between the two boundaries $q_A$ and $q_B$ is the probability (averaged over noise histories) that the point reaches the right boundary $q_B$ before the left boundary $q_A$. It can be shown that, for the dynamics considered here, the committor satisfies the adjoint time-independent Smoluchowski equation and can be expressed as \cite{TS_HUMMER1,TS_ONSAGER}
\begin{equation}
\label{eq:pBofq}
p_B(q)=\frac{\int_{q_A}^{q} \exp[\beta V(q')]\,dq'}{\int_{q_A}^{q_B} \exp[\beta V(q')]\,dq'}
\end{equation}
Hence, the committor at $q$ is the area under the curve $\exp[\beta V(q')]$ between $q_A$ and $q$, shown as shaded blue region in fig.~\ref{fig:F1d}, normalized by the area between $q_A$ and $q_B$. Accordingly, the committor grows monotonically from $0$ at $q_A$ to 1 at $q_B$. The inflection point of the committor is located at $q^*$, the position of the barrier. Due to the asymmetry of the potential, however, the point where the committor is $1/2$, {\it i.e.}, the point $q^{1/2}$ for which the area under the curve $\exp[\beta V(q')]$ between $q_A$ and $q^{1/2}$ is equal to the area between $q^{1/2}$ and $q_B$, is at a different position. For the example shown in fig.~\ref{fig:F1d}, basin $B$ is wider than basin $A$ and therefore $q^{1/2}$  lies at the right of the barrier top. Hence, in general, the isocommittor surface for $p_B=1/2$ differs from the hypersurface corresponding to the top of the free energy barrier. It is also worth noting that the committor-$1/2$ surface is not necessarily the dividing surface that maximizes the transmission coefficient.

\subsection{Committor distribution and transition state ensemble}
\label{subsec:committorTSE}
\begin{figure}
\begin{center}
\includegraphics[clip=,width=0.95\columnwidth]{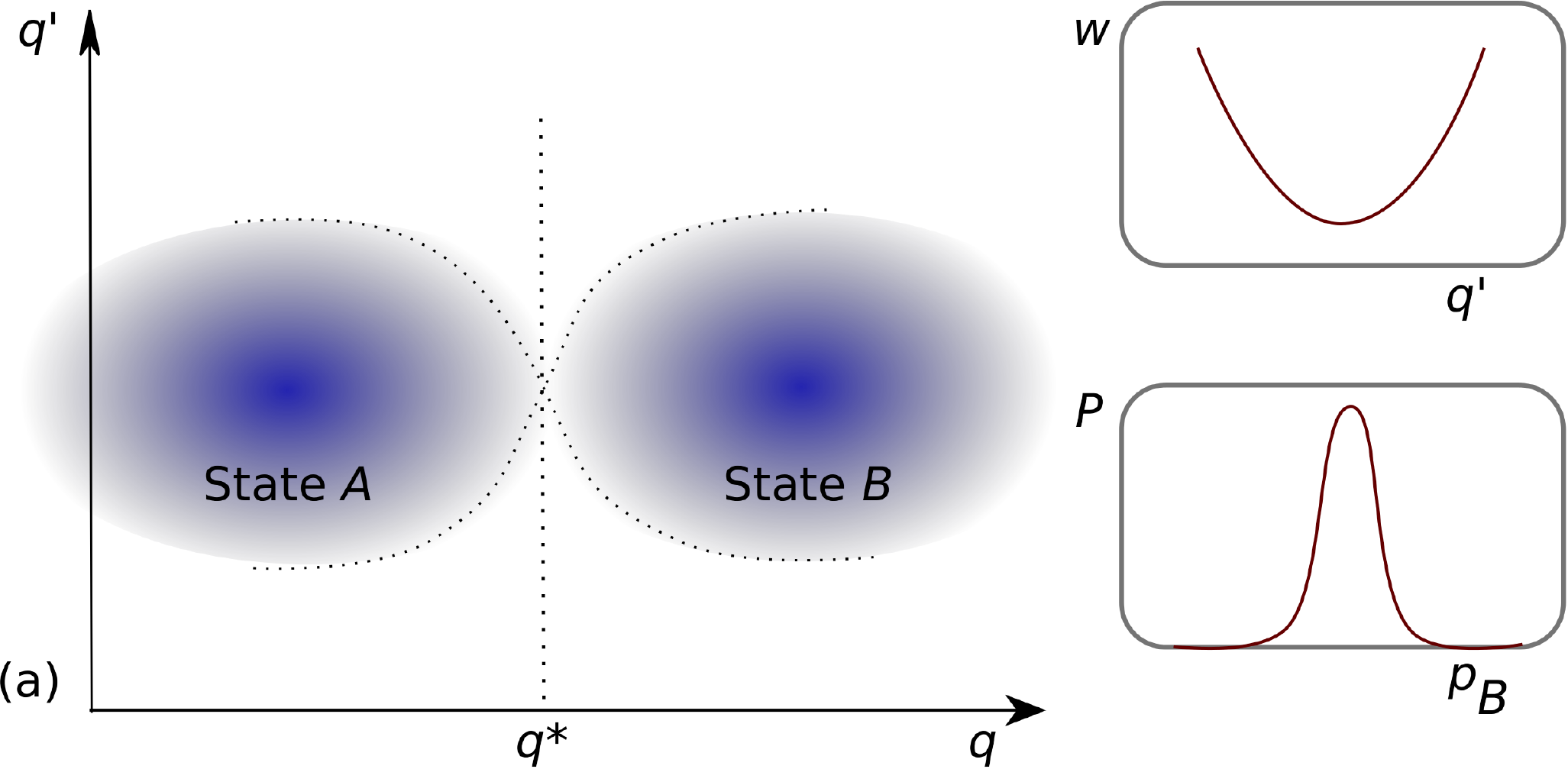}     

\vspace{4mm}

\includegraphics[clip=,width=0.95\columnwidth]{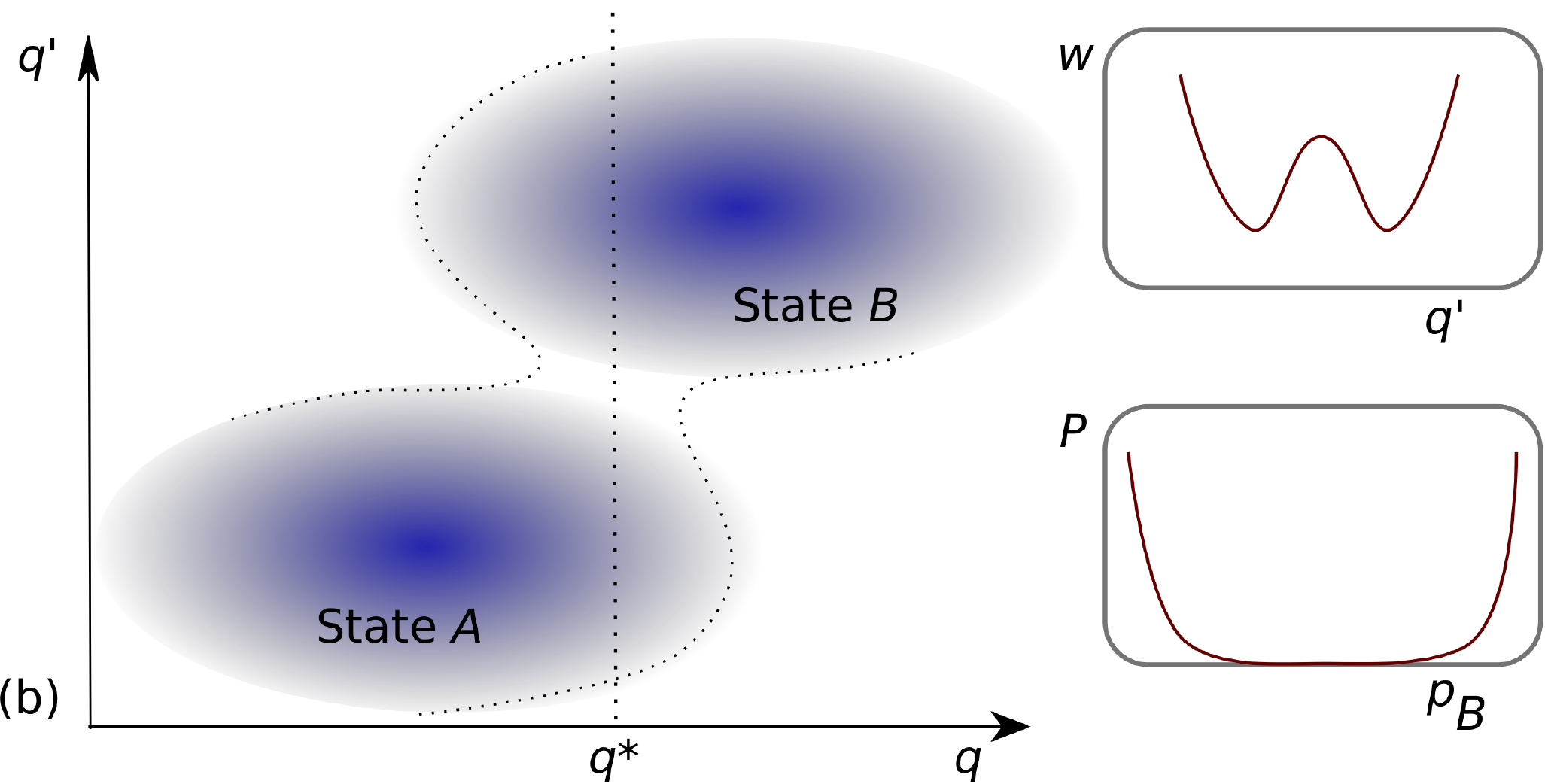}  

\vspace{4mm}

\includegraphics[clip=,width=0.95\columnwidth]{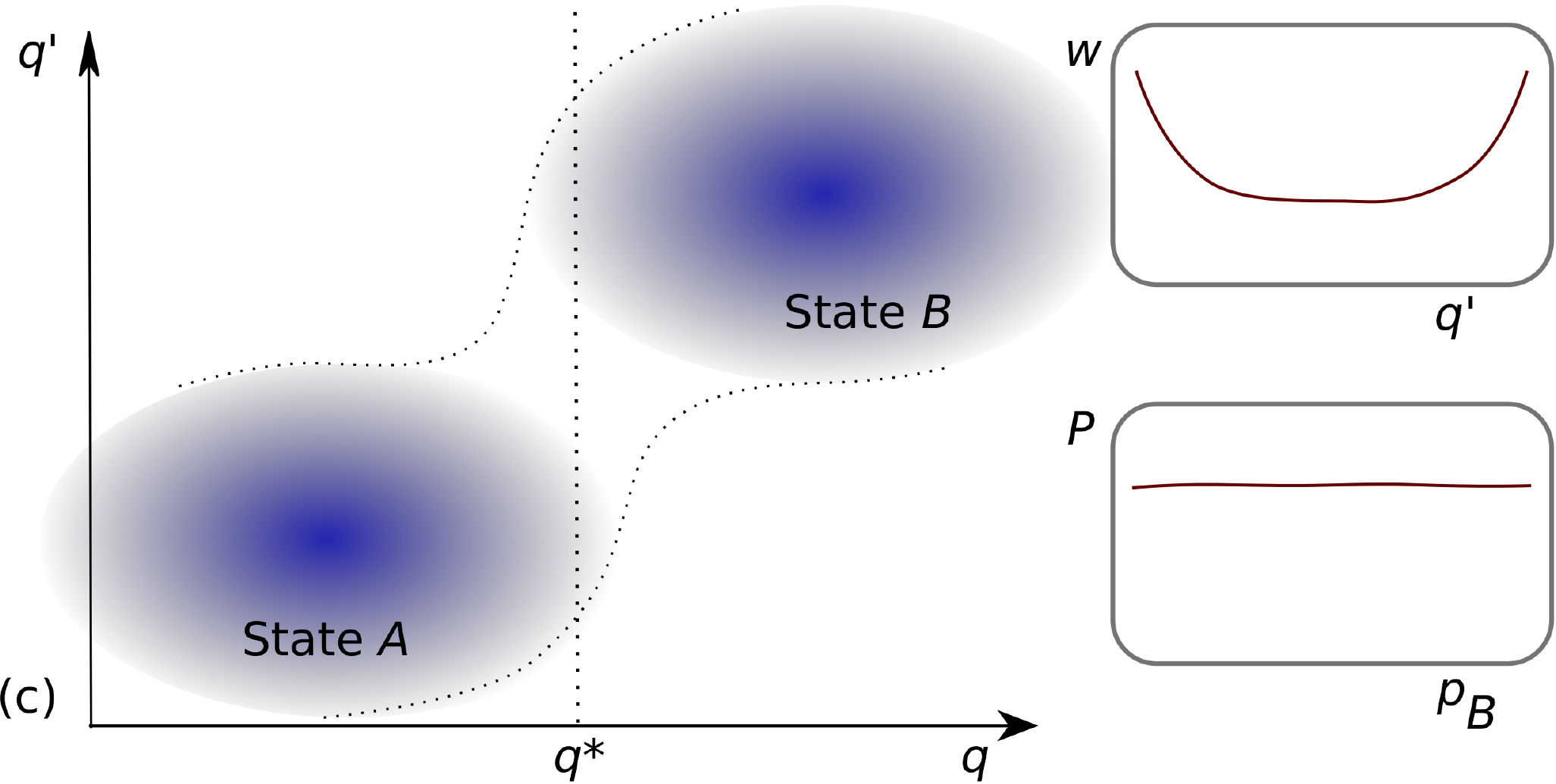}  
\caption{\label{fig:committor_distribution} Different two-dimensional free energy landscapes $w(q,q')$ as function of the coordinates $q$ and $q'$ alongside with the free energy $w(q^*,q')$ for $q$ fixed at $q=q^*$ and the corresponding committor function. (a) Here, the coordinate $q$ correctly describes the transition and there is no barrier in the orthogonal direction $q'$. In this case, the committor distribution is peaked at $p_B\approx 1/2$. (b) The transition involves the crossing of a barrier in the direction of $q'$ as can be seen in the double well shape of $w(q^*,q')$. Since configurations with $q=q^*$ are most likely located  near states $A$ and $B$, the committor has two peaks, one at 0 and one at 1. (c) The situation is similar to that shown in panel (b), but here the barrier in $q'$-direction is flat and thought to be crossed diffusively. This leads to a flat committor distribution.}
\end{center}
\end{figure}

Since for a good reaction coordinate its value determines the value of the committor, isosurfaces of the committor and the reaction coordinate should coincide. More specifically, provided $p_B(x)=p_B[q(x)]$ holds, the hypersurface in configuration space defined by $q(x)={\rm const}$ is identical to the hypersurface defined by requiring that the committor has the value $p_B({\rm const})$ \cite{E:2005}. One can easily test whether this is the case for a particular reaction coordinate $q(x)$. To do that, one samples configurations from the equilibrium distribution with the constraint that the reaction coordinate has the particular value $q^*$. For instance, in the case of crystallization, one would prepare a set of configurations sampled from the equilibrium distribution using an MC simulation with the additional condition that the crystalline nucleus has a particular size. For all configurations generated in this way one then estimates the committor by shooting off trajectories from the configurations and finally histograms the committor values. Mathematically, the committor distribution $P(p_B)$ resulting from this procedure can be written as
\begin{equation}
\label{eq:comm1}
P(p_B)=\langle \delta [p_B-p_B(x)] \rangle_{q(x)=q^*}
\end{equation}
where $\langle \cdots \rangle_{q(x)=q^*}$ denotes an equilibrium average restricted by the condition $q(x)=q^*$. If $q$ is a good reaction coordinate, the value of $q$ completely specifies the value of the committor $p_B$, such that all configurations with $q(x)=q^*$ yield the same committor. As a consequence, the committor distribution $P(p_B)$ is a delta-like peak located at $p_B(q^*)$. If, on the other hand, the postulated reaction coordinate is not appropriate for the reaction, configurations with the same $q$ may have very different values of the committor. In this case, the committor distribution $P(p_B)$ is smeared out. Thus, the width of the committor distribution can be viewed as a measure of the quality of the reaction coordinate. 

Three possible scenarios resulting in different committor distributions are illustrated in fig.~\ref{fig:committor_distribution}. For the free energy landscape $w(q, q')$ shown in panel (a) as a function of two variables $q$ and $q'$, the variable $q$ is a valid reaction coordinate and during a transition from $A$ to $B$ no barrier needs to be crossed in the direction of $q'$. Configurations sampled with the constraint that $q$ has the value $q^*$ lie at the top of the barrier separating $A$ and $B$ and are true transition states. This is reflected in a committor distribution peaked at $p_B\approx 1/2$. In the situation depicted in panel (b), the coordinate $q$ fails to provide a complete description of the transition and the coordinate $q'$ plays an important role too. For $q=q^*$ there is a barrier in direction of $q'$ such that configurations with $q=q^*$ are most likely located away from the barrier deep in the basins of attraction of $A$ or $B$. As a consequence, the committor distribution $P(p_B)$ has peaks at $0$ and $1$, respectively. Such a bimodal shape of $P(p_B)$ is indicative of a poor reaction coordinate $q$. In the scenario displayed in panel (c), the barrier crossing involves diffusion in direction of $q'$ on a flat free energy profile $w(q^*, q')$ leading to a flat committor distribution. Also in this case, $q$ neglects an important aspect of the transition and a good reaction coordinate needs to include $q'$ too.

Committor distributions were introduced to analyze kinetic pathways of ionic dissociation \cite{Geissler_ion_1999} and have subsequently also been used to study the mechanism of crystallization \cite{moroni:2005,lechner:2011,Lechner_RPE:2011}. For the crystallization of LJ particles \cite{moroni:2005} and soft spheres \cite{lechner:2011}, this analysis clearly indicates that the size of the crystalline embryo is not sufficient to describe the crystallization mechanism and other variables play an important role. Furthermore, comparison of configurations with different values of the committor may help to identify the variables determining the value of the committor. In particular, it may be helpful to analyze configurations with committor $p_B=1/2$ sampled from transition pathways. Such a collection of transition states is called the {\em transition state ensemble} (TSE). The transition states are the points where transition trajectories cross the isocommittor surface corresponding to a committor value of $p_B\approx 1/2$. Members of the TSE are not distributed uniformly on the isocommittor-$1/2$ surface but are located where transition pathways preferentially cross this surface. In the case of high friction dynamics, states belonging to TSE are distributed according to their equilibrium distribution restricted to the isocommittor-$1/2$ surface \cite{E:2005}. In a transition path sampling simulation the transition path ensemble can be determined by computing the committor on points collected from transition pathways. Since along a transition path the committor changes continuously from 0 to 1, there is always at least one (but usually several) configurations on the path for which $p_B\approx 1/2$. For the freezing of LJ particles \cite{moroni:2005}, an inspection of the TSE revealed that besides size also shape and structure play an important role for the transition. In particular, critical clusters are either (1) small and compact with an fcc core and a bcc-structured surface or (2) large and less packed with a more bcc-like structure. 

\subsection{Likelihood Maximization} 
\label{subsec:maximumLikelihood}

As mentioned before, one central goal in the study of rare event processes, and in particular of crystallization, is to identify the variables that characterize the transition and to separate them from irrelevant degrees of freedom that can be regarded as random noise. In other words, we would like to find a good reaction coordinate for the transition. Several statistical approaches have been developed for this purpose including the Bayesian statistics approach of Hummer and Best \cite{TS_HUMMER1,Best:2005}, or the artificial neural network method of Ma and Dinner \cite{ma:2005}. Here we will focus on the likelihood maximization method of Peters and Trout \cite{peters:2006,peters:2007,peters:2010,peters:2013}, which is particularly suitable to be combined with path sampling simulations and has been used to study the mechanism of crystallization \cite{beckham:2011,jungblut:2013a,Leitold_2014,JPCM_Leitold_string_2015}. 

The basic idea of the method of Peters and Trout is to construct a very general reaction coordinate $q(x, \alpha)$ which is a function of the configuration $x$ and a set of $M$ free parameters $\alpha=\left\{\alpha_1, \alpha_2, \cdots, \alpha_M \right\}$. These parameters are then tuned to maximize the probability to observe a particular data set. The data set can consist of the outcome of the trajectories used in committor calculations or of the sequence of rejections and acceptances in a path sampling simulation carried out with aimless shooting, a path sampling algorithm specifically designed to work in combination with likelihood maximization \cite{peters:2006}. To be more specific, consider a set of configurations $\left\{x_1, x_2, \cdots, x_P\right\}$, for instance selected from pathways connecting stable states $A$ and $B$. From each of these configurations one or several test trajectories are started with random momenta and followed until they reach either region $A$ or region $B$.  The information about the outcome of these test trajectories, {\it i.e.}, whether they go to $A$ or $B$, is used to optimize the parameters of the reaction coordinate. To do that one first assumes that the committor is a sigmoidal function of the reaction coordinate,
\begin{equation}
\label{eq:ml1}
p_B[q(x, \alpha), q^*, \eta]=\frac{1}{2}\left [1+\tanh([ q(x, \alpha) -q^*]\eta) \right],
\end{equation}
where $q^*$ and $\eta$ are two parameters that need to be optimized in addition to the parameter set $\alpha$. This model of the committor reflects the expectation that the committor grows smoothly from zero to one over a certain range of the reaction coordinate whose width can be tuned with the parameter $\eta$. In the above equation, $q^*$ is the value of the reaction coordinate for which the committor $p_B=1/2$. Given this model for the committor, the likelihood to observe a certain set of outcomes of the test trajectories can be written as 
\begin{equation}
L(\alpha, q^*, \eta)={\prod_{k(B)}} p_B[q(x_k)] {\prod_{k'(A)}} (1-p_B[q(x_{k'})]), 
\label{eq:likelihood}
\end{equation}
where we have exploited the fact that the test trajectories are statistically independent from each other. In the above equation, the first product on the right hand side runs over all trajectories starting from configurations $x_k$ that end in region $B$ and the second product runs over all trajectories ending up in $A$. The likelihood $L(\alpha, q^*, \eta)$ is a measure for the compatibility of the proposed committor model with the observed data. Maximizing this likelihood (or, in practice, its logarithm) with respect to the parameters  $\alpha$, $q^*$ and $\eta$ then yields the committor model, and hence the reaction coordinate $q(x, \alpha)$, that best explains the outcomes of the test trajectories. 

The ability of the maximum likelihood procedure sketched out above to yield a valid prediction of the committor depends on how the model reaction coordinate $q(x)$ is constructed. One possibility to do that is to define a number of collective coordinates $q_i(x)$, for which one suspects that they might play a role in the description of the mechanism of the transition under study, and to combine them linearly,
\begin{equation}
\label{eq:lm2}
q(x, \alpha)=\sum_i^M \alpha_i q_i(x).
\end{equation}
One can also define more general forms of the reaction coordinate, for instance by including also products of the collective variables, thereby increasing the number of free parameters. A further improvement consists in defining the reaction coordinate by a string in the space of collective variables \cite{JCP_Lechner_string_2010,JPCM_Leitold_string_2015}. In this approach, the reaction coordinate is constructed by projecting configurations on a piece-wise linear string defined by a set of points in collective variable space. Through this non-linear mapping the reaction coordinate is very flexible, such that it is better able to capture complex reaction mechanism.  

Likelihood maximization has been applied to identify mechanistic details of several rare transitions including nucleation in the Ising model \cite{peters:2006}, structural solid-solid transitions of terephtalic acid \cite{Beckham:2007} and the folding of a polymer chain with short range attractions \cite{Leitold_2014}. The method has also been used to study crystallization  \cite{beckham:2011,jungblut:2013a}. In particular, Beckham and Peters employed likelihood maximization to identify a scalar reaction coordinate for the liquid-solid transition of Lennard-Jonesium \cite{beckham:2011}. By analyzing trajectories sampled with aimless shooting and testing several structural and geometric candidate reaction coordinates, they established that the product $nQ_6^{\rm cl}$ of the size $n$ of the critical cluster and its structure as determined by the order parameter $Q_6^{\rm cl}$ evaluated for the cluster provides the best description of the nucleation process. Thus, both information about the size of the crystalline embryo as well as its degree of crystallinity need to be included in the definition of the reaction coordinate. The validity of the obtained reaction coordinate was confirmed by inspecting committor distributions. Interestingly, global structural indicators calculated for the entire system rather than only for the crystalline cluster proved to be very poor reaction coordinates.

As explained previously in section~\ref{subsec:orderParameter}, local crystallinity is usually probed with the help of Steinhardt bond order parameters  \cite{steinhardt:1983, tenwolde:1996}. Based on this order parameter one can decide, whether a particular particle sits in a crystalline or in a liquid environment. By performing a cluster analysis of neighboring crystalline particles, one can group the crystalline particles into connected clusters, several of which may exist in the system at a given time. The size of the largest of these crystalline clusters is then often used as reaction coordinate for the crystallization. The procedure to identify the largest crystalline cluster depends on several parameters. For the most popular definition of crystallinity \cite{tenwolde:1996}, these parameters are the thresholds for the nearest neighbor distance, the strength of crystalline bonds, and the number of crystalline connections, $\{d_{\rm th}, s_{\rm th}, n_{\rm th}\}$. These parameters are chosen according to some physical but nevertheless rather arbitrary criteria. 

An alternative way to choose the parameters entering the definition of local crystallinity is by likelihood maximization \cite{jungblut:2013a}. To do that, one considers the size $n(x, \alpha)$ of the largest crystalline cluster in a configuration $x$ as reaction coordinate for crystallization. Here, $\alpha = \{d_{\rm th}, s_{\rm th}, n_{\rm th}\}$ denotes the set of parameters used on the definition of the crystallinity as specified explicitly in section~\ref{subsec:orderParameter}. One then looks at the specific values of the parameters $\alpha$ for which the size of the largest crystalline cluster performs best as reaction coordinate. In other words, one searches for the parameter values that maximize the likelihood function of eq.~(\ref{eq:likelihood}) for a set of test trajectories generated for the crystallization under study. Application of this optimization procedure to the freezing of a moderately supercooled Lennard-Jones liquid \cite{jungblut:2013a} indicated that the optimum parameter set found by likelihood maximization differs considerably from the parameters that are commonly used \cite{tenwolde:1996,beckham:2011}. While the quality of the reaction coordinate is less sensitive to the next-neighbor distance, the thresholds for the number and strength of the crystalline connections should be chosen carefully. In particular, the quality of the largest cluster size as a reaction coordinate can be improved with respect to the standard definition by applying a stricter criterion in the definition of the crystalline bonds. However, as shown by analyzing committor distributions, even with the optimal definition the size of the largest crystalline cluster does not completely capture the crystallization mechanism \cite{jungblut:2013a}. Hence, to obtain a good reaction coordinate for crystallization, other variables must be included too as suggested earlier \cite{tenwolde:1999,lechner:2011}. 
 
\section{Applications}
\label{sec:applications}

\subsection{Hard sphere freezing}
\label{subsec:hs}

According to van der Waals theory, particles with a hard repulsive core and long-range attractive interactions display a first-order phase transition between a low density gas and a high density liquid. For purely repulsive hard sphere systems, however, a different question arises. It is clear that at low densities hard spheres exist as a disordered gas. At high densities near close packing, on the other hand, hard spheres must be arranged on a regular lattice such as the face-centered cubic or the hexagonal close-packed (hcp) lattices. Does this transition from a low density gas to a high density solid occur continuously of discontinuously? Or, in other words, is there a hard sphere freezing transition to a crystalline solid with long range order as first predicted by Kirkwood in 1951 \cite{kirkwood:1951}?  In the mid-1950s this was an important open question, but no theoretical tools were available to derive the properties of phase transitions from statistical mechanical principles \cite{vanhove:1957}. For instance, the virial expansion pioneered by Boltzmann does not give any indication about the existence of such a transition. So it was no surprise that at a symposium on ``The Many-Body Problem'' held 1957 at the Stevens Institute of Technology in Hoboken, New Jersey, during a round table discussion lead by G. E. Uhlenbeck on general topics of statistical mechanics a vote among prominent scientists (including several Nobel laureates) about this question was taken and it ended even \cite{uhlenbeck:1963}. The hesitation of half of the audience is understandable as it is indeed surprising that purely repulsive particles can form a stable crystal. The question was finally settled in favor of the existence of a first-order fluid-solid transition on the basis of now famous molecular dynamics simulations by Alder and Wainwright \cite{alder:1957} and of Monte Carlo simulations by Wood and Jacobson \cite{wood:1957}.

These simulations, as well as later ones by Hoover and Ree \cite{hoover:1968}, in which the entropy of both phases was computed, showed that for packing fractions in the range $\phi = 0.49 - 0.55$ (the packing fraction $\phi$ is the fraction of volume occupied by the hard spheres), a disordered fluid phase with packing fraction $\phi=0.49$ coexists with an ordered solid phase with packing fraction $\phi=0.55$. While for packing fractions below $\phi=0.49$ the fluid is the more stable form, at densities between $\phi=0.55$ and close packing occurring at $\phi=\pi /18\approx 0.74$ the hard sphere system exists as a solid. Both the solid and the fluid branch of the equation of state can be extended into the respective metastable region, indicating that the hard sphere freezing transition is indeed of first order.    

An interesting aspect of the hard sphere freezing transition is that it is purely entropic. To see what that means we need to consider the thermodynamics of the liquid-solid transition. We know that according to the Second Law of Thermodynamics a system with given volume $V$ and temperature $T$ exists in a state for which the Helmholtz free energy
\begin{equation}
\label{eq:Helmholtz}
F=U-TS
\end{equation}
is at a minimum. In other words, the thermodynamically stable phase is that one with the lower Helmholtz free energy. Other phases with higher free energy may be metastable but they tend to transform into the more stable phase provided that is kinetically possible. As stated in the famous inscription engraved on Boltzmann's tomb in Vienna's central cemetery,
\begin{equation}
\label{eq:entropy}
S=k \log W,
\end{equation}
the entropy $S$ is proportional to the logarithm of the number of microstates $W$ accessible to the system. (Incidentally, this equation, expressed by Boltzmann only in words, was explicitly written down later by Planck.) Boltzmann's expression for the entropy is the link between statistical mechanics and thermodynamics and it provides a precise prescription on how to calculate the entropy for a given microscopic model. The general intuition now is that for an ordered system a smaller number of states is accessible than for a disordered system and, hence, the entropy of the ordered system should be lower than that of the unordered one. Together with the expression of the Helmholtz free energy (see eq.~(\ref{eq:Helmholtz})), this suggests that for any particular substance the transition from the disordered liquid to the ordered solid can occur only if the lower entropy of the ordered crystal is compensated by a sufficiently large loss in energy. 

This scenario indeed describes some transitions, but in many cases this rather naive reasoning based on associating low/high entropy with apparent order/disorder is inaccurate, and it is entropy that drives the transition rather than energy. This is particularly evident for the hard sphere freezing transition. Since in the hard sphere system particles do not overlap due to the infinitely high interaction energy at contact, the total potential energy of the system vanishes for all possible configurations. Hence, the internal energy $U$ is just the kinetic energy and, as in the ideal gas, is a function of temperature only. Thus, along an isotherm, the only change in free energy is due to the changing entropy which can be determined with high accuracy from computer simulations (since the entropy is a measure for the available phase space volume and cannot be written as an ensemble average of a dynamical variable, advanced simulation techniques such as thermodynamic integration have to be used). While for packing fractions below $\phi=0.49$ the fluid has the higher entropy, for packing fractions above $\phi=0.55$ the solid is entropically favored and, hence, the thermodynamically stable phase. For intermediate packing fractions, the lower density fluid coexists with the higher density solid. The result that the solid, which we usually perceive as ``ordered'', has a higher entropy than the disordered liquid is slightly counter-intuitive and it demonstrates that one should be careful in relating high entropy with disorder (of course, if one {\em defines} disorder via the number of accessible states, no inconsistencies appear). In the case of the hard sphere system at high densities, the solid is the thermodynamically stable phase, because in the regular crystalline structure more configuration space is available.  It has become clear recently that other transitions also have very important entropic components including the isotropic-nematic transition of liquid crystals, the phase separation of binary mixtures, protein crystallization and entropic forces in general \cite{frenkel:1999,filion:2009}. 

While the early works of Alder and Wainwright and of Wood and Jacobsen, as well as  computer simulations performed later, unequivocally confirmed the existence of the hard sphere freezing transition, another related and seemingly simpler question proved much more persistent, namely that about the structure of the hard sphere solid. It took about forty more years to figure this out on the basis of computer simulations. At high densities,  the hard sphere solid is expected to exist in a structure that supports close packing such as the face-centered cubic or hexagonal close-packed structures, which differ in the particular way hexagonal close-packed layers of spheres are stacked on top of each other. The question now is: is the fcc or the hcp structure the more stable phase? Or, in other words, which one of the phases has the higher entropy? (Since in the hard sphere system all possible configurations have the same potential energy, the highest entropy corresponds to the lowest free energy.) While in both phases the first neighbor shell of each particle is identical, the fcc and hcp structures differ in their second neighbor shell and beyond. It is this difference that causes the entropy of the two phases to be different. Early analytic calculations based on a series expansion indicated that the hcp structure has a higher entropy than the fcc structure \cite{Stillinger:1968}. Computer simulations carried out in the 1960s, '70s and '80s gave inconclusive result which only demonstrated that the entropy difference per particle, if any, is very small compared to $k_{\rm B}$. Using the Frenkel-Ladd method to obtain the absolute free energy of a hard sphere system via thermodynamic integration from the Einstein crystal, a reference system with particles attached to lattice sites with harmonic springs, and sophisticated multicanonical sampling techniques it was finally shown in the late 1990s that the fcc structure has the highest entropy \cite{Frenkel:1997,Wilding:1997}. The free energy difference between fcc and hcp, which slightly increases as the density is raised from melting to close packing, has a value of about $10^{-3} k_{\rm B}T$ with error bars of the order of $10^{-4}  k_{\rm B}T$. The fcc structure has also a higher entropy than all other, random and periodic, stacking sequences \cite{Huse:1999}.  

More recently, computer simulations of hard sphere systems concentrated on the nucleation of the stable solid phase from the metastable fluid generated by a sudden compression. Contrary to the assumptions of classical nucleation theory, these computer simulations showed that already for packing fractions $ 0.42 <   \phi < 0.49$, well below the freezing transition of a hard sphere fluid, a structural precursor to the development of long-range order exists \cite{Truskett:1998}. It manifests itself as a shoulder in the second peak of the radial distribution function and is a consequence of crystalline domains commensurate with those in the crystal at the melting point ($\phi = 0.55$). Similar features have been experimentally observed for several simple liquids and traced back to the enhancement of local crystalline ordering in further simulations \cite{omalley:2005,schilling:2010} and analytically \cite{brader:2008}.

Although well studied, the behavior of the hard sphere systems still bears some mysteries. Due to advances in experimental techniques such as confocal microscopy \cite{gasser:2001}, colloidal systems with hard repulsive interactions can now be studied particle by particle and enable a direct comparison between experimental and simulation data. The critical size of the crystallizing droplet, the interfacial tension, as well as the shape and structure of the crystalline domains obtained in simulations and experiments were indeed very similar. The experimental nucleation rates, however, agree with the values measured in simulations only in the upper part of the coexistence region. The deviation increases with the decreasing volume fraction and amounts up to orders of magnitude. Currently, no conclusive explanation has been given for this observation, although several reasons have been proposed since Auer and Frenkel \cite{auer:2001a,auer:2005} first measured the nucleation rates in simulations. The list of possible issues includes polydispersity of the experimental particles \cite{auer:2001a}, which actually decreases the nucleation rates, gravitation effects \cite{russo:2012} and hydrodynamic interactions \cite{radu2014solvent}, both enhancing nucleation, as well as initial structural inhomogeneity of the supercooled fluid \cite{KawasakiTanaka:2010}. The validity of these arguments and some further explanations are discussed in the recent review by Palberg \cite{palberg:2014}, which also includes a comprehensive record of nucleation rates determined in various experiments and simulations. 

Furthermore, the existence and nature of the glass transition in a supercooled (monodisperse) hard sphere fluid is still under discussion. Since the relaxation times at high density become quite long, early simulations were inconclusive about whether the absence of the crystalline phase is an indication of the occurrence of the glass transition or just an artifact of the simulation time scales and finite sizes \cite{woodcock:1976,gordon:1976,woodcock:1978}. Later, the seminal experiment by Pusey and van Megen \cite{pusey:1986} localized the glassy state at $\phi > 0.58$, although there were spurious occurrences of the crystalline phase at the walls of the container and at the meniscus. Some crystalline domains were found even in the bulk, but the overall glassy behavior of the system was in line with mode coupling theory \cite{vanmegen:1993,vanmegen:1994}. In contrast, another experiment performed under microgravity revealed that evidently glassy samples crystallized rather rapidly, indicating that the nucleation of crystals may be hindered by gravity effects \cite{zhu:1997}. Similarly, an increase of the system size considered in computer simulations led to a decrease of crystallization times, which became shorter than the times needed to equilibrate the system in a metastable glassy state \cite{rintoul:1996}. On the other hand, a few recent investigations agree that the polydispersity of the experimental samples provides a very realistic explanation for the experimental evidence of a hard sphere glass, since even a very small degree of polydispersity decreases the crystallization rates by a considerable amount \cite{williams:2001,auer:2001a,schoepe:2006,zaccarelli:2009,pusey:2009}. There are, however, indications that the occurrence of the glassy state is correlated with the initial conditions \cite{xu:2010a,valeriani:2011,sanz:2011} and possibly with the presence of local crystalline precursors \cite{franke:2014}. To our knowledge, there is still no general agreement on the existence of the glassy state or the conditions for its occurrence in a {\em monodisperse} hard sphere fluid. It is however clear that, in computer simulations, the relaxation times strongly depend on the preparation conditions, which allows to construct an artificial protocol that produces a glassy state of a monodisperse supercooled hard sphere fluid by enhancing microstates prohibiting crystallization during cooling. The subsequent crystallization of such a state turns out to be rather fascinating and intricate \cite{valeriani:2011,sanz:2011,sanz:2014} due to the nontrivial correlation between dynamic heterogeneities and crystallizing domains. Investigations of hard sphere glasses are, however, by no means concluded and the freezing of hard spheres is still not completely understood.

\subsection{Water freezing}
\label{subsec:water}

The freezing of water and the formation of ice crystals in wet air are of great current interest due to the significance of atmospheric ice particles for the climate on earth \cite{Pruppacher_Klett,Hegg_Baker}. Due to this importance, the crystallization of water has been studied in greater detail than that of any other substance \cite{sear:2012}. Here, we therefore do not attempt to give an overview of this field, but rather concentrate on computer simulation studies of homogeneous crystal nucleation occurring in supercooled water. In particular, we will focus on the formation of ice Ih, the hexagonal ice polymorph that forms at atmospheric pressure. From experimental studies of freezing \cite{stoekel:2005,Shaw:2011,Sellberg2014} it is known that neat liquid water can be deeply supercooled down to temperatures as low as 227 K, corresponding to 46 degrees below freezing. Note, however, that homogeneous freezing rates for water are known experimentally only in the range from 235 to 242 K. For temperatures below 235 K, which is known as the homogeneous nucleation temperature, nucleation is too fast to be resolved experimentally. Above 242 K, on the other hand, nucleation is so slow that it does not occur on the time scale of available experiments. While experiments can give information on nucleation rates and their dependence on external conditions, they cannot resolve the atomistic mechanism of crystallization happening at very short length and time scales. Such information can only be obtained from computer simulations of freezing process. In the case of ice forming in water, such simulations are difficult because of the complications related to the development of accurate water models. For instance, the available empirical water models yield melting temperatures that range from 146 K to 274 K and the best agreement is obtained for the TIP4P/ice model \cite{JPCM_Vega_ice_2005}. But while obtaining quantitative reproduction of experimental data for phase diagrams and nucleation rates is rarely achieved due to the inaccuracies of the underlying models, useful microscopic insights can still be gained from computer simulations. 

At very low temperatures, the nucleation barrier is low and the crystallization of supercooled water can be studied with straightforward MD simulations. By running  simulations of 512 TIP4P water molecules for hundreds of nanoseconds at a temperature of 230 K and a slightly reduced density of 0.96 g/cm$^3$ (corresponding to a pressure of about -1000 bar), Ohmine and collaborators observed several spontaneous crystallization events in which hexagonal ice, ice Ih, formed \cite{Ohmine:2002}. Spontaneous crystallization was also observed in MD simulations \cite{Nature_Molinero_2011} of the mW model \cite{Molinero:2009}, in which each water molecule is represented as a single particle with short range anisotropic interactions designed to reproduce the properties of hydrogen bonds. These simulations \cite{Nature_Molinero_2011} indicate that mW water has a maximum crystallization rate at a temperature of 202 K (an analogous analysis carried out on the basis of classical nucleation theory for experimental data yields a maximum crystallization rate at 225 K). For temperatures that are higher, nucleation of a crystallite becomes rarer because the nucleation barrier grows, while at lower temperatures crystallization is slowed down because the supercooled liquid becomes increasingly glassy such that it cannot be equilibrated any more. Critical nuclei, determined using the MFPT method, were found to be between 90 and 120 molecules large in the temperature regime from 205-208 K. The critical nuclei have a broad shape distribution suggesting that the free energy of the solid-liquid interface is low at these conditions. Interestingly, nucleation occurred predominantly where large patches of four-coordinated water molecules formed spontaneously in the liquid \cite{Ohmine:2002,Nature_Molinero_2011}. 

In less strongly supercooled water the formation of a critical crystalline nucleus becomes so rare that the crystallization process cannot be studied any more with brute force MD simulations. Thus, Galli and collaborators simulated the crystallization of mW water in the temperature range from 220 to 240 K using forward flux sampling \cite{PCCP_Galli_2011}. In these simulations, the Steinhardt bond order parameter $q_6$ was used to detect crystalline molecules and specify the size of the crystalline nucleus necessary for the definition of the interfaces for the FFS simulations. The estimated nucleation rates showed a very strong temperature dependence ranging from $2.1\times 10^{25}$ m$^{-3}$s$^{-1}$ at 220 K to $1.7 \times 10^{-7}$ m$^{-3}$s$^{-1}$ at 240 K. In contrast, umbrella sampling simulations carried out by Reinhardt and Doye \cite{JCP_Doye_2013} resulted in nucleation rates that were up to 5 orders of magnitude higher than those of Galli and collaborators, pointing to the difficulties in calculating accurate nucleation rates for crystallization. In nanoscopic droplets of supercooled water, crystallization rates can be lower than in the bulk by several orders of magnitude due to the Laplace pressure in the nanodroplet \cite{NatComm_Donadio_2013}. 

In computational studies of crystallization it is an important task to accurately identify and classify local molecular structures. In particular, for a mechanistic analysis of the crystallization process and for simulations with a controlled bias (such as umbrella sampling or metadynamics simulations) it is crucial to be able to distinguish the various crystalline structures that may form. While various methods based on Steinhardt bond order parameters have been successfully applied to a variety of problems, they often yield unreliable results particularly for complex open structures and in the presence of elastic deformations and thermal fluctuations. For instance, it is notoriously difficult to distinguish local configurations of liquid water and of the various forms of crystalline and amorphous ice \cite{Ravi_Trout_2003}.  As discussed by Brukhno and collaborators \cite{JPCM_Handel_2008}, standard Steinhardt bond order parameters are ineffective in detecting hexagonal and cubic ice because oxygen atoms with different tetrahedral hydrogen bonding patterns occur. Recently, it was suggested to view the assignment of local structures as a pattern recognition problem that can be addressed with machine learning algorithms \cite{JCP_Geiger_2013}. It has been shown that an appropriately trained neural network can be used to accurately detect local ordered structures and distinguishing the phases of ice in a broad range of conditions. 

An alternative approach to determine homogeneous nucleation rates for water and study the properties of critical nuclei is the so-called ``seeding'' method \cite{JCP_LI_2005,JACS_Valeriani_2013,Vega_ice_2015}. In this method, a solid cluster is inserted into the supercooled liquid and briefly equilibrated. Then, several MD runs starting from this equilibrated configuration are carried out to determine the temperature at which the cluster neither grows nor shrinks. At this particular temperature the inserted crystalline cluster is the critical cluster. Assuming that CNT is valid and knowing the difference in chemical potential between the solid and the liquid phases, one can then determine the interfacial free energy at the given thermodynamic conditions. Furthermore, complementing this information with the diffusion coefficient for the cluster size evaluated at the critical size, one can determine the homogeneous nucleation rate. The advantage of the seeding method is that it allows to calculate nucleation rates at moderate undercooling, where the computational cost of other methods is prohibitive due to the large size of the critical cluster. The drawback is that the method relies on the validity of CNT and on the assumption that the initial seed indeed corresponds to the crystalline embryo that would form during the spontaneous crystallization process. Using the seeding method, Valeriani and collaborators \cite{JACS_Valeriani_2013} have studied the temperature dependence of the size of the critical cluster, finding that it varies from $\approx$8000 molecules at 15 K below melting to $\approx$600 molecules at 35 K below melting. The free energy barriers range from about 500 $k_{\rm B}T$ for the largest critical clusters to about $80$ $k_{\rm B}T$ for the smallest ones. Simulations carried out for several different water models indicate that the nucleation rate varies by 180 orders of magnitude between 35 K and 15 K below melting. For the TIP4P/2005 model (and to a slightly lesser degree also for the TIP4P/ice model) the computed nucleation rates are consistent with those of real water as known from experiments \cite{stoekel:2005}. 

An interesting question is whether the formation of hexagonal ice Ih proceeds directly or via the formation of cubic ice Ic first. Quigley and Rodger \cite{Quigley_Rodger_2008} have studied the free energetics of crystallization of TIP4P water by means of metadynamics and have found that at a temperature of 180K (TIP4P ice has a melting point of 232K) cubic rather than hexagonal ice forms preferentially. However, no such preference was observed by Radakrishnan and Trout \cite{Ravi_Trout_2003} for the same model at the same temperature using umbrella sampling MC simulations. Later simulations carried out for the TIP4P/ice model  \cite{JCP_Geiger_2013} found that cubic structures form only transiently at the surface of hexagonal crystallites, indicating that the occurrence of cubic ice might have to do with the particular collective variable used to drive the transition. Recent forward-flux sampling simulations carried out with the same model found transition states that are rich in cubic ice \cite{PNAS_Debenedetti_2015}. Long MD simulations carried out at 180 K for mW water \cite{JCP_Molinero_2010} yielded crystalline embryos consisting of a mixture of cubic and hexagonal regions. The critical nuclei identified in these simulations for these extreme conditions in the middle of water's so-called no-man's land consist of less then ten water molecules. A similar mixture of cubic and hexagonal ice was also found in the crystalline nuclei forming in FFS simulations of the mW model \cite{PCCP_Galli_2011}. 

\section{Summary and Outlook}
\label{sec:outlook}

In these lecture notes, we have sought to give an overview on computational methods to study self-assembly by nucleation and growth. For simplicity, we concentrated on homogeneous nucleation omitting a discussion of heterogeneous nucleation which occurs on surfaces and near impurities. We would like to stress, however, that the techniques discussed here can be applied equally well to heterogeneous nucleation, which is the main pathway to crystallization in nature and technology. As we have focused on the methodology employed in computer simulations of phase transformations, we have included only few illustrative examples.
Of course, many more computational studies of crystallization have been carried out shedding light on the freezing transition in a wide variety of substances including salts \cite{valeriani:2005}, hard polyhedra \cite{thapar:2014}, small organic molecules \cite{santiso:2011,shah_trout:2011} and proteins \cite{sear:2007,tenwolde:1997} to name but a few.

Although computer simulations have contributed significantly to our understanding of freezing mechanisms, much work remains to be done. For instance, recent experiments and simulations indicate that many crystallization processes from solution follow a non-classical nucleation pathway involving the formation of metastable pre-nucleation clusters \cite{baumgartner:2013,sear:2012,raitieri:2010,gebauer:2014}, which are liquid or solid precursors forming on the way to the final crystal. While the computer simulation methods presented here permit the efficient sampling of crystallization pathways and the calculation of nucleation rates, an accurate identification of the reaction coordinate is still difficult and a perfect reaction coordinate has not been found even for simple systems such as Lennard-Jonesium. It is clear that the size of the crystalline cluster and also its shape play an important role for the crystallization mechanism and need to be taken into account by describing a freezing transition. However, committor analysis, for instance carried out for the freezing of Lennard-Jones particles, indicates that additional but so far unknown degrees of freedom also contribute to the reaction coordinate. In the future, application of modern machine learning methods may lead to progress in this challenging problem.
 

\acknowledgments
This work was supported by the 
Austrian Science Fund (FWF) within the Project V 305-N27 as well as the SFB ViCoM (Grant F41).

\bibstyle{revtex}


\begin{thebibliography}{275}
\expandafter\ifx\csname natexlab\endcsname\relax\def\natexlab#1{#1}\fi
\expandafter\ifx\csname bibnamefont\endcsname\relax
  \def\bibnamefont#1{#1}\fi
\expandafter\ifx\csname bibfnamefont\endcsname\relax
  \def\bibfnamefont#1{#1}\fi
\expandafter\ifx\csname citenamefont\endcsname\relax
  \def\citenamefont#1{#1}\fi
\expandafter\ifx\csname url\endcsname\relax
  \def\url#1{\texttt{#1}}\fi
\expandafter\ifx\csname urlprefix\endcsname\relax\def\urlprefix{URL }\fi
\providecommand{\bibinfo}[2]{#2}
\providecommand{\eprint}[2][]{\url{#2}}

\bibitem[{\citenamefont{McPherson}(1991)}]{McPherson19911}
\bibinfo{author}{\bibfnamefont{A.}~\bibnamefont{McPherson}},
  \bibinfo{journal}{J. Cryst. Growth} \textbf{\bibinfo{volume}{110}},
  \bibinfo{pages}{1} (\bibinfo{year}{1991}).

\bibitem[{\citenamefont{{ten Wolde} and Frenkel}(1997)}]{tenwolde:1997}
\bibinfo{author}{\bibfnamefont{P.~R.} \bibnamefont{{ten Wolde}}}
  \bibnamefont{and} \bibinfo{author}{\bibfnamefont{D.}~\bibnamefont{Frenkel}},
  \bibinfo{journal}{Science} \textbf{\bibinfo{volume}{277}},
  \bibinfo{pages}{1975} (\bibinfo{year}{1997}).

\bibitem[{\citenamefont{Gasser}(2009)}]{gasser:2009}
\bibinfo{author}{\bibfnamefont{U.}~\bibnamefont{Gasser}}, \bibinfo{journal}{J.
  Phys.: Condens. Matter} \textbf{\bibinfo{volume}{21}},
  \bibinfo{pages}{203101} (\bibinfo{year}{2009}).

\bibitem[{\citenamefont{Thomas et~al.}(1994)\citenamefont{Thomas, Morfill,
  Demmel, Goree, Feuerbacher, and M{\"o}hlmann}}]{PhysRevLett.73.652}
\bibinfo{author}{\bibfnamefont{H.}~\bibnamefont{Thomas}},
  \bibinfo{author}{\bibfnamefont{G.~E.} \bibnamefont{Morfill}},
  \bibinfo{author}{\bibfnamefont{V.}~\bibnamefont{Demmel}},
  \bibinfo{author}{\bibfnamefont{J.}~\bibnamefont{Goree}},
  \bibinfo{author}{\bibfnamefont{B.}~\bibnamefont{Feuerbacher}},
  \bibnamefont{and}
  \bibinfo{author}{\bibfnamefont{D.}~\bibnamefont{M{\"o}hlmann}},
  \bibinfo{journal}{Phys. Rev. Lett.} \textbf{\bibinfo{volume}{73}},
  \bibinfo{pages}{652} (\bibinfo{year}{1994}).

\bibitem[{\citenamefont{Morfill and Ivlev}(2009)}]{RevModPhys.81.1353}
\bibinfo{author}{\bibfnamefont{G.~E.} \bibnamefont{Morfill}} \bibnamefont{and}
  \bibinfo{author}{\bibfnamefont{A.~V.} \bibnamefont{Ivlev}},
  \bibinfo{journal}{Rev. Mod. Phys.} \textbf{\bibinfo{volume}{81}},
  \bibinfo{pages}{1353} (\bibinfo{year}{2009}).

\bibitem[{\citenamefont{Fahrenheit}(1724)}]{Fahrenheit1724}
\bibinfo{author}{\bibfnamefont{D.~G.} \bibnamefont{Fahrenheit}},
  \bibinfo{journal}{Phil. Trans.} \textbf{\bibinfo{volume}{33}},
  \bibinfo{pages}{78} (\bibinfo{year}{1724}).

\bibitem[{\citenamefont{St{\"o}kel et~al.}(2005)\citenamefont{St{\"o}kel,
  Weidinger, Baumg{\"a}rtel, and Leisner}}]{stoekel:2005}
\bibinfo{author}{\bibfnamefont{P.}~\bibnamefont{St{\"o}kel}},
  \bibinfo{author}{\bibfnamefont{I.~M.} \bibnamefont{Weidinger}},
  \bibinfo{author}{\bibfnamefont{H.}~\bibnamefont{Baumg{\"a}rtel}},
  \bibnamefont{and} \bibinfo{author}{\bibfnamefont{T.}~\bibnamefont{Leisner}},
  \bibinfo{journal}{J. Phys. Chem. A} \textbf{\bibinfo{volume}{109}},
  \bibinfo{pages}{2540} (\bibinfo{year}{2005}).

\bibitem[{\citenamefont{Sellberg et~al.}(2014)\citenamefont{Sellberg, Huang,
  {McQueen}, Loh, Laksmono, Schlesinger, Sierra, Nordlund, Hampton, Starodub
  et~al.}}]{Sellberg2014}
\bibinfo{author}{\bibfnamefont{J.~A.} \bibnamefont{Sellberg}},
  \bibinfo{author}{\bibfnamefont{C.}~\bibnamefont{Huang}},
  \bibinfo{author}{\bibfnamefont{T.~A.} \bibnamefont{{McQueen}}},
  \bibinfo{author}{\bibfnamefont{N.~D.} \bibnamefont{Loh}},
  \bibinfo{author}{\bibfnamefont{H.}~\bibnamefont{Laksmono}},
  \bibinfo{author}{\bibfnamefont{D.}~\bibnamefont{Schlesinger}},
  \bibinfo{author}{\bibfnamefont{R.~G.} \bibnamefont{Sierra}},
  \bibinfo{author}{\bibfnamefont{D.}~\bibnamefont{Nordlund}},
  \bibinfo{author}{\bibfnamefont{C.~Y.} \bibnamefont{Hampton}},
  \bibinfo{author}{\bibfnamefont{D.}~\bibnamefont{Starodub}},
  \bibnamefont{et~al.}, \bibinfo{journal}{Nature}
  \textbf{\bibinfo{volume}{510}}, \bibinfo{pages}{381} (\bibinfo{year}{2014}).

\bibitem[{\citenamefont{Volmer and Weber}(1926)}]{Volmer:1926}
\bibinfo{author}{\bibfnamefont{M.}~\bibnamefont{Volmer}} \bibnamefont{and}
  \bibinfo{author}{\bibfnamefont{A.}~\bibnamefont{Weber}}, \bibinfo{journal}{Z.
  Phys. Chem.} \textbf{\bibinfo{volume}{119}}, \bibinfo{pages}{277}
  (\bibinfo{year}{1926}).

\bibitem[{\citenamefont{Becker and D{\"o}ring}(1935)}]{Becker:1935}
\bibinfo{author}{\bibfnamefont{R.}~\bibnamefont{Becker}} \bibnamefont{and}
  \bibinfo{author}{\bibfnamefont{W.}~\bibnamefont{D{\"o}ring}},
  \bibinfo{journal}{Ann. Phys. (Lepzig)} \textbf{\bibinfo{volume}{24}},
  \bibinfo{pages}{719} (\bibinfo{year}{1935}).

\bibitem[{\citenamefont{Turnbull and Fisher}(1949)}]{Turnbull:1949}
\bibinfo{author}{\bibfnamefont{D.}~\bibnamefont{Turnbull}} \bibnamefont{and}
  \bibinfo{author}{\bibfnamefont{J.~C.} \bibnamefont{Fisher}},
  \bibinfo{journal}{J. Chem. Phys.} \textbf{\bibinfo{volume}{17}},
  \bibinfo{pages}{71} (\bibinfo{year}{1949}).

\bibitem[{\citenamefont{Geiger and Dellago}(2013)}]{JCP_Geiger_2013}
\bibinfo{author}{\bibfnamefont{P.}~\bibnamefont{Geiger}} \bibnamefont{and}
  \bibinfo{author}{\bibfnamefont{C.}~\bibnamefont{Dellago}},
  \bibinfo{journal}{J. Chem. Phys.} \textbf{\bibinfo{volume}{139}},
  \bibinfo{pages}{164105} (\bibinfo{year}{2013}).

\bibitem[{\citenamefont{Cahn and Hilliard}(1958)}]{Cahn1958}
\bibinfo{author}{\bibfnamefont{J.~W.} \bibnamefont{Cahn}} \bibnamefont{and}
  \bibinfo{author}{\bibfnamefont{J.~E.} \bibnamefont{Hilliard}},
  \bibinfo{journal}{J. Chem. Phys.} \textbf{\bibinfo{volume}{28}},
  \bibinfo{pages}{258} (\bibinfo{year}{1958}).

\bibitem[{\citenamefont{Binder}(1987)}]{Binder_RepProgPhys_1987}
\bibinfo{author}{\bibfnamefont{K.}~\bibnamefont{Binder}},
  \bibinfo{journal}{Rep. Prog. Phys.} \textbf{\bibinfo{volume}{50}},
  \bibinfo{pages}{783} (\bibinfo{year}{1987}).

\bibitem[{\citenamefont{Frenkel and Smit}(2001)}]{Frenkel_Smit_book}
\bibinfo{author}{\bibfnamefont{D.}~\bibnamefont{Frenkel}} \bibnamefont{and}
  \bibinfo{author}{\bibfnamefont{B.}~\bibnamefont{Smit}},
  \emph{\bibinfo{title}{Understanding Molecular Simulation}}
  (\bibinfo{publisher}{Academic Press}, \bibinfo{address}{San Diego},
  \bibinfo{year}{2001}).

\bibitem[{\citenamefont{Allen and Tildesley}(1987)}]{Allen_Tildesley_book}
\bibinfo{author}{\bibfnamefont{M.}~\bibnamefont{Allen}} \bibnamefont{and}
  \bibinfo{author}{\bibfnamefont{D.}~\bibnamefont{Tildesley}},
  \emph{\bibinfo{title}{Computer Simulation of Liquids}}
  (\bibinfo{publisher}{Oxford Science Publications}, \bibinfo{address}{Oxford},
  \bibinfo{year}{1987}).

\bibitem[{\citenamefont{Salvalaglio et~al.}(2012)\citenamefont{Salvalaglio,
  Vetter, Giberti, Mazzotti, and Parrinello}}]{salvalaglio:2012}
\bibinfo{author}{\bibfnamefont{M.}~\bibnamefont{Salvalaglio}},
  \bibinfo{author}{\bibfnamefont{T.}~\bibnamefont{Vetter}},
  \bibinfo{author}{\bibfnamefont{F.}~\bibnamefont{Giberti}},
  \bibinfo{author}{\bibfnamefont{M.}~\bibnamefont{Mazzotti}}, \bibnamefont{and}
  \bibinfo{author}{\bibfnamefont{M.}~\bibnamefont{Parrinello}},
  \bibinfo{journal}{J. Am. Chem. Soc.} \textbf{\bibinfo{volume}{134}},
  \bibinfo{pages}{17221} (\bibinfo{year}{2012}).

\bibitem[{\citenamefont{Salvalaglio et~al.}(2013)\citenamefont{Salvalaglio,
  Vetter, Mazzotti, and Parrinello}}]{salvalaglio:2013}
\bibinfo{author}{\bibfnamefont{M.}~\bibnamefont{Salvalaglio}},
  \bibinfo{author}{\bibfnamefont{T.}~\bibnamefont{Vetter}},
  \bibinfo{author}{\bibfnamefont{M.}~\bibnamefont{Mazzotti}}, \bibnamefont{and}
  \bibinfo{author}{\bibfnamefont{M.}~\bibnamefont{Parrinello}},
  \bibinfo{journal}{Angew. Chem. Int. Ed.} \textbf{\bibinfo{volume}{52}},
  \bibinfo{pages}{13369} (\bibinfo{year}{2013}).

\bibitem[{\citenamefont{Baumgartner et~al.}(2013)\citenamefont{Baumgartner,
  Dey, Bomans, {Le Coadou}, Fratzl, Sommerdijk, and Faivre}}]{baumgartner:2013}
\bibinfo{author}{\bibfnamefont{J.}~\bibnamefont{Baumgartner}},
  \bibinfo{author}{\bibfnamefont{A.}~\bibnamefont{Dey}},
  \bibinfo{author}{\bibfnamefont{P.~H.~H.} \bibnamefont{Bomans}},
  \bibinfo{author}{\bibfnamefont{C.}~\bibnamefont{{Le Coadou}}},
  \bibinfo{author}{\bibfnamefont{P.}~\bibnamefont{Fratzl}},
  \bibinfo{author}{\bibfnamefont{N.~A. J.~M.} \bibnamefont{Sommerdijk}},
  \bibnamefont{and} \bibinfo{author}{\bibfnamefont{D.}~\bibnamefont{Faivre}},
  \bibinfo{journal}{Nature Materials} \textbf{\bibinfo{volume}{12}},
  \bibinfo{pages}{310} (\bibinfo{year}{2013}).

\bibitem[{\citenamefont{Gebauer et~al.}(2014)\citenamefont{Gebauer,
  Kellermeier, Gale, Bergstr{\"o}m, and C{\"o}lfen}}]{gebauer:2014}
\bibinfo{author}{\bibfnamefont{D.}~\bibnamefont{Gebauer}},
  \bibinfo{author}{\bibfnamefont{M.}~\bibnamefont{Kellermeier}},
  \bibinfo{author}{\bibfnamefont{J.~D.} \bibnamefont{Gale}},
  \bibinfo{author}{\bibfnamefont{L.}~\bibnamefont{Bergstr{\"o}m}},
  \bibnamefont{and}
  \bibinfo{author}{\bibfnamefont{H.}~\bibnamefont{C{\"o}lfen}},
  \bibinfo{journal}{Chem. Soc. Rev.} \textbf{\bibinfo{volume}{43}},
  \bibinfo{pages}{2348} (\bibinfo{year}{2014}).

\bibitem[{\citenamefont{Jacobs et~al.}(2015)\citenamefont{Jacobs, Reinhardt,
  and Frenkel}}]{Jacobs19052015}
\bibinfo{author}{\bibfnamefont{W.~M.} \bibnamefont{Jacobs}},
  \bibinfo{author}{\bibfnamefont{A.}~\bibnamefont{Reinhardt}},
  \bibnamefont{and} \bibinfo{author}{\bibfnamefont{D.}~\bibnamefont{Frenkel}},
  \bibinfo{journal}{Proc. Natl. Acad. Sci. USA} \textbf{\bibinfo{volume}{112}},
  \bibinfo{pages}{6313} (\bibinfo{year}{2015}).

\bibitem[{\citenamefont{Whitelam and Jack}(2015)}]{Whitelam_Jack_2015}
\bibinfo{author}{\bibfnamefont{S.}~\bibnamefont{Whitelam}} \bibnamefont{and}
  \bibinfo{author}{\bibfnamefont{R.~L.} \bibnamefont{Jack}},
  \bibinfo{journal}{Annu. Rev. Phys. Chem.} \textbf{\bibinfo{volume}{66}},
  \bibinfo{pages}{143} (\bibinfo{year}{2015}).

\bibitem[{\citenamefont{Debenedetti}(1997)}]{debenedetti}
\bibinfo{author}{\bibfnamefont{P.~G.} \bibnamefont{Debenedetti}},
  \emph{\bibinfo{title}{Metastable Liquids: Concepts and Principles}}
  (\bibinfo{publisher}{Princeton University Press},
  \bibinfo{address}{Princeton}, \bibinfo{year}{1997}).

\bibitem[{\citenamefont{Kashchiev}(2000)}]{kashchiev:2000}
\bibinfo{author}{\bibfnamefont{D.}~\bibnamefont{Kashchiev}},
  \emph{\bibinfo{title}{Nucleation: {B}asic {T}hory with {A}pplications}}
  (\bibinfo{publisher}{Butterworth-{H}einemann, Oxford}, \bibinfo{year}{2000}).

\bibitem[{\citenamefont{Gibbs}(1875)}]{gibbs:1875}
\bibinfo{author}{\bibfnamefont{J.~W.} \bibnamefont{Gibbs}},
  \bibinfo{journal}{Trans. Conn. Acad.} \textbf{\bibinfo{volume}{3}},
  \bibinfo{pages}{108} (\bibinfo{year}{1875}).

\bibitem[{\citenamefont{Gibbs}(1877)}]{gibbs:1877}
\bibinfo{author}{\bibfnamefont{J.~W.} \bibnamefont{Gibbs}},
  \bibinfo{journal}{Trans. Conn. Acad.} \textbf{\bibinfo{volume}{3}},
  \bibinfo{pages}{333} (\bibinfo{year}{1877}).

\bibitem[{\citenamefont{Farkas}(1927)}]{farkas:1927}
\bibinfo{author}{\bibfnamefont{L.}~\bibnamefont{Farkas}}, \bibinfo{journal}{Z.
  Phys. Chem.} \textbf{\bibinfo{volume}{125}}, \bibinfo{pages}{236}
  (\bibinfo{year}{1927}).

\bibitem[{\citenamefont{Stranski and Kaischew}(1934)}]{stranski:1934}
\bibinfo{author}{\bibfnamefont{I.~N.} \bibnamefont{Stranski}} \bibnamefont{and}
  \bibinfo{author}{\bibfnamefont{R.}~\bibnamefont{Kaischew}},
  \bibinfo{journal}{Z. Phys. Chem. B} \textbf{\bibinfo{volume}{26}},
  \bibinfo{pages}{100} (\bibinfo{year}{1934}).

\bibitem[{\citenamefont{Kaischew and
  Stranski}(1934{\natexlab{a}})}]{kaischew:1934}
\bibinfo{author}{\bibfnamefont{R.}~\bibnamefont{Kaischew}} \bibnamefont{and}
  \bibinfo{author}{\bibfnamefont{I.~N.} \bibnamefont{Stranski}},
  \bibinfo{journal}{Z. Phys. Chem. B} \textbf{\bibinfo{volume}{26}},
  \bibinfo{pages}{114} (\bibinfo{year}{1934}{\natexlab{a}}).

\bibitem[{\citenamefont{Kaischew and
  Stranski}(1934{\natexlab{b}})}]{kaischew:1934a}
\bibinfo{author}{\bibfnamefont{R.}~\bibnamefont{Kaischew}} \bibnamefont{and}
  \bibinfo{author}{\bibfnamefont{I.~N.} \bibnamefont{Stranski}},
  \bibinfo{journal}{Z. Phys. Chem.} \textbf{\bibinfo{volume}{170}},
  \bibinfo{pages}{295} (\bibinfo{year}{1934}{\natexlab{b}}).

\bibitem[{\citenamefont{Frenkel}(1939{\natexlab{a}})}]{frenkel:1939}
\bibinfo{author}{\bibfnamefont{J.}~\bibnamefont{Frenkel}}, \bibinfo{journal}{J.
  Chem. Phys.} \textbf{\bibinfo{volume}{7}}, \bibinfo{pages}{200}
  (\bibinfo{year}{1939}{\natexlab{a}}).

\bibitem[{\citenamefont{Frenkel}(1939{\natexlab{b}})}]{frenkel:1939a}
\bibinfo{author}{\bibfnamefont{J.}~\bibnamefont{Frenkel}}, \bibinfo{journal}{J.
  Chem. Phys.} \textbf{\bibinfo{volume}{7}}, \bibinfo{pages}{538}
  (\bibinfo{year}{1939}{\natexlab{b}}).

\bibitem[{\citenamefont{Zeldovich}(1942)}]{zeldovich:1942}
\bibinfo{author}{\bibfnamefont{J.~B.} \bibnamefont{Zeldovich}},
  \bibinfo{journal}{JETP} \textbf{\bibinfo{volume}{12}}, \bibinfo{pages}{525}
  (\bibinfo{year}{1942}).

\bibitem[{\citenamefont{Oxtoby}(1998)}]{Oxtoby:1998}
\bibinfo{author}{\bibfnamefont{D.~W.} \bibnamefont{Oxtoby}},
  \bibinfo{journal}{Acc. Chem. Res.} \textbf{\bibinfo{volume}{31}},
  \bibinfo{pages}{91} (\bibinfo{year}{1998}).

\bibitem[{\citenamefont{Sear}(2007)}]{sear:2007}
\bibinfo{author}{\bibfnamefont{R.~P.} \bibnamefont{Sear}}, \bibinfo{journal}{J.
  Phys.: Condens. Matter} \textbf{\bibinfo{volume}{19}},
  \bibinfo{pages}{033101} (\bibinfo{year}{2007}).

\bibitem[{\citenamefont{Sear}(2012)}]{sear:2012}
\bibinfo{author}{\bibfnamefont{R.~P.} \bibnamefont{Sear}},
  \bibinfo{journal}{Int. Mater. Rev.} \textbf{\bibinfo{volume}{57}},
  \bibinfo{pages}{328} (\bibinfo{year}{2012}).

\bibitem[{\citenamefont{Kelton and Greer}(2010)}]{kelton:2010}
\bibinfo{author}{\bibfnamefont{K.~F.} \bibnamefont{Kelton}} \bibnamefont{and}
  \bibinfo{author}{\bibfnamefont{A.~L.} \bibnamefont{Greer}},
  \emph{\bibinfo{title}{Nucleation in Condensed Matter}}
  (\bibinfo{publisher}{Elsevier}, \bibinfo{address}{Amsterdam},
  \bibinfo{year}{2010}).

\bibitem[{\citenamefont{Kramers}(1940)}]{kramers:1940}
\bibinfo{author}{\bibfnamefont{H.~A.} \bibnamefont{Kramers}},
  \bibinfo{journal}{Physica} \textbf{\bibinfo{volume}{7}}, \bibinfo{pages}{284}
  (\bibinfo{year}{1940}).

\bibitem[{\citenamefont{H{\"a}nggi et~al.}(1990)\citenamefont{H{\"a}nggi,
  Talkner, and Borkovec}}]{haenggi:1990}
\bibinfo{author}{\bibfnamefont{P.}~\bibnamefont{H{\"a}nggi}},
  \bibinfo{author}{\bibfnamefont{P.}~\bibnamefont{Talkner}}, \bibnamefont{and}
  \bibinfo{author}{\bibfnamefont{M.}~\bibnamefont{Borkovec}},
  \bibinfo{journal}{Rev. Mod. Phys.} \textbf{\bibinfo{volume}{62}},
  \bibinfo{pages}{251} (\bibinfo{year}{1990}).

\bibitem[{\citenamefont{Zwanzig}(2001)}]{zwanzig:2001}
\bibinfo{author}{\bibfnamefont{R.}~\bibnamefont{Zwanzig}},
  \emph{\bibinfo{title}{Nonequilibrium statistical mechanics}}
  (\bibinfo{publisher}{Oxford University Press}, \bibinfo{year}{2001}).

\bibitem[{\citenamefont{Pontryagin et~al.}(1933)\citenamefont{Pontryagin,
  Andronov, and Vitt}}]{pontryagin:1933}
\bibinfo{author}{\bibfnamefont{L.~S.} \bibnamefont{Pontryagin}},
  \bibinfo{author}{\bibfnamefont{A.~A.} \bibnamefont{Andronov}},
  \bibnamefont{and} \bibinfo{author}{\bibfnamefont{A.~A.} \bibnamefont{Vitt}},
  \bibinfo{journal}{JETP} \textbf{\bibinfo{volume}{3}}, \bibinfo{pages}{165}
  (\bibinfo{year}{1933}).

\bibitem[{\citenamefont{Jungblut and Dellago}(2011)}]{jungblut:2011}
\bibinfo{author}{\bibfnamefont{S.}~\bibnamefont{Jungblut}} \bibnamefont{and}
  \bibinfo{author}{\bibfnamefont{C.}~\bibnamefont{Dellago}},
  \bibinfo{journal}{J. Chem. Phys.} \textbf{\bibinfo{volume}{134}},
  \bibinfo{pages}{104501} (\bibinfo{year}{2011}).

\bibitem[{\citenamefont{Frenkel}(1946)}]{frenkel:1946}
\bibinfo{author}{\bibfnamefont{J.}~\bibnamefont{Frenkel}},
  \emph{\bibinfo{title}{Kinetic Theory of Liquids}}
  (\bibinfo{publisher}{Clarendon Press}, \bibinfo{address}{Oxford},
  \bibinfo{year}{1946}).

\bibitem[{\citenamefont{Espinosa et~al.}(2014)\citenamefont{Espinosa, Sanz,
  Valeriani, and Vega}}]{Vega_ice_2015}
\bibinfo{author}{\bibfnamefont{J.~R.} \bibnamefont{Espinosa}},
  \bibinfo{author}{\bibfnamefont{E.}~\bibnamefont{Sanz}},
  \bibinfo{author}{\bibfnamefont{C.}~\bibnamefont{Valeriani}},
  \bibnamefont{and} \bibinfo{author}{\bibfnamefont{C.}~\bibnamefont{Vega}},
  \bibinfo{journal}{J. Chem. Phys.} \textbf{\bibinfo{volume}{141}},
  \bibinfo{pages}{18C529} (\bibinfo{year}{2014}).

\bibitem[{\citenamefont{Bennett}(1977)}]{bennett:1977}
\bibinfo{author}{\bibfnamefont{C.~H.} \bibnamefont{Bennett}}, in
  \emph{\bibinfo{booktitle}{Algorithms for {C}hemical {C}omputations}}, edited
  by \bibinfo{editor}{\bibfnamefont{R.~E.} \bibnamefont{Christoffersen}}
  (\bibinfo{publisher}{American Chemical Society, Washington, D.C.},
  \bibinfo{year}{1977}), vol.~\bibinfo{volume}{46} of
  \emph{\bibinfo{series}{ACS Symposium Series}}, p.~\bibinfo{pages}{63}.

\bibitem[{\citenamefont{Chandler}(1978)}]{chandler:1978}
\bibinfo{author}{\bibfnamefont{D.}~\bibnamefont{Chandler}},
  \bibinfo{journal}{J. Chem. Phys.} \textbf{\bibinfo{volume}{68}},
  \bibinfo{pages}{2959} (\bibinfo{year}{1978}).

\bibitem[{\citenamefont{ten Wolde et~al.}(1996)\citenamefont{ten Wolde,
  {Ruiz-Montero}, and Frenkel}}]{tenwolde:1996}
\bibinfo{author}{\bibfnamefont{P.~R.} \bibnamefont{ten Wolde}},
  \bibinfo{author}{\bibfnamefont{M.~J.} \bibnamefont{{Ruiz-Montero}}},
  \bibnamefont{and} \bibinfo{author}{\bibfnamefont{D.}~\bibnamefont{Frenkel}},
  \bibinfo{journal}{J. Chem. Phys.} \textbf{\bibinfo{volume}{104}},
  \bibinfo{pages}{9932} (\bibinfo{year}{1996}).

\bibitem[{\citenamefont{{Ruiz-Montero}
  et~al.}(1997)\citenamefont{{Ruiz-Montero}, Frenkel, and
  Brey}}]{ruizmontero:1997}
\bibinfo{author}{\bibfnamefont{M.~J.} \bibnamefont{{Ruiz-Montero}}},
  \bibinfo{author}{\bibfnamefont{D.}~\bibnamefont{Frenkel}}, \bibnamefont{and}
  \bibinfo{author}{\bibfnamefont{J.~J.} \bibnamefont{Brey}},
  \bibinfo{journal}{Mol. Phys.} \textbf{\bibinfo{volume}{90}},
  \bibinfo{pages}{925} (\bibinfo{year}{1997}).

\bibitem[{\citenamefont{Auer and Frenkel}(2004)}]{auer:2004}
\bibinfo{author}{\bibfnamefont{S.}~\bibnamefont{Auer}} \bibnamefont{and}
  \bibinfo{author}{\bibfnamefont{D.}~\bibnamefont{Frenkel}},
  \bibinfo{journal}{J. Chem. Phys.} \textbf{\bibinfo{volume}{120}},
  \bibinfo{pages}{3015} (\bibinfo{year}{2004}).

\bibitem[{\citenamefont{Auer and Frenkel}(2001{\natexlab{a}})}]{auer:2001}
\bibinfo{author}{\bibfnamefont{S.}~\bibnamefont{Auer}} \bibnamefont{and}
  \bibinfo{author}{\bibfnamefont{D.}~\bibnamefont{Frenkel}},
  \bibinfo{journal}{Nature} \textbf{\bibinfo{volume}{409}},
  \bibinfo{pages}{1020} (\bibinfo{year}{2001}{\natexlab{a}}).

\bibitem[{\citenamefont{Mandell et~al.}(1977)\citenamefont{Mandell, McTague,
  and Rahman}}]{rahmanII:1977}
\bibinfo{author}{\bibfnamefont{M.~J.} \bibnamefont{Mandell}},
  \bibinfo{author}{\bibfnamefont{J.~P.} \bibnamefont{McTague}},
  \bibnamefont{and} \bibinfo{author}{\bibfnamefont{A.}~\bibnamefont{Rahman}},
  \bibinfo{journal}{J. Chem. Phys.} \textbf{\bibinfo{volume}{66}},
  \bibinfo{pages}{3070} (\bibinfo{year}{1977}).

\bibitem[{\citenamefont{Alexander and McTague}(1978)}]{alexander:1978}
\bibinfo{author}{\bibfnamefont{S.}~\bibnamefont{Alexander}} \bibnamefont{and}
  \bibinfo{author}{\bibfnamefont{J.~P.} \bibnamefont{McTague}},
  \bibinfo{journal}{Phys. Rev. Lett.} \textbf{\bibinfo{volume}{41}},
  \bibinfo{pages}{702} (\bibinfo{year}{1978}).

\bibitem[{\citenamefont{Moroni et~al.}(2005{\natexlab{a}})\citenamefont{Moroni,
  ten Wolde, and Bolhuis}}]{moroni:2005}
\bibinfo{author}{\bibfnamefont{D.}~\bibnamefont{Moroni}},
  \bibinfo{author}{\bibfnamefont{P.~R.} \bibnamefont{ten Wolde}},
  \bibnamefont{and} \bibinfo{author}{\bibfnamefont{P.~G.}
  \bibnamefont{Bolhuis}}, \bibinfo{journal}{Phys. Rev. Lett.}
  \textbf{\bibinfo{volume}{94}}, \bibinfo{pages}{235703}
  (\bibinfo{year}{2005}{\natexlab{a}}).

\bibitem[{\citenamefont{Wang et~al.}(2007)\citenamefont{Wang, Gould, and
  Klein}}]{wang:2007}
\bibinfo{author}{\bibfnamefont{H.}~\bibnamefont{Wang}},
  \bibinfo{author}{\bibfnamefont{H.}~\bibnamefont{Gould}}, \bibnamefont{and}
  \bibinfo{author}{\bibfnamefont{W.}~\bibnamefont{Klein}},
  \bibinfo{journal}{Phys. Rev. E} \textbf{\bibinfo{volume}{76}},
  \bibinfo{pages}{031604} (\bibinfo{year}{2007}).

\bibitem[{\citenamefont{Beckham and Peters}(2011)}]{beckham:2011}
\bibinfo{author}{\bibfnamefont{G.~T.} \bibnamefont{Beckham}} \bibnamefont{and}
  \bibinfo{author}{\bibfnamefont{B.}~\bibnamefont{Peters}},
  \bibinfo{journal}{J. Phys. Chem. Lett.} \textbf{\bibinfo{volume}{2}},
  \bibinfo{pages}{1133} (\bibinfo{year}{2011}).

\bibitem[{\citenamefont{Jungblut et~al.}(2013)\citenamefont{Jungblut,
  Singraber, and Dellago}}]{jungblut:2013a}
\bibinfo{author}{\bibfnamefont{S.}~\bibnamefont{Jungblut}},
  \bibinfo{author}{\bibfnamefont{A.}~\bibnamefont{Singraber}},
  \bibnamefont{and} \bibinfo{author}{\bibfnamefont{C.}~\bibnamefont{Dellago}},
  \bibinfo{journal}{Mol. Phys.} \textbf{\bibinfo{volume}{111}},
  \bibinfo{pages}{3527} (\bibinfo{year}{2013}).

\bibitem[{\citenamefont{Ostwald}(1897)}]{ostwald:1897}
\bibinfo{author}{\bibfnamefont{W.}~\bibnamefont{Ostwald}}, \bibinfo{journal}{Z.
  Phys. Chem.} \textbf{\bibinfo{volume}{22}}, \bibinfo{pages}{289}
  (\bibinfo{year}{1897}).

\bibitem[{\citenamefont{Stranski and Totomanow}(1933)}]{stranski:1933}
\bibinfo{author}{\bibfnamefont{I.~N.} \bibnamefont{Stranski}} \bibnamefont{and}
  \bibinfo{author}{\bibfnamefont{D.}~\bibnamefont{Totomanow}},
  \bibinfo{journal}{Z. Phys. Chem.} \textbf{\bibinfo{volume}{163}},
  \bibinfo{pages}{399} (\bibinfo{year}{1933}).

\bibitem[{\citenamefont{Oxtoby and Evans}(1988)}]{oxtoby:1988}
\bibinfo{author}{\bibfnamefont{D.~W.} \bibnamefont{Oxtoby}} \bibnamefont{and}
  \bibinfo{author}{\bibfnamefont{R.}~\bibnamefont{Evans}}, \bibinfo{journal}{J.
  Chem. Phys.} \textbf{\bibinfo{volume}{89}}, \bibinfo{pages}{7521}
  (\bibinfo{year}{1988}).

\bibitem[{\citenamefont{Tolman}(1949)}]{tolman:1949}
\bibinfo{author}{\bibfnamefont{R.~C.} \bibnamefont{Tolman}},
  \bibinfo{journal}{J. Chem. Phys.} \textbf{\bibinfo{volume}{17}},
  \bibinfo{pages}{333} (\bibinfo{year}{1949}).

\bibitem[{\citenamefont{ten Wolde and Frenkel}(1998)}]{tenwolde:1998}
\bibinfo{author}{\bibfnamefont{P.~R.} \bibnamefont{ten Wolde}}
  \bibnamefont{and} \bibinfo{author}{\bibfnamefont{D.}~\bibnamefont{Frenkel}},
  \bibinfo{journal}{J. Chem. Phys.} \textbf{\bibinfo{volume}{109}},
  \bibinfo{pages}{9901} (\bibinfo{year}{1998}).

\bibitem[{\citenamefont{Iwamatsu}(1994)}]{iwamatsu:1994}
\bibinfo{author}{\bibfnamefont{M.}~\bibnamefont{Iwamatsu}},
  \bibinfo{journal}{J. Phys.: Condens. Matter} \textbf{\bibinfo{volume}{6}},
  \bibinfo{pages}{L173} (\bibinfo{year}{1994}).

\bibitem[{\citenamefont{Gr{\'a}n{\'a}sy}(1998)}]{granasy:1998}
\bibinfo{author}{\bibfnamefont{L.}~\bibnamefont{Gr{\'a}n{\'a}sy}},
  \bibinfo{journal}{J. Chem. Phys.} \textbf{\bibinfo{volume}{109}},
  \bibinfo{pages}{9660} (\bibinfo{year}{1998}).

\bibitem[{\citenamefont{Anisimov and Pierre}(2008)}]{anisimov:2008}
\bibinfo{author}{\bibfnamefont{M.~A.} \bibnamefont{Anisimov}} \bibnamefont{and}
  \bibinfo{author}{\bibfnamefont{H.~J.~S.} \bibnamefont{Pierre}},
  \bibinfo{journal}{Phys. Rev. E} \textbf{\bibinfo{volume}{78}},
  \bibinfo{pages}{011105} (\bibinfo{year}{2008}).

\bibitem[{\citenamefont{Tr{\"o}ster et~al.}(2012)\citenamefont{Tr{\"o}ster,
  Oettel, Block, Virnau, and Binder}}]{troester:2012}
\bibinfo{author}{\bibfnamefont{A.}~\bibnamefont{Tr{\"o}ster}},
  \bibinfo{author}{\bibfnamefont{M.}~\bibnamefont{Oettel}},
  \bibinfo{author}{\bibfnamefont{B.}~\bibnamefont{Block}},
  \bibinfo{author}{\bibfnamefont{P.}~\bibnamefont{Virnau}}, \bibnamefont{and}
  \bibinfo{author}{\bibfnamefont{K.}~\bibnamefont{Binder}},
  \bibinfo{journal}{J. Chem. Phys.} \textbf{\bibinfo{volume}{136}},
  \bibinfo{pages}{064709} (\bibinfo{year}{2012}).

\bibitem[{\citenamefont{Tr{\"o}ster and Binder}(2012)}]{troester:2012a}
\bibinfo{author}{\bibfnamefont{A.}~\bibnamefont{Tr{\"o}ster}} \bibnamefont{and}
  \bibinfo{author}{\bibfnamefont{K.}~\bibnamefont{Binder}},
  \bibinfo{journal}{J. Phys.: Condens. Matter} \textbf{\bibinfo{volume}{24}},
  \bibinfo{pages}{284107} (\bibinfo{year}{2012}).

\bibitem[{\citenamefont{Rull and Toxvaerd}(1983)}]{rull:1983}
\bibinfo{author}{\bibfnamefont{L.~F.} \bibnamefont{Rull}} \bibnamefont{and}
  \bibinfo{author}{\bibfnamefont{S.}~\bibnamefont{Toxvaerd}},
  \bibinfo{journal}{J. Chem. Phys.} \textbf{\bibinfo{volume}{78}},
  \bibinfo{pages}{3273} (\bibinfo{year}{1983}).

\bibitem[{\citenamefont{Broughton and Gilmer}(1986)}]{broughton:1986}
\bibinfo{author}{\bibfnamefont{J.~Q.} \bibnamefont{Broughton}}
  \bibnamefont{and} \bibinfo{author}{\bibfnamefont{G.~H.}
  \bibnamefont{Gilmer}}, \bibinfo{journal}{J. Chem. Phys.}
  \textbf{\bibinfo{volume}{84}}, \bibinfo{pages}{5759} (\bibinfo{year}{1986}).

\bibitem[{\citenamefont{Davidchack and Laird}(2003)}]{davidchack:2003}
\bibinfo{author}{\bibfnamefont{R.~L.} \bibnamefont{Davidchack}}
  \bibnamefont{and} \bibinfo{author}{\bibfnamefont{B.~B.} \bibnamefont{Laird}},
  \bibinfo{journal}{J. Chem. Phys.} \textbf{\bibinfo{volume}{118}},
  \bibinfo{pages}{7651} (\bibinfo{year}{2003}).

\bibitem[{\citenamefont{Davidchack and Laird}(2005)}]{davidchack:2005}
\bibinfo{author}{\bibfnamefont{R.~L.} \bibnamefont{Davidchack}}
  \bibnamefont{and} \bibinfo{author}{\bibfnamefont{B.~B.} \bibnamefont{Laird}},
  \bibinfo{journal}{Phys. Rev. Lett.} \textbf{\bibinfo{volume}{94}},
  \bibinfo{pages}{086102} (\bibinfo{year}{2005}).

\bibitem[{\citenamefont{Lundrigan and {Saika-Voivod}}(2009)}]{lundrigan:2009}
\bibinfo{author}{\bibfnamefont{S.~E.~M.} \bibnamefont{Lundrigan}}
  \bibnamefont{and}
  \bibinfo{author}{\bibfnamefont{I.}~\bibnamefont{{Saika-Voivod}}},
  \bibinfo{journal}{J. Chem. Phys.} \textbf{\bibinfo{volume}{131}},
  \bibinfo{pages}{104503} (\bibinfo{year}{2009}).

\bibitem[{\citenamefont{Trudu et~al.}(2006)\citenamefont{Trudu, Donadio, and
  Parrinello}}]{trudu:2006}
\bibinfo{author}{\bibfnamefont{F.}~\bibnamefont{Trudu}},
  \bibinfo{author}{\bibfnamefont{D.}~\bibnamefont{Donadio}}, \bibnamefont{and}
  \bibinfo{author}{\bibfnamefont{M.}~\bibnamefont{Parrinello}},
  \bibinfo{journal}{Phys. Rev. Lett.} \textbf{\bibinfo{volume}{97}},
  \bibinfo{pages}{105701} (\bibinfo{year}{2006}).

\bibitem[{\citenamefont{Prestipino et~al.}(2012)\citenamefont{Prestipino, Laio,
  and Tosatti}}]{prestipino:2012}
\bibinfo{author}{\bibfnamefont{S.}~\bibnamefont{Prestipino}},
  \bibinfo{author}{\bibfnamefont{A.}~\bibnamefont{Laio}}, \bibnamefont{and}
  \bibinfo{author}{\bibfnamefont{E.}~\bibnamefont{Tosatti}},
  \bibinfo{journal}{Phys. Rev. Lett.} \textbf{\bibinfo{volume}{108}},
  \bibinfo{pages}{225701} (\bibinfo{year}{2012}).

\bibitem[{\citenamefont{Prestipino et~al.}(2013)\citenamefont{Prestipino, Laio,
  and Tosatti}}]{prestipino:2013}
\bibinfo{author}{\bibfnamefont{S.}~\bibnamefont{Prestipino}},
  \bibinfo{author}{\bibfnamefont{A.}~\bibnamefont{Laio}}, \bibnamefont{and}
  \bibinfo{author}{\bibfnamefont{E.}~\bibnamefont{Tosatti}},
  \bibinfo{journal}{J. Chem. Phys.} \textbf{\bibinfo{volume}{138}},
  \bibinfo{pages}{064508} (\bibinfo{year}{2013}).

\bibitem[{\citenamefont{Prestipino et~al.}(2014)\citenamefont{Prestipino, Laio,
  and Tosatti}}]{prestipino:2014}
\bibinfo{author}{\bibfnamefont{S.}~\bibnamefont{Prestipino}},
  \bibinfo{author}{\bibfnamefont{A.}~\bibnamefont{Laio}}, \bibnamefont{and}
  \bibinfo{author}{\bibfnamefont{E.}~\bibnamefont{Tosatti}},
  \bibinfo{journal}{J. Chem. Phys.} \textbf{\bibinfo{volume}{140}},
  \bibinfo{pages}{094501} (\bibinfo{year}{2014}).

\bibitem[{\citenamefont{Landau and Binder}(2000)}]{LandauBinderBook:2000}
\bibinfo{author}{\bibfnamefont{D.~P.} \bibnamefont{Landau}} \bibnamefont{and}
  \bibinfo{author}{\bibfnamefont{K.}~\bibnamefont{Binder}},
  \emph{\bibinfo{title}{A {G}uide to {M}onte {C}arlo {S}imulations in
  {S}tatistical {P}hysics}} (\bibinfo{publisher}{Cambridge {U}niversity
  {P}ress}, \bibinfo{year}{2000}).

\bibitem[{\citenamefont{Panagiotopoulos}(2000)}]{panagiotopoulos:2000}
\bibinfo{author}{\bibfnamefont{A.~Z.} \bibnamefont{Panagiotopoulos}},
  \bibinfo{journal}{J. Phys.: Condens. Matter} \textbf{\bibinfo{volume}{12}},
  \bibinfo{pages}{R25} (\bibinfo{year}{2000}).

\bibitem[{\citenamefont{Rickman and {LeSar}}(2002)}]{rickman:2002}
\bibinfo{author}{\bibfnamefont{J.~M.} \bibnamefont{Rickman}} \bibnamefont{and}
  \bibinfo{author}{\bibfnamefont{R.}~\bibnamefont{{LeSar}}},
  \bibinfo{journal}{Annu. Rev. Mater. Res.} \textbf{\bibinfo{volume}{32}},
  \bibinfo{pages}{195} (\bibinfo{year}{2002}).

\bibitem[{\citenamefont{Vega et~al.}(2008)\citenamefont{Vega, Sanz, Abascal,
  and Noya}}]{vega:2008}
\bibinfo{author}{\bibfnamefont{C.}~\bibnamefont{Vega}},
  \bibinfo{author}{\bibfnamefont{E.}~\bibnamefont{Sanz}},
  \bibinfo{author}{\bibfnamefont{J.~L.~F.} \bibnamefont{Abascal}},
  \bibnamefont{and} \bibinfo{author}{\bibfnamefont{E.~G.} \bibnamefont{Noya}},
  \bibinfo{journal}{J. Phys.: Condens. Matter} \textbf{\bibinfo{volume}{20}},
  \bibinfo{pages}{153101} (\bibinfo{year}{2008}).

\bibitem[{\citenamefont{Mobley and Klimovich}(2012)}]{mobley:2012}
\bibinfo{author}{\bibfnamefont{D.~L.} \bibnamefont{Mobley}} \bibnamefont{and}
  \bibinfo{author}{\bibfnamefont{P.~V.} \bibnamefont{Klimovich}},
  \bibinfo{journal}{J. Chem. Phys.} \textbf{\bibinfo{volume}{137}},
  \bibinfo{pages}{230901} (\bibinfo{year}{2012}).

\bibitem[{\citenamefont{Sweatman}(2015)}]{sweatman:2015}
\bibinfo{author}{\bibfnamefont{M.~B.} \bibnamefont{Sweatman}},
  \bibinfo{journal}{Mol. Phys.} p.
  \bibinfo{pages}{10.1080/00268976.2015.1005704} (\bibinfo{year}{2015}).

\bibitem[{\citenamefont{Panagiotopoulos}(1987)}]{panagitopoulos:1987}
\bibinfo{author}{\bibfnamefont{A.~Z.} \bibnamefont{Panagiotopoulos}},
  \bibinfo{journal}{Mol. Phys.} \textbf{\bibinfo{volume}{61}},
  \bibinfo{pages}{813} (\bibinfo{year}{1987}).

\bibitem[{\citenamefont{Panagiotopoulos
  et~al.}(1988)\citenamefont{Panagiotopoulos, Quirke, Stapleton, and
  Tildesley}}]{panagitopoulos:1988}
\bibinfo{author}{\bibfnamefont{A.~Z.} \bibnamefont{Panagiotopoulos}},
  \bibinfo{author}{\bibfnamefont{N.}~\bibnamefont{Quirke}},
  \bibinfo{author}{\bibfnamefont{M.}~\bibnamefont{Stapleton}},
  \bibnamefont{and} \bibinfo{author}{\bibfnamefont{D.~J.}
  \bibnamefont{Tildesley}}, \bibinfo{journal}{Mol. Phys.}
  \textbf{\bibinfo{volume}{63}}, \bibinfo{pages}{527} (\bibinfo{year}{1988}).

\bibitem[{\citenamefont{Kerrache et~al.}(2008)\citenamefont{Kerrache, Horbach,
  and Binder}}]{kerrache:2008}
\bibinfo{author}{\bibfnamefont{A.}~\bibnamefont{Kerrache}},
  \bibinfo{author}{\bibfnamefont{J.}~\bibnamefont{Horbach}}, \bibnamefont{and}
  \bibinfo{author}{\bibfnamefont{K.}~\bibnamefont{Binder}},
  \bibinfo{journal}{EPL} \textbf{\bibinfo{volume}{81}}, \bibinfo{pages}{58001}
  (\bibinfo{year}{2008}).

\bibitem[{\citenamefont{Pedersen}(2013)}]{pedersen:2013}
\bibinfo{author}{\bibfnamefont{U.~R.} \bibnamefont{Pedersen}},
  \bibinfo{journal}{J. Chem. Phys.} \textbf{\bibinfo{volume}{139}},
  \bibinfo{pages}{104102} (\bibinfo{year}{2013}).

\bibitem[{\citenamefont{Schilling and Schmid}(2009)}]{schilling:2009}
\bibinfo{author}{\bibfnamefont{T.}~\bibnamefont{Schilling}} \bibnamefont{and}
  \bibinfo{author}{\bibfnamefont{F.}~\bibnamefont{Schmid}},
  \bibinfo{journal}{J. Chem. Phys.} \textbf{\bibinfo{volume}{131}},
  \bibinfo{pages}{231102} (\bibinfo{year}{2009}).

\bibitem[{\citenamefont{Frenkel and Ladd}(1984)}]{frenkel:1984}
\bibinfo{author}{\bibfnamefont{D.}~\bibnamefont{Frenkel}} \bibnamefont{and}
  \bibinfo{author}{\bibfnamefont{A.~J.~C.} \bibnamefont{Ladd}},
  \bibinfo{journal}{J. Chem. Phys.} \textbf{\bibinfo{volume}{81}},
  \bibinfo{pages}{3188} (\bibinfo{year}{1984}).

\bibitem[{\citenamefont{Polson et~al.}(2000)\citenamefont{Polson, Trizac,
  Pronk, and Frenkel}}]{polson:2000}
\bibinfo{author}{\bibfnamefont{J.~M.} \bibnamefont{Polson}},
  \bibinfo{author}{\bibfnamefont{E.}~\bibnamefont{Trizac}},
  \bibinfo{author}{\bibfnamefont{S.}~\bibnamefont{Pronk}}, \bibnamefont{and}
  \bibinfo{author}{\bibfnamefont{D.}~\bibnamefont{Frenkel}},
  \bibinfo{journal}{J. Chem. Phys.} \textbf{\bibinfo{volume}{112}},
  \bibinfo{pages}{5339} (\bibinfo{year}{2000}).

\bibitem[{\citenamefont{Wilms et~al.}(2012)\citenamefont{Wilms, Wilding, and
  Binder}}]{wilms:2012}
\bibinfo{author}{\bibfnamefont{D.}~\bibnamefont{Wilms}},
  \bibinfo{author}{\bibfnamefont{N.~B.} \bibnamefont{Wilding}},
  \bibnamefont{and} \bibinfo{author}{\bibfnamefont{K.}~\bibnamefont{Binder}},
  \bibinfo{journal}{Phys. Rev. E} \textbf{\bibinfo{volume}{85}},
  \bibinfo{pages}{056703} (\bibinfo{year}{2012}).

\bibitem[{\citenamefont{Bruce et~al.}(1997{\natexlab{a}})\citenamefont{Bruce,
  Wilding, and Ackland}}]{bruce:1997}
\bibinfo{author}{\bibfnamefont{A.~D.} \bibnamefont{Bruce}},
  \bibinfo{author}{\bibfnamefont{N.~B.} \bibnamefont{Wilding}},
  \bibnamefont{and} \bibinfo{author}{\bibfnamefont{G.~J.}
  \bibnamefont{Ackland}}, \bibinfo{journal}{Phys. Rev. Lett.}
  \textbf{\bibinfo{volume}{79}}, \bibinfo{pages}{3002}
  (\bibinfo{year}{1997}{\natexlab{a}}).

\bibitem[{\citenamefont{Bruce et~al.}(2000)\citenamefont{Bruce, Jackson,
  Ackland, and Wilding}}]{bruce:2000}
\bibinfo{author}{\bibfnamefont{A.~D.} \bibnamefont{Bruce}},
  \bibinfo{author}{\bibfnamefont{A.~N.} \bibnamefont{Jackson}},
  \bibinfo{author}{\bibfnamefont{G.~J.} \bibnamefont{Ackland}},
  \bibnamefont{and} \bibinfo{author}{\bibfnamefont{N.~B.}
  \bibnamefont{Wilding}}, \bibinfo{journal}{Phys. Rev. E}
  \textbf{\bibinfo{volume}{61}}, \bibinfo{pages}{906} (\bibinfo{year}{2000}).

\bibitem[{\citenamefont{Wilding and Bruce}(2000)}]{wilding:2000}
\bibinfo{author}{\bibfnamefont{N.~B.} \bibnamefont{Wilding}} \bibnamefont{and}
  \bibinfo{author}{\bibfnamefont{A.~D.} \bibnamefont{Bruce}},
  \bibinfo{journal}{Phys. Rev. Lett.} \textbf{\bibinfo{volume}{85}},
  \bibinfo{pages}{5138} (\bibinfo{year}{2000}).

\bibitem[{\citenamefont{Jackson et~al.}(2002)\citenamefont{Jackson, Bruce, and
  Ackland}}]{jackson:2002}
\bibinfo{author}{\bibfnamefont{A.~N.} \bibnamefont{Jackson}},
  \bibinfo{author}{\bibfnamefont{A.~D.} \bibnamefont{Bruce}}, \bibnamefont{and}
  \bibinfo{author}{\bibfnamefont{G.~J.} \bibnamefont{Ackland}},
  \bibinfo{journal}{Phys. Rev. E} \textbf{\bibinfo{volume}{65}},
  \bibinfo{pages}{036710} (\bibinfo{year}{2002}).

\bibitem[{\citenamefont{Binder}(2008)}]{binder:2008}
\bibinfo{author}{\bibfnamefont{K.}~\bibnamefont{Binder}},
  \bibinfo{journal}{Eur. Phys. J. B} \textbf{\bibinfo{volume}{64}},
  \bibinfo{pages}{307} (\bibinfo{year}{2008}).

\bibitem[{\citenamefont{Torrie and Valleau}(1974)}]{torrie:1974}
\bibinfo{author}{\bibfnamefont{G.~M.} \bibnamefont{Torrie}} \bibnamefont{and}
  \bibinfo{author}{\bibfnamefont{J.~P.} \bibnamefont{Valleau}},
  \bibinfo{journal}{Chem. Phys. Lett.} \textbf{\bibinfo{volume}{28}},
  \bibinfo{pages}{578} (\bibinfo{year}{1974}).

\bibitem[{\citenamefont{Berg and Neuhaus}(1992)}]{berg:1992}
\bibinfo{author}{\bibfnamefont{B.~A.} \bibnamefont{Berg}} \bibnamefont{and}
  \bibinfo{author}{\bibfnamefont{T.}~\bibnamefont{Neuhaus}},
  \bibinfo{journal}{Phys. Rev. Lett.} \textbf{\bibinfo{volume}{68}},
  \bibinfo{pages}{9} (\bibinfo{year}{1992}).

\bibitem[{\citenamefont{Laio and Parrinello}(2002)}]{laio:2002}
\bibinfo{author}{\bibfnamefont{A.}~\bibnamefont{Laio}} \bibnamefont{and}
  \bibinfo{author}{\bibfnamefont{M.}~\bibnamefont{Parrinello}},
  \bibinfo{journal}{Proc. Natl. Acad. Sci. USA} \textbf{\bibinfo{volume}{99}},
  \bibinfo{pages}{12562} (\bibinfo{year}{2002}).

\bibitem[{\citenamefont{Barducci et~al.}(2008)\citenamefont{Barducci, Bussi,
  and Parrinello}}]{barducci:2008}
\bibinfo{author}{\bibfnamefont{A.}~\bibnamefont{Barducci}},
  \bibinfo{author}{\bibfnamefont{G.}~\bibnamefont{Bussi}}, \bibnamefont{and}
  \bibinfo{author}{\bibfnamefont{M.}~\bibnamefont{Parrinello}},
  \bibinfo{journal}{Phys. Rev. Lett.} \textbf{\bibinfo{volume}{100}},
  \bibinfo{pages}{020603} (\bibinfo{year}{2008}).

\bibitem[{\citenamefont{Wang and Landau}(2001{\natexlab{a}})}]{wang:2001}
\bibinfo{author}{\bibfnamefont{F.}~\bibnamefont{Wang}} \bibnamefont{and}
  \bibinfo{author}{\bibfnamefont{D.~P.} \bibnamefont{Landau}},
  \bibinfo{journal}{Phys. Rev. Lett.} \textbf{\bibinfo{volume}{86}},
  \bibinfo{pages}{2050} (\bibinfo{year}{2001}{\natexlab{a}}).

\bibitem[{\citenamefont{Wang and Landau}(2001{\natexlab{b}})}]{wang:2001a}
\bibinfo{author}{\bibfnamefont{F.}~\bibnamefont{Wang}} \bibnamefont{and}
  \bibinfo{author}{\bibfnamefont{D.~P.} \bibnamefont{Landau}},
  \bibinfo{journal}{Phys. Rev. E} \textbf{\bibinfo{volume}{64}},
  \bibinfo{pages}{056101} (\bibinfo{year}{2001}{\natexlab{b}}).

\bibitem[{\citenamefont{Landau et~al.}(2004)\citenamefont{Landau, Tsai, and
  Exler}}]{landau:2004}
\bibinfo{author}{\bibfnamefont{D.~P.} \bibnamefont{Landau}},
  \bibinfo{author}{\bibfnamefont{S.-H.} \bibnamefont{Tsai}}, \bibnamefont{and}
  \bibinfo{author}{\bibfnamefont{M.}~\bibnamefont{Exler}},
  \bibinfo{journal}{Am. J. Phys.} \textbf{\bibinfo{volume}{72}},
  \bibinfo{pages}{1294} (\bibinfo{year}{2004}).

\bibitem[{\citenamefont{Ferrenberg and Swendsen}(1988)}]{ferrenberg:1988}
\bibinfo{author}{\bibfnamefont{A.~M.} \bibnamefont{Ferrenberg}}
  \bibnamefont{and} \bibinfo{author}{\bibfnamefont{R.~H.}
  \bibnamefont{Swendsen}}, \bibinfo{journal}{Phys. Rev. Lett.}
  \textbf{\bibinfo{volume}{61}}, \bibinfo{pages}{2635} (\bibinfo{year}{1988}).

\bibitem[{\citenamefont{Ferrenberg and Swendsen}(1989)}]{ferrenberg:1989}
\bibinfo{author}{\bibfnamefont{A.~M.} \bibnamefont{Ferrenberg}}
  \bibnamefont{and} \bibinfo{author}{\bibfnamefont{R.~H.}
  \bibnamefont{Swendsen}}, \bibinfo{journal}{Phys. Rev. Lett.}
  \textbf{\bibinfo{volume}{63}}, \bibinfo{pages}{1195} (\bibinfo{year}{1989}).

\bibitem[{\citenamefont{Swendsen and Wang}(1986)}]{swendsen:1986}
\bibinfo{author}{\bibfnamefont{R.~H.} \bibnamefont{Swendsen}} \bibnamefont{and}
  \bibinfo{author}{\bibfnamefont{J.-S.} \bibnamefont{Wang}},
  \bibinfo{journal}{Phys. Rev. Lett.} \textbf{\bibinfo{volume}{57}},
  \bibinfo{pages}{2607} (\bibinfo{year}{1986}).

\bibitem[{\citenamefont{Earl and Deem}(2005)}]{earl:2005}
\bibinfo{author}{\bibfnamefont{D.~J.} \bibnamefont{Earl}} \bibnamefont{and}
  \bibinfo{author}{\bibfnamefont{M.~W.} \bibnamefont{Deem}},
  \bibinfo{journal}{Phys. Chem. Chem. Phys.} \textbf{\bibinfo{volume}{7}},
  \bibinfo{pages}{3910} (\bibinfo{year}{2005}).

\bibitem[{\citenamefont{Virnau and M{\"u}ller}(2004)}]{virnau:2004}
\bibinfo{author}{\bibfnamefont{P.}~\bibnamefont{Virnau}} \bibnamefont{and}
  \bibinfo{author}{\bibfnamefont{M.}~\bibnamefont{M{\"u}ller}},
  \bibinfo{journal}{J. Chem. Phys.} \textbf{\bibinfo{volume}{120}},
  \bibinfo{pages}{10925} (\bibinfo{year}{2004}).

\bibitem[{\citenamefont{Chopra et~al.}(2006)\citenamefont{Chopra, M{\"u}ller,
  and de~Pablo}}]{chopra:2006}
\bibinfo{author}{\bibfnamefont{M.}~\bibnamefont{Chopra}},
  \bibinfo{author}{\bibfnamefont{M.}~\bibnamefont{M{\"u}ller}},
  \bibnamefont{and} \bibinfo{author}{\bibfnamefont{J.~J.}
  \bibnamefont{de~Pablo}}, \bibinfo{journal}{J. Chem. Phys.}
  \textbf{\bibinfo{volume}{124}}, \bibinfo{pages}{134102}
  (\bibinfo{year}{2006}).

\bibitem[{\citenamefont{Sethna}(2006)}]{SethnaBook:2006}
\bibinfo{author}{\bibfnamefont{J.~P.} \bibnamefont{Sethna}},
  \emph{\bibinfo{title}{Statistical Mechanics: Entropy, Order Parameters, and
  Complexity}} (\bibinfo{publisher}{Oxford {U}niversity {P}ress},
  \bibinfo{year}{2006}).

\bibitem[{\citenamefont{Santiso and Trout}(2011)}]{santiso:2011}
\bibinfo{author}{\bibfnamefont{E.~E.} \bibnamefont{Santiso}} \bibnamefont{and}
  \bibinfo{author}{\bibfnamefont{B.~L.} \bibnamefont{Trout}},
  \bibinfo{journal}{J. Chem. Phys.} \textbf{\bibinfo{volume}{134}},
  \bibinfo{pages}{064109} (\bibinfo{year}{2011}).

\bibitem[{\citenamefont{Keys et~al.}(2011)\citenamefont{Keys, Iacovella, and
  Glotzer}}]{keys:2011}
\bibinfo{author}{\bibfnamefont{A.~S.} \bibnamefont{Keys}},
  \bibinfo{author}{\bibfnamefont{C.~R.} \bibnamefont{Iacovella}},
  \bibnamefont{and} \bibinfo{author}{\bibfnamefont{S.~C.}
  \bibnamefont{Glotzer}}, \bibinfo{journal}{J. Comput. Phys.}
  \textbf{\bibinfo{volume}{230}}, \bibinfo{pages}{6438} (\bibinfo{year}{2011}).

\bibitem[{\citenamefont{Steinhardt et~al.}(1983)\citenamefont{Steinhardt,
  Nelson, and Ronchetti}}]{steinhardt:1983}
\bibinfo{author}{\bibfnamefont{P.~J.} \bibnamefont{Steinhardt}},
  \bibinfo{author}{\bibfnamefont{D.~R.} \bibnamefont{Nelson}},
  \bibnamefont{and}
  \bibinfo{author}{\bibfnamefont{M.}~\bibnamefont{Ronchetti}},
  \bibinfo{journal}{Phys. Rev. B} \textbf{\bibinfo{volume}{28}},
  \bibinfo{pages}{784} (\bibinfo{year}{1983}).

\bibitem[{\citenamefont{ten Wolde et~al.}(1995)\citenamefont{ten Wolde,
  {Ruiz-Montero}, and Frenkel}}]{tenwolde:1995}
\bibinfo{author}{\bibfnamefont{P.~R.} \bibnamefont{ten Wolde}},
  \bibinfo{author}{\bibfnamefont{M.~J.} \bibnamefont{{Ruiz-Montero}}},
  \bibnamefont{and} \bibinfo{author}{\bibfnamefont{D.}~\bibnamefont{Frenkel}},
  \bibinfo{journal}{Phys. Rev. Lett.} \textbf{\bibinfo{volume}{75}},
  \bibinfo{pages}{2714} (\bibinfo{year}{1995}).

\bibitem[{\citenamefont{Mandell et~al.}(1976)\citenamefont{Mandell, McTague,
  and Rahman}}]{rahmanI:1977}
\bibinfo{author}{\bibfnamefont{M.~J.} \bibnamefont{Mandell}},
  \bibinfo{author}{\bibfnamefont{J.~P.} \bibnamefont{McTague}},
  \bibnamefont{and} \bibinfo{author}{\bibfnamefont{A.}~\bibnamefont{Rahman}},
  \bibinfo{journal}{J. Chem. Phys.} \textbf{\bibinfo{volume}{64}},
  \bibinfo{pages}{3699} (\bibinfo{year}{1976}).

\bibitem[{\citenamefont{Gasser et~al.}(2001)\citenamefont{Gasser, Weeks,
  Schofield, Pusey, and Weitz}}]{gasser:2001}
\bibinfo{author}{\bibfnamefont{U.}~\bibnamefont{Gasser}},
  \bibinfo{author}{\bibfnamefont{E.~R.} \bibnamefont{Weeks}},
  \bibinfo{author}{\bibfnamefont{A.}~\bibnamefont{Schofield}},
  \bibinfo{author}{\bibfnamefont{P.~N.} \bibnamefont{Pusey}}, \bibnamefont{and}
  \bibinfo{author}{\bibfnamefont{D.~A.} \bibnamefont{Weitz}},
  \bibinfo{journal}{Science} \textbf{\bibinfo{volume}{292}},
  \bibinfo{pages}{258} (\bibinfo{year}{2001}).

\bibitem[{\citenamefont{Herlach et~al.}(2010)\citenamefont{Herlach, Klassen,
  Wette, and {Holland-Moritz}}}]{herlach:2010}
\bibinfo{author}{\bibfnamefont{D.~M.} \bibnamefont{Herlach}},
  \bibinfo{author}{\bibfnamefont{I.}~\bibnamefont{Klassen}},
  \bibinfo{author}{\bibfnamefont{P.}~\bibnamefont{Wette}}, \bibnamefont{and}
  \bibinfo{author}{\bibfnamefont{D.}~\bibnamefont{{Holland-Moritz}}},
  \bibinfo{journal}{J. Phys.: Condens. Matter} \textbf{\bibinfo{volume}{22}},
  \bibinfo{pages}{153101} (\bibinfo{year}{2010}).

\bibitem[{\citenamefont{Lechner and Dellago}(2008)}]{lechner:2008}
\bibinfo{author}{\bibfnamefont{W.}~\bibnamefont{Lechner}} \bibnamefont{and}
  \bibinfo{author}{\bibfnamefont{C.}~\bibnamefont{Dellago}},
  \bibinfo{journal}{J. Chem. Phys.} \textbf{\bibinfo{volume}{129}},
  \bibinfo{pages}{114707} (\bibinfo{year}{2008}).

\bibitem[{\citenamefont{Lechner
  et~al.}(2011{\natexlab{a}})\citenamefont{Lechner, Dellago, and
  Bolhuis}}]{lechner:2011}
\bibinfo{author}{\bibfnamefont{W.}~\bibnamefont{Lechner}},
  \bibinfo{author}{\bibfnamefont{C.}~\bibnamefont{Dellago}}, \bibnamefont{and}
  \bibinfo{author}{\bibfnamefont{P.~G.} \bibnamefont{Bolhuis}},
  \bibinfo{journal}{Phys. Rev. Lett.} \textbf{\bibinfo{volume}{106}},
  \bibinfo{pages}{085701} (\bibinfo{year}{2011}{\natexlab{a}}).

\bibitem[{\citenamefont{Chushak and Bartell}(2000)}]{chushak:2000}
\bibinfo{author}{\bibfnamefont{Y.}~\bibnamefont{Chushak}} \bibnamefont{and}
  \bibinfo{author}{\bibfnamefont{L.~S.} \bibnamefont{Bartell}},
  \bibinfo{journal}{J. Phys. Chem. A} \textbf{\bibinfo{volume}{104}},
  \bibinfo{pages}{9328} (\bibinfo{year}{2000}).

\bibitem[{\citenamefont{Leyssale et~al.}(2003)\citenamefont{Leyssale,
  Delhommelle, and Millot}}]{leyssale:2003}
\bibinfo{author}{\bibfnamefont{J.-M.} \bibnamefont{Leyssale}},
  \bibinfo{author}{\bibfnamefont{J.}~\bibnamefont{Delhommelle}},
  \bibnamefont{and} \bibinfo{author}{\bibfnamefont{C.}~\bibnamefont{Millot}},
  \bibinfo{journal}{Chem. Phys. Lett.} \textbf{\bibinfo{volume}{375}},
  \bibinfo{pages}{612} (\bibinfo{year}{2003}).

\bibitem[{\citenamefont{Valeriani et~al.}(2005)\citenamefont{Valeriani, Sanz,
  and Frenkel}}]{valeriani:2005}
\bibinfo{author}{\bibfnamefont{C.}~\bibnamefont{Valeriani}},
  \bibinfo{author}{\bibfnamefont{E.}~\bibnamefont{Sanz}}, \bibnamefont{and}
  \bibinfo{author}{\bibfnamefont{D.}~\bibnamefont{Frenkel}},
  \bibinfo{journal}{J. Chem. Phys.} \textbf{\bibinfo{volume}{122}},
  \bibinfo{pages}{194501} (\bibinfo{year}{2005}).

\bibitem[{\citenamefont{{van Meel} et~al.}(2008)\citenamefont{{van Meel}, Page,
  Sear, and Frenkel}}]{vanmeel:2008}
\bibinfo{author}{\bibfnamefont{J.~A.} \bibnamefont{{van Meel}}},
  \bibinfo{author}{\bibfnamefont{A.~J.} \bibnamefont{Page}},
  \bibinfo{author}{\bibfnamefont{R.~P.} \bibnamefont{Sear}}, \bibnamefont{and}
  \bibinfo{author}{\bibfnamefont{D.}~\bibnamefont{Frenkel}},
  \bibinfo{journal}{J. Chem. Phys.} \textbf{\bibinfo{volume}{129}},
  \bibinfo{pages}{204505} (\bibinfo{year}{2008}).

\bibitem[{\citenamefont{Bokeloh et~al.}(2011)\citenamefont{Bokeloh, Rozas,
  Horbach, and Wilde}}]{bokeloh:2011}
\bibinfo{author}{\bibfnamefont{J.}~\bibnamefont{Bokeloh}},
  \bibinfo{author}{\bibfnamefont{R.~E.} \bibnamefont{Rozas}},
  \bibinfo{author}{\bibfnamefont{J.}~\bibnamefont{Horbach}}, \bibnamefont{and}
  \bibinfo{author}{\bibfnamefont{G.}~\bibnamefont{Wilde}},
  \bibinfo{journal}{Phys. Rev. Lett.} \textbf{\bibinfo{volume}{107}},
  \bibinfo{pages}{145701} (\bibinfo{year}{2011}).

\bibitem[{\citenamefont{Russo and Tanaka}(2012)}]{russo:2012}
\bibinfo{author}{\bibfnamefont{J.}~\bibnamefont{Russo}} \bibnamefont{and}
  \bibinfo{author}{\bibfnamefont{H.}~\bibnamefont{Tanaka}},
  \bibinfo{journal}{Sci. Rep.} \textbf{\bibinfo{volume}{2}},
  \bibinfo{pages}{505} (\bibinfo{year}{2012}).

\bibitem[{\citenamefont{ten Wolde and Frenkel}(1999)}]{tenwolde:1999}
\bibinfo{author}{\bibfnamefont{P.~R.} \bibnamefont{ten Wolde}}
  \bibnamefont{and} \bibinfo{author}{\bibfnamefont{D.}~\bibnamefont{Frenkel}},
  \bibinfo{journal}{Phys. Chem. Chem. Phys.} \textbf{\bibinfo{volume}{1}},
  \bibinfo{pages}{2191} (\bibinfo{year}{1999}).

\bibitem[{\citenamefont{Klumov}(2010)}]{klumov:2010}
\bibinfo{author}{\bibfnamefont{B.~A.} \bibnamefont{Klumov}},
  \bibinfo{journal}{Physics -- Uspekhi} \textbf{\bibinfo{volume}{53}},
  \bibinfo{pages}{1053} (\bibinfo{year}{2010}).

\bibitem[{\citenamefont{Schilling et~al.}(2010)\citenamefont{Schilling,
  Sch{\"o}pe, Oettel, Opletal, and Snook}}]{schilling:2010}
\bibinfo{author}{\bibfnamefont{T.}~\bibnamefont{Schilling}},
  \bibinfo{author}{\bibfnamefont{H.~J.} \bibnamefont{Sch{\"o}pe}},
  \bibinfo{author}{\bibfnamefont{M.}~\bibnamefont{Oettel}},
  \bibinfo{author}{\bibfnamefont{G.}~\bibnamefont{Opletal}}, \bibnamefont{and}
  \bibinfo{author}{\bibfnamefont{I.}~\bibnamefont{Snook}},
  \bibinfo{journal}{Phys. Rev. Lett.} \textbf{\bibinfo{volume}{105}},
  \bibinfo{pages}{025701} (\bibinfo{year}{2010}).

\bibitem[{\citenamefont{Kawasaki and Onuki}(2011)}]{kawasaki:2011}
\bibinfo{author}{\bibfnamefont{T.}~\bibnamefont{Kawasaki}} \bibnamefont{and}
  \bibinfo{author}{\bibfnamefont{A.}~\bibnamefont{Onuki}}, \bibinfo{journal}{J.
  Chem. Phys.} \textbf{\bibinfo{volume}{135}}, \bibinfo{pages}{174109}
  (\bibinfo{year}{2011}).

\bibitem[{\citenamefont{Frenkel}(2013)}]{Frenkel:2013}
\bibinfo{author}{\bibfnamefont{D.}~\bibnamefont{Frenkel}}, in
  \emph{\bibinfo{booktitle}{Proceedings of the International School of Physics
  ``Enrico Fermi'', Course CLXXXIV ``Physics of Complex Colloids''}}, edited by
  \bibinfo{editor}{\bibfnamefont{C.}~\bibnamefont{Bechinger}},
  \bibinfo{editor}{\bibfnamefont{F.}~\bibnamefont{Sciortino}},
  \bibnamefont{and} \bibinfo{editor}{\bibfnamefont{P.}~\bibnamefont{Ziherl}}
  (\bibinfo{publisher}{Societ{\'a} Italiana di Fisica}, \bibinfo{year}{2013}),
  p. \bibinfo{pages}{195}.

\bibitem[{\citenamefont{Wedekind et~al.}(2007)\citenamefont{Wedekind, Strey,
  and Reguera}}]{wedekind:2007}
\bibinfo{author}{\bibfnamefont{J.}~\bibnamefont{Wedekind}},
  \bibinfo{author}{\bibfnamefont{R.}~\bibnamefont{Strey}}, \bibnamefont{and}
  \bibinfo{author}{\bibfnamefont{D.}~\bibnamefont{Reguera}},
  \bibinfo{journal}{J. Chem. Phys.} \textbf{\bibinfo{volume}{126}},
  \bibinfo{pages}{134103} (\bibinfo{year}{2007}).

\bibitem[{\citenamefont{Wedekind and Reguera}(2008)}]{wedekind:2008}
\bibinfo{author}{\bibfnamefont{J.}~\bibnamefont{Wedekind}} \bibnamefont{and}
  \bibinfo{author}{\bibfnamefont{D.}~\bibnamefont{Reguera}},
  \bibinfo{journal}{J. Phys. Chem. B} \textbf{\bibinfo{volume}{112}},
  \bibinfo{pages}{11060} (\bibinfo{year}{2008}).

\bibitem[{\citenamefont{Mokshin and Galimzyanov}(2014)}]{mokshin:2014}
\bibinfo{author}{\bibfnamefont{A.~V.} \bibnamefont{Mokshin}} \bibnamefont{and}
  \bibinfo{author}{\bibfnamefont{B.~N.} \bibnamefont{Galimzyanov}},
  \bibinfo{journal}{J. Chem. Phys.} \textbf{\bibinfo{volume}{140}},
  \bibinfo{pages}{024104} (\bibinfo{year}{2014}).

\bibitem[{\citenamefont{Shneidman}(2014)}]{shneidman:2014}
\bibinfo{author}{\bibfnamefont{V.~A.} \bibnamefont{Shneidman}},
  \bibinfo{journal}{J. Chem. Phys.} \textbf{\bibinfo{volume}{141}},
  \bibinfo{pages}{051101} (\bibinfo{year}{2014}).

\bibitem[{\citenamefont{Jungblut and
  Dellago}(2015{\natexlab{a}})}]{jungblut:2015}
\bibinfo{author}{\bibfnamefont{S.}~\bibnamefont{Jungblut}} \bibnamefont{and}
  \bibinfo{author}{\bibfnamefont{C.}~\bibnamefont{Dellago}},
  \bibinfo{journal}{J. Chem. Phys.} \textbf{\bibinfo{volume}{142}},
  \bibinfo{pages}{064103} (\bibinfo{year}{2015}{\natexlab{a}}).

\bibitem[{\citenamefont{Baidakov and Tipeev}(2011)}]{baidakov:2011a}
\bibinfo{author}{\bibfnamefont{V.~G.} \bibnamefont{Baidakov}} \bibnamefont{and}
  \bibinfo{author}{\bibfnamefont{A.~O.} \bibnamefont{Tipeev}},
  \bibinfo{journal}{Thermochimica Acta} \textbf{\bibinfo{volume}{522}},
  \bibinfo{pages}{14} (\bibinfo{year}{2011}).

\bibitem[{\citenamefont{Chkonia et~al.}(2009)\citenamefont{Chkonia, W{\"o}lk,
  Strey, Wedekind, and Reguera}}]{chkonia:2009}
\bibinfo{author}{\bibfnamefont{G.}~\bibnamefont{Chkonia}},
  \bibinfo{author}{\bibfnamefont{J.}~\bibnamefont{W{\"o}lk}},
  \bibinfo{author}{\bibfnamefont{R.}~\bibnamefont{Strey}},
  \bibinfo{author}{\bibfnamefont{J.}~\bibnamefont{Wedekind}}, \bibnamefont{and}
  \bibinfo{author}{\bibfnamefont{D.}~\bibnamefont{Reguera}},
  \bibinfo{journal}{J. Chem. Phys.} \textbf{\bibinfo{volume}{130}},
  \bibinfo{pages}{064505} (\bibinfo{year}{2009}).

\bibitem[{\citenamefont{Talkner}(1987)}]{talkner:1987}
\bibinfo{author}{\bibfnamefont{P.}~\bibnamefont{Talkner}}, \bibinfo{journal}{Z.
  Phys. B -- Condens. Matter} \textbf{\bibinfo{volume}{68}},
  \bibinfo{pages}{201} (\bibinfo{year}{1987}).

\bibitem[{\citenamefont{Eyring}(1935)}]{eyring:1935}
\bibinfo{author}{\bibfnamefont{H.}~\bibnamefont{Eyring}}, \bibinfo{journal}{J.
  Chem. Phys.} \textbf{\bibinfo{volume}{3}}, \bibinfo{pages}{107}
  (\bibinfo{year}{1935}).

\bibitem[{\citenamefont{Eyring}(1938)}]{eyring:1938}
\bibinfo{author}{\bibfnamefont{H.}~\bibnamefont{Eyring}},
  \bibinfo{journal}{Trans. Faraday Soc.} \textbf{\bibinfo{volume}{34}},
  \bibinfo{pages}{41} (\bibinfo{year}{1938}).

\bibitem[{\citenamefont{Wigner}(1938)}]{wigner:1938}
\bibinfo{author}{\bibfnamefont{E.}~\bibnamefont{Wigner}},
  \bibinfo{journal}{Trans. Faraday Soc.} \textbf{\bibinfo{volume}{34}},
  \bibinfo{pages}{29} (\bibinfo{year}{1938}).

\bibitem[{\citenamefont{Pollak and Talkner}(2005)}]{pollak:2005}
\bibinfo{author}{\bibfnamefont{E.}~\bibnamefont{Pollak}} \bibnamefont{and}
  \bibinfo{author}{\bibfnamefont{P.}~\bibnamefont{Talkner}},
  \bibinfo{journal}{CHAOS} \textbf{\bibinfo{volume}{15}},
  \bibinfo{pages}{026116} (\bibinfo{year}{2005}).

\bibitem[{\citenamefont{Pollak}(1986)}]{pollak:1986}
\bibinfo{author}{\bibfnamefont{E.}~\bibnamefont{Pollak}}, \bibinfo{journal}{J.
  Chem. Phys.} \textbf{\bibinfo{volume}{85}}, \bibinfo{pages}{865}
  (\bibinfo{year}{1986}).

\bibitem[{\citenamefont{Keck}(1960)}]{keck:1960}
\bibinfo{author}{\bibfnamefont{J.~C.} \bibnamefont{Keck}}, \bibinfo{journal}{J.
  Chem. Phys.} \textbf{\bibinfo{volume}{32}}, \bibinfo{pages}{1035}
  (\bibinfo{year}{1960}).

\bibitem[{\citenamefont{Mullen et~al.}(2014)\citenamefont{Mullen, Shea, and
  Peters}}]{mullen:2014}
\bibinfo{author}{\bibfnamefont{R.~G.} \bibnamefont{Mullen}},
  \bibinfo{author}{\bibfnamefont{J.-E.} \bibnamefont{Shea}}, \bibnamefont{and}
  \bibinfo{author}{\bibfnamefont{B.}~\bibnamefont{Peters}},
  \bibinfo{journal}{J. Chem. Phys.} \textbf{\bibinfo{volume}{140}},
  \bibinfo{pages}{041104} (\bibinfo{year}{2014}).

\bibitem[{\citenamefont{Dellago et~al.}(1998)\citenamefont{Dellago, Bolhuis,
  Csaijka, and Chandler}}]{dellago:1998}
\bibinfo{author}{\bibfnamefont{C.}~\bibnamefont{Dellago}},
  \bibinfo{author}{\bibfnamefont{P.~G.} \bibnamefont{Bolhuis}},
  \bibinfo{author}{\bibfnamefont{F.~S.} \bibnamefont{Csaijka}},
  \bibnamefont{and} \bibinfo{author}{\bibfnamefont{D.}~\bibnamefont{Chandler}},
  \bibinfo{journal}{J. Chem. Phys.} \textbf{\bibinfo{volume}{108}},
  \bibinfo{pages}{1964} (\bibinfo{year}{1998}).

\bibitem[{\citenamefont{Bolhuis et~al.}(1998)\citenamefont{Bolhuis, Dellago,
  and Chandler}}]{bolhuis:1998}
\bibinfo{author}{\bibfnamefont{P.~G.} \bibnamefont{Bolhuis}},
  \bibinfo{author}{\bibfnamefont{C.}~\bibnamefont{Dellago}}, \bibnamefont{and}
  \bibinfo{author}{\bibfnamefont{D.}~\bibnamefont{Chandler}},
  \bibinfo{journal}{Faraday Discuss.} \textbf{\bibinfo{volume}{110}},
  \bibinfo{pages}{412} (\bibinfo{year}{1998}).

\bibitem[{\citenamefont{Bolhuis et~al.}(2000)\citenamefont{Bolhuis, Dellago,
  Geissler, and Chandler}}]{dellago:2000}
\bibinfo{author}{\bibfnamefont{P.~G.} \bibnamefont{Bolhuis}},
  \bibinfo{author}{\bibfnamefont{C.}~\bibnamefont{Dellago}},
  \bibinfo{author}{\bibfnamefont{P.~L.} \bibnamefont{Geissler}},
  \bibnamefont{and} \bibinfo{author}{\bibfnamefont{D.}~\bibnamefont{Chandler}},
  \bibinfo{journal}{J. Phys.: Condens. Matter} \textbf{\bibinfo{volume}{12}},
  \bibinfo{pages}{A147} (\bibinfo{year}{2000}).

\bibitem[{\citenamefont{Dellago et~al.}(2002)\citenamefont{Dellago, Bolhuis,
  and Geissler}}]{dellago:2002}
\bibinfo{author}{\bibfnamefont{C.}~\bibnamefont{Dellago}},
  \bibinfo{author}{\bibfnamefont{P.~G.} \bibnamefont{Bolhuis}},
  \bibnamefont{and} \bibinfo{author}{\bibfnamefont{P.~L.}
  \bibnamefont{Geissler}}, \bibinfo{journal}{Adv. Chem. Phys.}
  \textbf{\bibinfo{volume}{123}}, \bibinfo{pages}{1} (\bibinfo{year}{2002}).

\bibitem[{\citenamefont{Bolhuis et~al.}(2002)\citenamefont{Bolhuis, Chandler,
  Dellago, and Geissler}}]{bolhuis:2002}
\bibinfo{author}{\bibfnamefont{P.~G.} \bibnamefont{Bolhuis}},
  \bibinfo{author}{\bibfnamefont{D.}~\bibnamefont{Chandler}},
  \bibinfo{author}{\bibfnamefont{C.}~\bibnamefont{Dellago}}, \bibnamefont{and}
  \bibinfo{author}{\bibfnamefont{P.~L.} \bibnamefont{Geissler}},
  \bibinfo{journal}{Annu. Rev. Phys. Chem.} \textbf{\bibinfo{volume}{53}},
  \bibinfo{pages}{291} (\bibinfo{year}{2002}).

\bibitem[{\citenamefont{Dellago and Bolhuis}(2009)}]{dellago:2009}
\bibinfo{author}{\bibfnamefont{C.}~\bibnamefont{Dellago}} \bibnamefont{and}
  \bibinfo{author}{\bibfnamefont{P.~G.} \bibnamefont{Bolhuis}}, in
  \emph{\bibinfo{booktitle}{Advanced Computer Simulation Approaches for Soft
  Matter Sciences III}}, edited by
  \bibinfo{editor}{\bibfnamefont{C.}~\bibnamefont{Holm}} \bibnamefont{and}
  \bibinfo{editor}{\bibfnamefont{K.}~\bibnamefont{Kremer}}
  (\bibinfo{publisher}{Springer-Verlag}, \bibinfo{address}{Berlin Heidelberg},
  \bibinfo{year}{2009}), vol. \bibinfo{volume}{221} of
  \emph{\bibinfo{series}{Advances in Polymer Science}}, p.
  \bibinfo{pages}{167}.

\bibitem[{\citenamefont{Dellago et~al.}(1999)\citenamefont{Dellago, Bolhuis,
  and Chandler}}]{dellago:1999}
\bibinfo{author}{\bibfnamefont{C.}~\bibnamefont{Dellago}},
  \bibinfo{author}{\bibfnamefont{P.~G.} \bibnamefont{Bolhuis}},
  \bibnamefont{and} \bibinfo{author}{\bibfnamefont{D.}~\bibnamefont{Chandler}},
  \bibinfo{journal}{J. Chem. Phys.} \textbf{\bibinfo{volume}{110}},
  \bibinfo{pages}{6617} (\bibinfo{year}{1999}).

\bibitem[{\citenamefont{van Erp et~al.}(2003)\citenamefont{van Erp, Moroni, and
  Bolhuis}}]{vanerp:2003}
\bibinfo{author}{\bibfnamefont{T.~S.} \bibnamefont{van Erp}},
  \bibinfo{author}{\bibfnamefont{D.}~\bibnamefont{Moroni}}, \bibnamefont{and}
  \bibinfo{author}{\bibfnamefont{P.~G.} \bibnamefont{Bolhuis}},
  \bibinfo{journal}{J. Chem. Phys.} \textbf{\bibinfo{volume}{118}},
  \bibinfo{pages}{7762} (\bibinfo{year}{2003}).

\bibitem[{\citenamefont{van Erp and Bolhuis}(2005)}]{vanerp:2005}
\bibinfo{author}{\bibfnamefont{T.~S.} \bibnamefont{van Erp}} \bibnamefont{and}
  \bibinfo{author}{\bibfnamefont{P.~G.} \bibnamefont{Bolhuis}},
  \bibinfo{journal}{J. Comput. Phys.} \textbf{\bibinfo{volume}{205}},
  \bibinfo{pages}{157} (\bibinfo{year}{2005}).

\bibitem[{\citenamefont{van Erp}(2007)}]{vanerp:2007}
\bibinfo{author}{\bibfnamefont{T.~S.} \bibnamefont{van Erp}},
  \bibinfo{journal}{Phys. Rev. Lett.} \textbf{\bibinfo{volume}{98}},
  \bibinfo{pages}{268301} (\bibinfo{year}{2007}).

\bibitem[{\citenamefont{Bolhuis}(2008)}]{bolhuis:2008}
\bibinfo{author}{\bibfnamefont{P.~G.} \bibnamefont{Bolhuis}},
  \bibinfo{journal}{J. Chem. Phys.} \textbf{\bibinfo{volume}{129}},
  \bibinfo{pages}{114108} (\bibinfo{year}{2008}).

\bibitem[{\citenamefont{Moroni et~al.}(2004)\citenamefont{Moroni, Bolhuis, and
  van Erp}}]{moroni:2004}
\bibinfo{author}{\bibfnamefont{D.}~\bibnamefont{Moroni}},
  \bibinfo{author}{\bibfnamefont{P.~G.} \bibnamefont{Bolhuis}},
  \bibnamefont{and} \bibinfo{author}{\bibfnamefont{T.~S.} \bibnamefont{van
  Erp}}, \bibinfo{journal}{J. Chem. Phys.} \textbf{\bibinfo{volume}{120}},
  \bibinfo{pages}{4055} (\bibinfo{year}{2004}).

\bibitem[{\citenamefont{Moroni et~al.}(2005{\natexlab{b}})\citenamefont{Moroni,
  van Erp, and Bolhuis}}]{moroni:2005a}
\bibinfo{author}{\bibfnamefont{D.}~\bibnamefont{Moroni}},
  \bibinfo{author}{\bibfnamefont{T.~S.} \bibnamefont{van Erp}},
  \bibnamefont{and} \bibinfo{author}{\bibfnamefont{P.~G.}
  \bibnamefont{Bolhuis}}, \bibinfo{journal}{Phys. Rev. E}
  \textbf{\bibinfo{volume}{71}}, \bibinfo{pages}{056709}
  (\bibinfo{year}{2005}{\natexlab{b}}).

\bibitem[{\citenamefont{Geissler and Dellago}(2004)}]{geissler:2004}
\bibinfo{author}{\bibfnamefont{P.~L.} \bibnamefont{Geissler}} \bibnamefont{and}
  \bibinfo{author}{\bibfnamefont{C.}~\bibnamefont{Dellago}},
  \bibinfo{journal}{J. Phys.Chem. B} \textbf{\bibinfo{volume}{108}},
  \bibinfo{pages}{6667} (\bibinfo{year}{2004}).

\bibitem[{\citenamefont{Allen et~al.}(2005)\citenamefont{Allen, Warren, and ten
  Wolde}}]{allen:2005}
\bibinfo{author}{\bibfnamefont{R.~J.} \bibnamefont{Allen}},
  \bibinfo{author}{\bibfnamefont{P.~B.} \bibnamefont{Warren}},
  \bibnamefont{and} \bibinfo{author}{\bibfnamefont{P.~R.} \bibnamefont{ten
  Wolde}}, \bibinfo{journal}{Phys. Rev. Lett.} \textbf{\bibinfo{volume}{94}},
  \bibinfo{pages}{018104} (\bibinfo{year}{2005}).

\bibitem[{\citenamefont{Allen et~al.}(2006)\citenamefont{Allen, Frenkel, and
  ten Wolde}}]{allen:2006}
\bibinfo{author}{\bibfnamefont{R.~J.} \bibnamefont{Allen}},
  \bibinfo{author}{\bibfnamefont{D.}~\bibnamefont{Frenkel}}, \bibnamefont{and}
  \bibinfo{author}{\bibfnamefont{P.~R.} \bibnamefont{ten Wolde}},
  \bibinfo{journal}{J. Chem. Phys.} \textbf{\bibinfo{volume}{124}},
  \bibinfo{pages}{024102} (\bibinfo{year}{2006}).

\bibitem[{\citenamefont{Borrero and Escobedo}(2007)}]{borrero:2007}
\bibinfo{author}{\bibfnamefont{E.~E.} \bibnamefont{Borrero}} \bibnamefont{and}
  \bibinfo{author}{\bibfnamefont{F.~A.} \bibnamefont{Escobedo}},
  \bibinfo{journal}{J. Chem. Phys.} \textbf{\bibinfo{volume}{127}},
  \bibinfo{pages}{164101} (\bibinfo{year}{2007}).

\bibitem[{\citenamefont{Allen et~al.}(2009)\citenamefont{Allen, Valeriani, and
  ten Wolde}}]{allen:2009}
\bibinfo{author}{\bibfnamefont{R.~J.} \bibnamefont{Allen}},
  \bibinfo{author}{\bibfnamefont{C.}~\bibnamefont{Valeriani}},
  \bibnamefont{and} \bibinfo{author}{\bibfnamefont{P.~R.} \bibnamefont{ten
  Wolde}}, \bibinfo{journal}{J. Phys.: Condens. Matter}
  \textbf{\bibinfo{volume}{21}}, \bibinfo{pages}{463102}
  (\bibinfo{year}{2009}).

\bibitem[{\citenamefont{Escobedo et~al.}(2009)\citenamefont{Escobedo, Borrero,
  and Araque}}]{escobedo:2009}
\bibinfo{author}{\bibfnamefont{F.~A.} \bibnamefont{Escobedo}},
  \bibinfo{author}{\bibfnamefont{E.~E.} \bibnamefont{Borrero}},
  \bibnamefont{and} \bibinfo{author}{\bibfnamefont{J.~C.}
  \bibnamefont{Araque}}, \bibinfo{journal}{J. Phys.: Condens. Matter}
  \textbf{\bibinfo{volume}{21}}, \bibinfo{pages}{333101}
  (\bibinfo{year}{2009}).

\bibitem[{\citenamefont{{van Erp}}(2012)}]{vanerp:2012}
\bibinfo{author}{\bibfnamefont{T.~S.} \bibnamefont{{van Erp}}},
  \bibinfo{journal}{Adv. Chem Phys.} \textbf{\bibinfo{volume}{151}},
  \bibinfo{pages}{27} (\bibinfo{year}{2012}).

\bibitem[{\citenamefont{Bolhuis and Dellago}(2015)}]{bolhuis:2015}
\bibinfo{author}{\bibfnamefont{P.~G.} \bibnamefont{Bolhuis}} \bibnamefont{and}
  \bibinfo{author}{\bibfnamefont{C.}~\bibnamefont{Dellago}},
  \bibinfo{journal}{Eur. Phys. J. Special Topics} pp.
  \bibinfo{pages}{doi:10.1140/epjst/e2015--02419--6} (\bibinfo{year}{2015}).

\bibitem[{\citenamefont{Vlugt and Smit}(2001)}]{vlugt:2001}
\bibinfo{author}{\bibfnamefont{T.~J.~H.} \bibnamefont{Vlugt}} \bibnamefont{and}
  \bibinfo{author}{\bibfnamefont{B.}~\bibnamefont{Smit}},
  \bibinfo{journal}{Phys. Chem. Comm.} \textbf{\bibinfo{volume}{2}},
  \bibinfo{pages}{1} (\bibinfo{year}{2001}).

\bibitem[{\citenamefont{Mullen et~al.}(2015)\citenamefont{Mullen, Shea, and
  Peters}}]{mullen:2015}
\bibinfo{author}{\bibfnamefont{R.~G.} \bibnamefont{Mullen}},
  \bibinfo{author}{\bibfnamefont{J.-E.} \bibnamefont{Shea}}, \bibnamefont{and}
  \bibinfo{author}{\bibfnamefont{B.}~\bibnamefont{Peters}},
  \bibinfo{journal}{J. Chem. Theory Comput.} p.
  \bibinfo{pages}{10.1021/acs.jctc.5b00032} (\bibinfo{year}{2015}).

\bibitem[{\citenamefont{Bolhuis}(2003)}]{bolhuis:2003}
\bibinfo{author}{\bibfnamefont{P.~G.} \bibnamefont{Bolhuis}},
  \bibinfo{journal}{J. Phys.: Condens. Matter} \textbf{\bibinfo{volume}{15}},
  \bibinfo{pages}{S113} (\bibinfo{year}{2003}).

\bibitem[{\citenamefont{{Miller III} and Predescu}(2007)}]{miller:2007}
\bibinfo{author}{\bibfnamefont{T.~F.} \bibnamefont{{Miller III}}}
  \bibnamefont{and} \bibinfo{author}{\bibfnamefont{C.}~\bibnamefont{Predescu}},
  \bibinfo{journal}{J. Chem. Phys.} \textbf{\bibinfo{volume}{126}},
  \bibinfo{pages}{144102} (\bibinfo{year}{2007}).

\bibitem[{\citenamefont{Gr{\"u}nwald et~al.}(2008)\citenamefont{Gr{\"u}nwald,
  Dellago, and Geissler}}]{gruenwald:2008}
\bibinfo{author}{\bibfnamefont{M.}~\bibnamefont{Gr{\"u}nwald}},
  \bibinfo{author}{\bibfnamefont{C.}~\bibnamefont{Dellago}}, \bibnamefont{and}
  \bibinfo{author}{\bibfnamefont{P.~L.} \bibnamefont{Geissler}},
  \bibinfo{journal}{J. Chem. Phys.} \textbf{\bibinfo{volume}{129}},
  \bibinfo{pages}{194101} (\bibinfo{year}{2008}).

\bibitem[{\citenamefont{Chopra et~al.}(2008)\citenamefont{Chopra, Malshe,
  Reddy, and {de Pablo}}}]{chopra:2008}
\bibinfo{author}{\bibfnamefont{M.}~\bibnamefont{Chopra}},
  \bibinfo{author}{\bibfnamefont{R.}~\bibnamefont{Malshe}},
  \bibinfo{author}{\bibfnamefont{A.~S.} \bibnamefont{Reddy}}, \bibnamefont{and}
  \bibinfo{author}{\bibfnamefont{J.~J.} \bibnamefont{{de Pablo}}},
  \bibinfo{journal}{J. Chem. Phys.} \textbf{\bibinfo{volume}{128}},
  \bibinfo{pages}{144104} (\bibinfo{year}{2008}).

\bibitem[{\citenamefont{Peters and Trout}(2006)}]{peters:2006}
\bibinfo{author}{\bibfnamefont{B.}~\bibnamefont{Peters}} \bibnamefont{and}
  \bibinfo{author}{\bibfnamefont{B.~L.} \bibnamefont{Trout}},
  \bibinfo{journal}{J. Chem. Phys.} \textbf{\bibinfo{volume}{125}},
  \bibinfo{pages}{054108} (\bibinfo{year}{2006}).

\bibitem[{\citenamefont{Peters et~al.}(2007)\citenamefont{Peters, Beckham, and
  Trout}}]{peters:2007}
\bibinfo{author}{\bibfnamefont{B.}~\bibnamefont{Peters}},
  \bibinfo{author}{\bibfnamefont{G.~T.} \bibnamefont{Beckham}},
  \bibnamefont{and} \bibinfo{author}{\bibfnamefont{B.~L.} \bibnamefont{Trout}},
  \bibinfo{journal}{J. Chem. Phys.} \textbf{\bibinfo{volume}{127}},
  \bibinfo{pages}{034109} (\bibinfo{year}{2007}).

\bibitem[{\citenamefont{Rogal and Bolhuis}(2008)}]{rogal:2008}
\bibinfo{author}{\bibfnamefont{J.}~\bibnamefont{Rogal}} \bibnamefont{and}
  \bibinfo{author}{\bibfnamefont{P.~G.} \bibnamefont{Bolhuis}},
  \bibinfo{journal}{J. Chem. Phys.} \textbf{\bibinfo{volume}{129}},
  \bibinfo{pages}{224107} (\bibinfo{year}{2008}).

\bibitem[{\citenamefont{Borrero and Dellago}(2010)}]{borrero:2010}
\bibinfo{author}{\bibfnamefont{E.~E.} \bibnamefont{Borrero}} \bibnamefont{and}
  \bibinfo{author}{\bibfnamefont{C.}~\bibnamefont{Dellago}},
  \bibinfo{journal}{J. Chem. Phys.} \textbf{\bibinfo{volume}{133}},
  \bibinfo{pages}{134112} (\bibinfo{year}{2010}).

\bibitem[{\citenamefont{Guttenberg et~al.}(2012)\citenamefont{Guttenberg,
  Dinner, and Weare}}]{guttenberg:2012}
\bibinfo{author}{\bibfnamefont{N.}~\bibnamefont{Guttenberg}},
  \bibinfo{author}{\bibfnamefont{A.~R.} \bibnamefont{Dinner}},
  \bibnamefont{and} \bibinfo{author}{\bibfnamefont{J.}~\bibnamefont{Weare}},
  \bibinfo{journal}{J. Chem. Phys.} \textbf{\bibinfo{volume}{136}},
  \bibinfo{pages}{234103} (\bibinfo{year}{2012}).

\bibitem[{\citenamefont{Borrero et~al.}(2011)\citenamefont{Borrero, Weinwurm,
  and Dellago}}]{borrero:2011}
\bibinfo{author}{\bibfnamefont{E.~E.} \bibnamefont{Borrero}},
  \bibinfo{author}{\bibfnamefont{M.}~\bibnamefont{Weinwurm}}, \bibnamefont{and}
  \bibinfo{author}{\bibfnamefont{C.}~\bibnamefont{Dellago}},
  \bibinfo{journal}{J. Chem. Phys.} \textbf{\bibinfo{volume}{134}},
  \bibinfo{pages}{244118} (\bibinfo{year}{2011}).

\bibitem[{\citenamefont{Du and Bolhuis}(2013)}]{du:2013}
\bibinfo{author}{\bibfnamefont{W.-N.} \bibnamefont{Du}} \bibnamefont{and}
  \bibinfo{author}{\bibfnamefont{P.~G.} \bibnamefont{Bolhuis}},
  \bibinfo{journal}{J. Chem. Phys.} \textbf{\bibinfo{volume}{139}},
  \bibinfo{pages}{044105} (\bibinfo{year}{2013}).

\bibitem[{\citenamefont{Swenson and Bolhuis}(2014)}]{swenson:2014}
\bibinfo{author}{\bibfnamefont{D.~W.~H.} \bibnamefont{Swenson}}
  \bibnamefont{and} \bibinfo{author}{\bibfnamefont{P.~G.}
  \bibnamefont{Bolhuis}}, \bibinfo{journal}{J. Chem. Phys.}
  \textbf{\bibinfo{volume}{141}}, \bibinfo{pages}{044101}
  (\bibinfo{year}{2014}).

\bibitem[{\citenamefont{Becker et~al.}(2012)\citenamefont{Becker, Allen, and
  {ten Wolde}}}]{becker:2012}
\bibinfo{author}{\bibfnamefont{N.~B.} \bibnamefont{Becker}},
  \bibinfo{author}{\bibfnamefont{R.~J.} \bibnamefont{Allen}}, \bibnamefont{and}
  \bibinfo{author}{\bibfnamefont{P.~R.} \bibnamefont{{ten Wolde}}},
  \bibinfo{journal}{J. Chem. Phys.} \textbf{\bibinfo{volume}{136}},
  \bibinfo{pages}{174118} (\bibinfo{year}{2012}).

\bibitem[{\citenamefont{Becker and {ten Wolde}}(2012)}]{becker:2012a}
\bibinfo{author}{\bibfnamefont{N.~B.} \bibnamefont{Becker}} \bibnamefont{and}
  \bibinfo{author}{\bibfnamefont{P.~R.} \bibnamefont{{ten Wolde}}},
  \bibinfo{journal}{J. Chem. Phys.} \textbf{\bibinfo{volume}{136}},
  \bibinfo{pages}{174119} (\bibinfo{year}{2012}).

\bibitem[{\citenamefont{Valeriani et~al.}(2007)\citenamefont{Valeriani, Allen,
  Morelli, Frenkel, and {ten Wolde}}}]{valeriani:2007}
\bibinfo{author}{\bibfnamefont{C.}~\bibnamefont{Valeriani}},
  \bibinfo{author}{\bibfnamefont{R.~J.} \bibnamefont{Allen}},
  \bibinfo{author}{\bibfnamefont{M.~J.} \bibnamefont{Morelli}},
  \bibinfo{author}{\bibfnamefont{D.}~\bibnamefont{Frenkel}}, \bibnamefont{and}
  \bibinfo{author}{\bibfnamefont{P.~R.} \bibnamefont{{ten Wolde}}},
  \bibinfo{journal}{J. Chem. Phys.} \textbf{\bibinfo{volume}{127}},
  \bibinfo{pages}{114109} (\bibinfo{year}{2007}).

\bibitem[{\citenamefont{Borrero and Escobedo}(2008)}]{borrero:2008}
\bibinfo{author}{\bibfnamefont{E.~E.} \bibnamefont{Borrero}} \bibnamefont{and}
  \bibinfo{author}{\bibfnamefont{F.~A.} \bibnamefont{Escobedo}},
  \bibinfo{journal}{J. Chem. Phys.} \textbf{\bibinfo{volume}{129}},
  \bibinfo{pages}{024115} (\bibinfo{year}{2008}).

\bibitem[{\citenamefont{Faradjian and Elber}(2004)}]{faradjian:2004}
\bibinfo{author}{\bibfnamefont{A.~K.} \bibnamefont{Faradjian}}
  \bibnamefont{and} \bibinfo{author}{\bibfnamefont{R.}~\bibnamefont{Elber}},
  \bibinfo{journal}{J. Chem. Phys.} \textbf{\bibinfo{volume}{120}},
  \bibinfo{pages}{10880} (\bibinfo{year}{2004}).

\bibitem[{\citenamefont{West et~al.}(2007)\citenamefont{West, Elber, and
  Shalloway}}]{west:2007}
\bibinfo{author}{\bibfnamefont{A.~M.~A.} \bibnamefont{West}},
  \bibinfo{author}{\bibfnamefont{R.}~\bibnamefont{Elber}}, \bibnamefont{and}
  \bibinfo{author}{\bibfnamefont{D.}~\bibnamefont{Shalloway}},
  \bibinfo{journal}{J. Chem. Phys.} \textbf{\bibinfo{volume}{126}},
  \bibinfo{pages}{145104} (\bibinfo{year}{2007}).

\bibitem[{\citenamefont{Huber and Kim}(1996)}]{huber:1996}
\bibinfo{author}{\bibfnamefont{G.~A.} \bibnamefont{Huber}} \bibnamefont{and}
  \bibinfo{author}{\bibfnamefont{S.}~\bibnamefont{Kim}},
  \bibinfo{journal}{Biophys. J.} \textbf{\bibinfo{volume}{70}},
  \bibinfo{pages}{97} (\bibinfo{year}{1996}).

\bibitem[{\citenamefont{Zhang et~al.}(2007)\citenamefont{Zhang, Jasnow, and
  Zuckerman}}]{zhangB:2007}
\bibinfo{author}{\bibfnamefont{B.~W.} \bibnamefont{Zhang}},
  \bibinfo{author}{\bibfnamefont{D.}~\bibnamefont{Jasnow}}, \bibnamefont{and}
  \bibinfo{author}{\bibfnamefont{D.~M.} \bibnamefont{Zuckerman}},
  \bibinfo{journal}{Proc. Natl. Acad. Sci. USA} \textbf{\bibinfo{volume}{104}},
  \bibinfo{pages}{18043} (\bibinfo{year}{2007}).

\bibitem[{\citenamefont{Zhang et~al.}(2010)\citenamefont{Zhang, Jasnow, and
  Zuckerman}}]{zhangB:2010}
\bibinfo{author}{\bibfnamefont{B.~W.} \bibnamefont{Zhang}},
  \bibinfo{author}{\bibfnamefont{D.}~\bibnamefont{Jasnow}}, \bibnamefont{and}
  \bibinfo{author}{\bibfnamefont{D.~M.} \bibnamefont{Zuckerman}},
  \bibinfo{journal}{J. Chem. Phys.} \textbf{\bibinfo{volume}{132}},
  \bibinfo{pages}{054107} (\bibinfo{year}{2010}).

\bibitem[{\citenamefont{Warmflash et~al.}(2007)\citenamefont{Warmflash,
  Bhimalapuram, and Dinner}}]{warmflash:2007}
\bibinfo{author}{\bibfnamefont{A.}~\bibnamefont{Warmflash}},
  \bibinfo{author}{\bibfnamefont{P.}~\bibnamefont{Bhimalapuram}},
  \bibnamefont{and} \bibinfo{author}{\bibfnamefont{A.~R.}
  \bibnamefont{Dinner}}, \bibinfo{journal}{J. Chem. Phys.}
  \textbf{\bibinfo{volume}{127}}, \bibinfo{pages}{154112}
  (\bibinfo{year}{2007}).

\bibitem[{\citenamefont{Wales}(2002)}]{wales:2002}
\bibinfo{author}{\bibfnamefont{D.~J.} \bibnamefont{Wales}},
  \bibinfo{journal}{Mol. Phys.} \textbf{\bibinfo{volume}{100}},
  \bibinfo{pages}{3285} (\bibinfo{year}{2002}).

\bibitem[{\citenamefont{Wales}(2005)}]{wales:2005}
\bibinfo{author}{\bibfnamefont{D.~J.} \bibnamefont{Wales}},
  \bibinfo{journal}{Phys. Biol.} \textbf{\bibinfo{volume}{2}},
  \bibinfo{pages}{S86} (\bibinfo{year}{2005}).

\bibitem[{\citenamefont{Adams et~al.}(2010)\citenamefont{Adams, Sander, and
  Ziff}}]{adams:2010}
\bibinfo{author}{\bibfnamefont{D.~A.} \bibnamefont{Adams}},
  \bibinfo{author}{\bibfnamefont{L.~M.} \bibnamefont{Sander}},
  \bibnamefont{and} \bibinfo{author}{\bibfnamefont{R.~M.} \bibnamefont{Ziff}},
  \bibinfo{journal}{J. Chem. Phys.} \textbf{\bibinfo{volume}{133}},
  \bibinfo{pages}{124103} (\bibinfo{year}{2010}).

\bibitem[{\citenamefont{Williams}(2013)}]{williams:2013}
\bibinfo{author}{\bibfnamefont{S.~R.} \bibnamefont{Williams}},
  \bibinfo{journal}{Phys. Rev. E} \textbf{\bibinfo{volume}{88}},
  \bibinfo{pages}{043301} (\bibinfo{year}{2013}).

\bibitem[{\citenamefont{Berryman and Schilling}(2010)}]{berryman:2010}
\bibinfo{author}{\bibfnamefont{J.~T.} \bibnamefont{Berryman}} \bibnamefont{and}
  \bibinfo{author}{\bibfnamefont{T.}~\bibnamefont{Schilling}},
  \bibinfo{journal}{J. Chem. Phys.} \textbf{\bibinfo{volume}{133}},
  \bibinfo{pages}{244101} (\bibinfo{year}{2010}).

\bibitem[{\citenamefont{Baidakov et~al.}(2011)\citenamefont{Baidakov, Bobrov,
  and Teterin}}]{baidakov:2011b}
\bibinfo{author}{\bibfnamefont{V.~G.} \bibnamefont{Baidakov}},
  \bibinfo{author}{\bibfnamefont{K.~S.} \bibnamefont{Bobrov}},
  \bibnamefont{and} \bibinfo{author}{\bibfnamefont{A.~S.}
  \bibnamefont{Teterin}}, \bibinfo{journal}{J. Chem. Phys.}
  \textbf{\bibinfo{volume}{135}}, \bibinfo{pages}{054512}
  (\bibinfo{year}{2011}).

\bibitem[{\citenamefont{Jungblut and
  Dellago}(2015{\natexlab{b}})}]{jungblut:2015a}
\bibinfo{author}{\bibfnamefont{S.}~\bibnamefont{Jungblut}} \bibnamefont{and}
  \bibinfo{author}{\bibfnamefont{C.}~\bibnamefont{Dellago}},
  \bibinfo{journal}{Mol. Phys.} p.
  \bibinfo{pages}{doi:10.1080/00268976.2015.1038326}
  (\bibinfo{year}{2015}{\natexlab{b}}).

\bibitem[{\citenamefont{Onsager}(1938)}]{TS_ONSAGER}
\bibinfo{author}{\bibfnamefont{L.}~\bibnamefont{Onsager}},
  \bibinfo{journal}{Phys. Rev.} \textbf{\bibinfo{volume}{54}},
  \bibinfo{pages}{554} (\bibinfo{year}{1938}).

\bibitem[{\citenamefont{Ryter}(1987)}]{RYTER}
\bibinfo{author}{\bibfnamefont{D.}~\bibnamefont{Ryter}},
  \bibinfo{journal}{Physica A} \textbf{\bibinfo{volume}{142}},
  \bibinfo{pages}{103} (\bibinfo{year}{1987}).

\bibitem[{\citenamefont{Berezhkovskii and Szabo}(2006)}]{SZABO2}
\bibinfo{author}{\bibfnamefont{A.}~\bibnamefont{Berezhkovskii}}
  \bibnamefont{and} \bibinfo{author}{\bibfnamefont{A.}~\bibnamefont{Szabo}},
  \bibinfo{journal}{J. Chem. Phys.} \textbf{\bibinfo{volume}{125}},
  \bibinfo{pages}{104902} (\bibinfo{year}{2006}).

\bibitem[{\citenamefont{Klosek et~al.}(1991)\citenamefont{Klosek, Matkowsky,
  and Schuss}}]{TS_SCHUSS}
\bibinfo{author}{\bibfnamefont{M.~M.} \bibnamefont{Klosek}},
  \bibinfo{author}{\bibfnamefont{B.~J.} \bibnamefont{Matkowsky}},
  \bibnamefont{and} \bibinfo{author}{\bibfnamefont{Z.}~\bibnamefont{Schuss}},
  \bibinfo{journal}{Ber. Bunsenges. Phys. Chem.} \textbf{\bibinfo{volume}{95}},
  \bibinfo{pages}{331} (\bibinfo{year}{1991}).

\bibitem[{\citenamefont{Pollak et~al.}(1994)\citenamefont{Pollak,
  Berezhkovskii, and Schuss}}]{POLLAK}
\bibinfo{author}{\bibfnamefont{E.}~\bibnamefont{Pollak}},
  \bibinfo{author}{\bibfnamefont{A.~M.} \bibnamefont{Berezhkovskii}},
  \bibnamefont{and} \bibinfo{author}{\bibfnamefont{Z.}~\bibnamefont{Schuss}},
  \bibinfo{journal}{J. Chem. Phys.} \textbf{\bibinfo{volume}{100}},
  \bibinfo{pages}{334} (\bibinfo{year}{1994}).

\bibitem[{\citenamefont{Talkner}(1994)}]{TALKNER}
\bibinfo{author}{\bibfnamefont{P.}~\bibnamefont{Talkner}},
  \bibinfo{journal}{Chem. Phys.} \textbf{\bibinfo{volume}{180}},
  \bibinfo{pages}{199} (\bibinfo{year}{1994}).

\bibitem[{\citenamefont{Du et~al.}(1998)\citenamefont{Du, Pande, Grosberg,
  Tanaka, and Shakhnovich}}]{TS_PANDE}
\bibinfo{author}{\bibfnamefont{R.}~\bibnamefont{Du}},
  \bibinfo{author}{\bibfnamefont{V.~S.} \bibnamefont{Pande}},
  \bibinfo{author}{\bibfnamefont{A.~Y.} \bibnamefont{Grosberg}},
  \bibinfo{author}{\bibfnamefont{T.}~\bibnamefont{Tanaka}}, \bibnamefont{and}
  \bibinfo{author}{\bibfnamefont{E.~S.} \bibnamefont{Shakhnovich}},
  \bibinfo{journal}{J. Chem. Phys.} \textbf{\bibinfo{volume}{108}},
  \bibinfo{pages}{334} (\bibinfo{year}{1998}).

\bibitem[{\citenamefont{Vanden-Eijnden}(2006)}]{Eric_TPT_Erice}
\bibinfo{author}{\bibfnamefont{E.}~\bibnamefont{Vanden-Eijnden}},
  \bibinfo{journal}{Lect. Notes Phys.} \textbf{\bibinfo{volume}{703}},
  \bibinfo{pages}{439} (\bibinfo{year}{2006}).

\bibitem[{\citenamefont{E and Vanden-Eijnden}(2006)}]{TPT_2006}
\bibinfo{author}{\bibfnamefont{W.}~\bibnamefont{E}} \bibnamefont{and}
  \bibinfo{author}{\bibfnamefont{E.}~\bibnamefont{Vanden-Eijnden}},
  \bibinfo{journal}{J. Stat. Phys.} \textbf{\bibinfo{volume}{123}},
  \bibinfo{pages}{503} (\bibinfo{year}{2006}).

\bibitem[{\citenamefont{Honeycutt and Andersen}(1984)}]{Honeycutt:1984}
\bibinfo{author}{\bibfnamefont{J.~D.} \bibnamefont{Honeycutt}}
  \bibnamefont{and} \bibinfo{author}{\bibfnamefont{H.~C.}
  \bibnamefont{Andersen}}, \bibinfo{journal}{Chem. Phys. Lett.}
  \textbf{\bibinfo{volume}{108}}, \bibinfo{pages}{535} (\bibinfo{year}{1984}).

\bibitem[{\citenamefont{Honeycutt and Andersen}(1986)}]{Honeycutt:1986}
\bibinfo{author}{\bibfnamefont{J.~D.} \bibnamefont{Honeycutt}}
  \bibnamefont{and} \bibinfo{author}{\bibfnamefont{H.~C.}
  \bibnamefont{Andersen}}, \bibinfo{journal}{J. Phys. Chem.}
  \textbf{\bibinfo{volume}{90}}, \bibinfo{pages}{1585} (\bibinfo{year}{1986}).

\bibitem[{\citenamefont{E et~al.}(2005)\citenamefont{E, Ren, and
  Vanden-Eijnden}}]{E:2005}
\bibinfo{author}{\bibfnamefont{W.}~\bibnamefont{E}},
  \bibinfo{author}{\bibfnamefont{W.}~\bibnamefont{Ren}}, \bibnamefont{and}
  \bibinfo{author}{\bibfnamefont{E.}~\bibnamefont{Vanden-Eijnden}},
  \bibinfo{journal}{Chem. Phys. Lett.} \textbf{\bibinfo{volume}{413}},
  \bibinfo{pages}{242} (\bibinfo{year}{2005}).

\bibitem[{\citenamefont{Best and Hummer}(2005)}]{Best:2005}
\bibinfo{author}{\bibfnamefont{R.~B.} \bibnamefont{Best}} \bibnamefont{and}
  \bibinfo{author}{\bibfnamefont{G.}~\bibnamefont{Hummer}},
  \bibinfo{journal}{Proc. Natl. Acad. Sci. USA} \textbf{\bibinfo{volume}{102}},
  \bibinfo{pages}{6732} (\bibinfo{year}{2005}).

\bibitem[{\citenamefont{Hummer}(2004)}]{TS_HUMMER1}
\bibinfo{author}{\bibfnamefont{G.}~\bibnamefont{Hummer}}, \bibinfo{journal}{J.
  Chem. Phys.} \textbf{\bibinfo{volume}{120}}, \bibinfo{pages}{516}
  (\bibinfo{year}{2004}).

\bibitem[{\citenamefont{Geissler et~al.}(1999)\citenamefont{Geissler, Dellago,
  and Chandler}}]{Geissler_ion_1999}
\bibinfo{author}{\bibfnamefont{P.~L.} \bibnamefont{Geissler}},
  \bibinfo{author}{\bibfnamefont{C.}~\bibnamefont{Dellago}}, \bibnamefont{and}
  \bibinfo{author}{\bibfnamefont{D.}~\bibnamefont{Chandler}},
  \bibinfo{journal}{J. Phys. Chem. B} \textbf{\bibinfo{volume}{103}},
  \bibinfo{pages}{3706} (\bibinfo{year}{1999}).

\bibitem[{\citenamefont{Lechner
  et~al.}(2011{\natexlab{b}})\citenamefont{Lechner, Dellago, and
  Bolhuis}}]{Lechner_RPE:2011}
\bibinfo{author}{\bibfnamefont{W.}~\bibnamefont{Lechner}},
  \bibinfo{author}{\bibfnamefont{C.}~\bibnamefont{Dellago}}, \bibnamefont{and}
  \bibinfo{author}{\bibfnamefont{P.~G.} \bibnamefont{Bolhuis}},
  \bibinfo{journal}{J. Chem. Phys.} \textbf{\bibinfo{volume}{135}},
  \bibinfo{pages}{154110} (\bibinfo{year}{2011}{\natexlab{b}}).

\bibitem[{\citenamefont{Ma and Dinner}(2005)}]{ma:2005}
\bibinfo{author}{\bibfnamefont{A.}~\bibnamefont{Ma}} \bibnamefont{and}
  \bibinfo{author}{\bibfnamefont{A.~R.} \bibnamefont{Dinner}},
  \bibinfo{journal}{J. Phys. Chem. B} \textbf{\bibinfo{volume}{109}},
  \bibinfo{pages}{6769} (\bibinfo{year}{2005}).

\bibitem[{\citenamefont{Peters}(2010)}]{peters:2010}
\bibinfo{author}{\bibfnamefont{B.}~\bibnamefont{Peters}},
  \bibinfo{journal}{Mol. Sim.} \textbf{\bibinfo{volume}{36}},
  \bibinfo{pages}{1265} (\bibinfo{year}{2010}).

\bibitem[{\citenamefont{Peters et~al.}(2013)\citenamefont{Peters, Bolhuis,
  Mullen, and Shea}}]{peters:2013}
\bibinfo{author}{\bibfnamefont{B.}~\bibnamefont{Peters}},
  \bibinfo{author}{\bibfnamefont{P.~G.} \bibnamefont{Bolhuis}},
  \bibinfo{author}{\bibfnamefont{R.~G.} \bibnamefont{Mullen}},
  \bibnamefont{and} \bibinfo{author}{\bibfnamefont{J.}~\bibnamefont{Shea}},
  \bibinfo{journal}{J. Chem. Phys.} \textbf{\bibinfo{volume}{138}},
  \bibinfo{pages}{054106} (\bibinfo{year}{2013}).

\bibitem[{\citenamefont{Leitold and Dellago}(2014)}]{Leitold_2014}
\bibinfo{author}{\bibfnamefont{C.}~\bibnamefont{Leitold}} \bibnamefont{and}
  \bibinfo{author}{\bibfnamefont{C.}~\bibnamefont{Dellago}},
  \bibinfo{journal}{J. Chem. Phys.} \textbf{\bibinfo{volume}{141}},
  \bibinfo{pages}{134901} (\bibinfo{year}{2014}).

\bibitem[{\citenamefont{Leitold et~al.}(2015)\citenamefont{Leitold, Lechner,
  and Dellago}}]{JPCM_Leitold_string_2015}
\bibinfo{author}{\bibfnamefont{C.}~\bibnamefont{Leitold}},
  \bibinfo{author}{\bibfnamefont{W.}~\bibnamefont{Lechner}}, \bibnamefont{and}
  \bibinfo{author}{\bibfnamefont{C.}~\bibnamefont{Dellago}},
  \bibinfo{journal}{J. Phys.: Condens. Matter} \textbf{\bibinfo{volume}{27}},
  \bibinfo{pages}{194126} (\bibinfo{year}{2015}).

\bibitem[{\citenamefont{Lechner et~al.}(2010)\citenamefont{Lechner, Rogal,
  Juraszek, Ensing, and Bolhuis}}]{JCP_Lechner_string_2010}
\bibinfo{author}{\bibfnamefont{W.}~\bibnamefont{Lechner}},
  \bibinfo{author}{\bibfnamefont{J.}~\bibnamefont{Rogal}},
  \bibinfo{author}{\bibfnamefont{J.}~\bibnamefont{Juraszek}},
  \bibinfo{author}{\bibfnamefont{B.}~\bibnamefont{Ensing}}, \bibnamefont{and}
  \bibinfo{author}{\bibfnamefont{P.~G.} \bibnamefont{Bolhuis}},
  \bibinfo{journal}{J. Chem. Phys.} \textbf{\bibinfo{volume}{133}},
  \bibinfo{pages}{174110} (\bibinfo{year}{2010}).

\bibitem[{\citenamefont{Beckham et~al.}(2007)\citenamefont{Beckham, Peters,
  Starbuck, Variankaval, and Trout}}]{Beckham:2007}
\bibinfo{author}{\bibfnamefont{G.~T.} \bibnamefont{Beckham}},
  \bibinfo{author}{\bibfnamefont{B.}~\bibnamefont{Peters}},
  \bibinfo{author}{\bibfnamefont{C.}~\bibnamefont{Starbuck}},
  \bibinfo{author}{\bibfnamefont{N.}~\bibnamefont{Variankaval}},
  \bibnamefont{and} \bibinfo{author}{\bibfnamefont{B.~L.} \bibnamefont{Trout}},
  \bibinfo{journal}{J. Am. Chem. Soc.} \textbf{\bibinfo{volume}{129}},
  \bibinfo{pages}{4714} (\bibinfo{year}{2007}).

\bibitem[{\citenamefont{Kirkwood}(1951)}]{kirkwood:1951}
\bibinfo{author}{\bibfnamefont{J.~G.} \bibnamefont{Kirkwood}}, in
  \emph{\bibinfo{booktitle}{Phase Transitions in Solids}}, edited by
  \bibinfo{editor}{\bibfnamefont{R.}~\bibnamefont{Smoluchowski}},
  \bibinfo{editor}{\bibfnamefont{J.~E.} \bibnamefont{Mayer}}, \bibnamefont{and}
  \bibinfo{editor}{\bibfnamefont{W.~A.} \bibnamefont{Weyl}}
  (\bibinfo{publisher}{Wiley, New York}, \bibinfo{year}{1951}),
  p.~\bibinfo{pages}{67}.

\bibitem[{\citenamefont{van Hove}(1957)}]{vanhove:1957}
\bibinfo{author}{\bibfnamefont{L.}~\bibnamefont{van Hove}},
  \bibinfo{journal}{Rev. Mod. Phys.} \textbf{\bibinfo{volume}{29}},
  \bibinfo{pages}{200} (\bibinfo{year}{1957}).

\bibitem[{\citenamefont{Uhlenbeck}(1963)}]{uhlenbeck:1963}
\bibinfo{author}{\bibfnamefont{G.~E.} \bibnamefont{Uhlenbeck}}, in
  \emph{\bibinfo{booktitle}{The many body problem}}, edited by
  \bibinfo{editor}{\bibfnamefont{J.~K.} \bibnamefont{Percus}}
  (\bibinfo{publisher}{Interscience Publishers / John Wiley, London},
  \bibinfo{year}{1963}), p. \bibinfo{pages}{498}.

\bibitem[{\citenamefont{Alder and Wainwright}(1957)}]{alder:1957}
\bibinfo{author}{\bibfnamefont{B.~J.} \bibnamefont{Alder}} \bibnamefont{and}
  \bibinfo{author}{\bibfnamefont{T.~E.} \bibnamefont{Wainwright}},
  \bibinfo{journal}{J. Chem. Phys.} \textbf{\bibinfo{volume}{27}},
  \bibinfo{pages}{1208} (\bibinfo{year}{1957}).

\bibitem[{\citenamefont{Wood and Jacobson}(1957)}]{wood:1957}
\bibinfo{author}{\bibfnamefont{W.~W.} \bibnamefont{Wood}} \bibnamefont{and}
  \bibinfo{author}{\bibfnamefont{J.~D.} \bibnamefont{Jacobson}},
  \bibinfo{journal}{J. Chem. Phys.} \textbf{\bibinfo{volume}{27}},
  \bibinfo{pages}{1207} (\bibinfo{year}{1957}).

\bibitem[{\citenamefont{Hoover and Ree}(1968)}]{hoover:1968}
\bibinfo{author}{\bibfnamefont{W.~G.} \bibnamefont{Hoover}} \bibnamefont{and}
  \bibinfo{author}{\bibfnamefont{F.~H.} \bibnamefont{Ree}},
  \bibinfo{journal}{J. Chem. Phys.} \textbf{\bibinfo{volume}{49}},
  \bibinfo{pages}{3609} (\bibinfo{year}{1968}).

\bibitem[{\citenamefont{Frenkel}(1999)}]{frenkel:1999}
\bibinfo{author}{\bibfnamefont{D.}~\bibnamefont{Frenkel}},
  \bibinfo{journal}{Physica A} \textbf{\bibinfo{volume}{263}},
  \bibinfo{pages}{26} (\bibinfo{year}{1999}).

\bibitem[{\citenamefont{Filion and Dijkstra}(2009)}]{filion:2009}
\bibinfo{author}{\bibfnamefont{L.}~\bibnamefont{Filion}} \bibnamefont{and}
  \bibinfo{author}{\bibfnamefont{M.}~\bibnamefont{Dijkstra}},
  \bibinfo{journal}{Phys. Rev. E} \textbf{\bibinfo{volume}{79}},
  \bibinfo{pages}{046714} (\bibinfo{year}{2009}).

\bibitem[{\citenamefont{Rudd et~al.}(1968)\citenamefont{Rudd, Salsburg, Yu, and
  Stillinger}}]{Stillinger:1968}
\bibinfo{author}{\bibfnamefont{W.~G.} \bibnamefont{Rudd}},
  \bibinfo{author}{\bibfnamefont{Z.~W.} \bibnamefont{Salsburg}},
  \bibinfo{author}{\bibfnamefont{A.~P.} \bibnamefont{Yu}}, \bibnamefont{and}
  \bibinfo{author}{\bibfnamefont{F.~H.} \bibnamefont{Stillinger}},
  \bibinfo{journal}{J. Chem. Phys.} \textbf{\bibinfo{volume}{49}},
  \bibinfo{pages}{4857} (\bibinfo{year}{1968}).

\bibitem[{\citenamefont{Bolhuis et~al.}(1997)\citenamefont{Bolhuis, Frenkel,
  Mau, and Huse}}]{Frenkel:1997}
\bibinfo{author}{\bibfnamefont{P.~G.} \bibnamefont{Bolhuis}},
  \bibinfo{author}{\bibfnamefont{D.}~\bibnamefont{Frenkel}},
  \bibinfo{author}{\bibfnamefont{S.-C.} \bibnamefont{Mau}}, \bibnamefont{and}
  \bibinfo{author}{\bibfnamefont{D.~A.} \bibnamefont{Huse}},
  \bibinfo{journal}{Nature} \textbf{\bibinfo{volume}{49}}, \bibinfo{pages}{388}
  (\bibinfo{year}{1997}).

\bibitem[{\citenamefont{Bruce et~al.}(1997{\natexlab{b}})\citenamefont{Bruce,
  Wilding, and Ackland}}]{Wilding:1997}
\bibinfo{author}{\bibfnamefont{A.~D.} \bibnamefont{Bruce}},
  \bibinfo{author}{\bibfnamefont{N.~B.} \bibnamefont{Wilding}},
  \bibnamefont{and} \bibinfo{author}{\bibfnamefont{G.~J.}
  \bibnamefont{Ackland}}, \bibinfo{journal}{Phys. Rev. Lett.}
  \textbf{\bibinfo{volume}{79}}, \bibinfo{pages}{3002}
  (\bibinfo{year}{1997}{\natexlab{b}}).

\bibitem[{\citenamefont{Mau and Huse}(1999)}]{Huse:1999}
\bibinfo{author}{\bibfnamefont{S.-C.} \bibnamefont{Mau}} \bibnamefont{and}
  \bibinfo{author}{\bibfnamefont{D.~A.} \bibnamefont{Huse}},
  \bibinfo{journal}{Phys. Rev. E} \textbf{\bibinfo{volume}{59}},
  \bibinfo{pages}{4396} (\bibinfo{year}{1999}).

\bibitem[{\citenamefont{Truskett et~al.}(1998)\citenamefont{Truskett, Torquato,
  Sastry, Debenedetti, and Stillinger}}]{Truskett:1998}
\bibinfo{author}{\bibfnamefont{T.~M.} \bibnamefont{Truskett}},
  \bibinfo{author}{\bibfnamefont{S.}~\bibnamefont{Torquato}},
  \bibinfo{author}{\bibfnamefont{S.}~\bibnamefont{Sastry}},
  \bibinfo{author}{\bibfnamefont{P.~G.} \bibnamefont{Debenedetti}},
  \bibnamefont{and} \bibinfo{author}{\bibfnamefont{F.~H.}
  \bibnamefont{Stillinger}}, \bibinfo{journal}{Phys. Rev. E}
  \textbf{\bibinfo{volume}{58}}, \bibinfo{pages}{3083} (\bibinfo{year}{1998}).

\bibitem[{\citenamefont{O'Malley and Snook}(2005)}]{omalley:2005}
\bibinfo{author}{\bibfnamefont{B.}~\bibnamefont{O'Malley}} \bibnamefont{and}
  \bibinfo{author}{\bibfnamefont{I.}~\bibnamefont{Snook}}, \bibinfo{journal}{J.
  Chem. Phys.} \textbf{\bibinfo{volume}{123}}, \bibinfo{pages}{054511}
  (\bibinfo{year}{2005}).

\bibitem[{\citenamefont{Brader}(2008)}]{brader:2008}
\bibinfo{author}{\bibfnamefont{J.~M.} \bibnamefont{Brader}},
  \bibinfo{journal}{J. Chem. Phys.} \textbf{\bibinfo{volume}{128}},
  \bibinfo{pages}{104503} (\bibinfo{year}{2008}).

\bibitem[{\citenamefont{Auer and Frenkel}(2001{\natexlab{b}})}]{auer:2001a}
\bibinfo{author}{\bibfnamefont{S.}~\bibnamefont{Auer}} \bibnamefont{and}
  \bibinfo{author}{\bibfnamefont{D.}~\bibnamefont{Frenkel}},
  \bibinfo{journal}{Nature} \textbf{\bibinfo{volume}{413}},
  \bibinfo{pages}{711} (\bibinfo{year}{2001}{\natexlab{b}}).

\bibitem[{\citenamefont{Auer and Frenkel}(2005)}]{auer:2005}
\bibinfo{author}{\bibfnamefont{S.}~\bibnamefont{Auer}} \bibnamefont{and}
  \bibinfo{author}{\bibfnamefont{D.}~\bibnamefont{Frenkel}}, in
  \emph{\bibinfo{booktitle}{Advanced Computer Simulation Approaches for Soft
  Matter Sciences I}}, edited by
  \bibinfo{editor}{\bibfnamefont{C.}~\bibnamefont{Holm}} \bibnamefont{and}
  \bibinfo{editor}{\bibfnamefont{K.}~\bibnamefont{Kremer}}
  (\bibinfo{publisher}{Springer-Verlag}, \bibinfo{address}{Berlin Heidelberg},
  \bibinfo{year}{2005}), vol. \bibinfo{volume}{173} of
  \emph{\bibinfo{series}{Advances in Polymer Science}}, p.
  \bibinfo{pages}{149}.

\bibitem[{\citenamefont{Radu and Schilling}(2014)}]{radu2014solvent}
\bibinfo{author}{\bibfnamefont{M.}~\bibnamefont{Radu}} \bibnamefont{and}
  \bibinfo{author}{\bibfnamefont{T.}~\bibnamefont{Schilling}},
  \bibinfo{journal}{EPL (Europhys. Lett.)} \textbf{\bibinfo{volume}{105}},
  \bibinfo{pages}{26001} (\bibinfo{year}{2014}).

\bibitem[{\citenamefont{Kawasaki and Tanaka}(2010)}]{KawasakiTanaka:2010}
\bibinfo{author}{\bibfnamefont{T.}~\bibnamefont{Kawasaki}} \bibnamefont{and}
  \bibinfo{author}{\bibfnamefont{H.}~\bibnamefont{Tanaka}},
  \bibinfo{journal}{Proc. Natl. Acad. Sci. USA} \textbf{\bibinfo{volume}{107}},
  \bibinfo{pages}{14036} (\bibinfo{year}{2010}).

\bibitem[{\citenamefont{Palberg}(2014)}]{palberg:2014}
\bibinfo{author}{\bibfnamefont{T.}~\bibnamefont{Palberg}}, \bibinfo{journal}{J.
  Phys.: Condens. Matter} \textbf{\bibinfo{volume}{26}},
  \bibinfo{pages}{333101} (\bibinfo{year}{2014}).

\bibitem[{\citenamefont{Woodcock}(1976)}]{woodcock:1976}
\bibinfo{author}{\bibfnamefont{L.~V.} \bibnamefont{Woodcock}},
  \bibinfo{journal}{J. Chem. Soc., Faraday Trans. 2}
  \textbf{\bibinfo{volume}{72}}, \bibinfo{pages}{1667} (\bibinfo{year}{1976}).

\bibitem[{\citenamefont{Gordon et~al.}(1976)\citenamefont{Gordon, Gibbs, and
  Fleming}}]{gordon:1976}
\bibinfo{author}{\bibfnamefont{J.~M.} \bibnamefont{Gordon}},
  \bibinfo{author}{\bibfnamefont{J.~H.} \bibnamefont{Gibbs}}, \bibnamefont{and}
  \bibinfo{author}{\bibfnamefont{P.~D.} \bibnamefont{Fleming}},
  \bibinfo{journal}{J. Chem. Phys.} \textbf{\bibinfo{volume}{65}},
  \bibinfo{pages}{2771} (\bibinfo{year}{1976}).

\bibitem[{\citenamefont{Woodcock}(1978)}]{woodcock:1978}
\bibinfo{author}{\bibfnamefont{L.~V.} \bibnamefont{Woodcock}},
  \bibinfo{journal}{J. Chem. Soc., Faraday Trans. 2}
  \textbf{\bibinfo{volume}{74}}, \bibinfo{pages}{11} (\bibinfo{year}{1978}).

\bibitem[{\citenamefont{Pusey and {van Megen}}(1986)}]{pusey:1986}
\bibinfo{author}{\bibfnamefont{P.~N.} \bibnamefont{Pusey}} \bibnamefont{and}
  \bibinfo{author}{\bibfnamefont{W.}~\bibnamefont{{van Megen}}},
  \bibinfo{journal}{Nature} \textbf{\bibinfo{volume}{320}},
  \bibinfo{pages}{340} (\bibinfo{year}{1986}).

\bibitem[{\citenamefont{{van Megen} and Underwood}(1993)}]{vanmegen:1993}
\bibinfo{author}{\bibfnamefont{W.}~\bibnamefont{{van Megen}}} \bibnamefont{and}
  \bibinfo{author}{\bibfnamefont{S.~M.} \bibnamefont{Underwood}},
  \bibinfo{journal}{Phys. Rev. Lett.} \textbf{\bibinfo{volume}{70}},
  \bibinfo{pages}{2766} (\bibinfo{year}{1993}).

\bibitem[{\citenamefont{{van Megen} and Underwood}(1994)}]{vanmegen:1994}
\bibinfo{author}{\bibfnamefont{W.}~\bibnamefont{{van Megen}}} \bibnamefont{and}
  \bibinfo{author}{\bibfnamefont{S.~M.} \bibnamefont{Underwood}},
  \bibinfo{journal}{Phys. Rev. E} \textbf{\bibinfo{volume}{49}},
  \bibinfo{pages}{4206} (\bibinfo{year}{1994}).

\bibitem[{\citenamefont{Zhu et~al.}(1997)\citenamefont{Zhu, Li, Rogers, Meyer,
  Ottewill, {STS-73 Space Shuttle Crew}, Russel, and Chaikin}}]{zhu:1997}
\bibinfo{author}{\bibfnamefont{J.}~\bibnamefont{Zhu}},
  \bibinfo{author}{\bibfnamefont{M.}~\bibnamefont{Li}},
  \bibinfo{author}{\bibfnamefont{R.}~\bibnamefont{Rogers}},
  \bibinfo{author}{\bibfnamefont{W.}~\bibnamefont{Meyer}},
  \bibinfo{author}{\bibfnamefont{R.~H.} \bibnamefont{Ottewill}},
  \bibinfo{author}{\bibnamefont{{STS-73 Space Shuttle Crew}}},
  \bibinfo{author}{\bibfnamefont{W.~B.} \bibnamefont{Russel}},
  \bibnamefont{and} \bibinfo{author}{\bibfnamefont{P.~M.}
  \bibnamefont{Chaikin}}, \bibinfo{journal}{Nature}
  \textbf{\bibinfo{volume}{387}}, \bibinfo{pages}{883} (\bibinfo{year}{1997}).

\bibitem[{\citenamefont{Rintoul and Torquato}(1996)}]{rintoul:1996}
\bibinfo{author}{\bibfnamefont{M.~D.} \bibnamefont{Rintoul}} \bibnamefont{and}
  \bibinfo{author}{\bibfnamefont{S.}~\bibnamefont{Torquato}},
  \bibinfo{journal}{J. Chem. Phys.} \textbf{\bibinfo{volume}{105}},
  \bibinfo{pages}{9258} (\bibinfo{year}{1996}).

\bibitem[{\citenamefont{Williams et~al.}(2001)\citenamefont{Williams, Snook,
  and {van Megen}}}]{williams:2001}
\bibinfo{author}{\bibfnamefont{S.~R.} \bibnamefont{Williams}},
  \bibinfo{author}{\bibfnamefont{I.~K.} \bibnamefont{Snook}}, \bibnamefont{and}
  \bibinfo{author}{\bibfnamefont{W.}~\bibnamefont{{van Megen}}},
  \bibinfo{journal}{Phys. Rev. E} \textbf{\bibinfo{volume}{64}},
  \bibinfo{pages}{021506} (\bibinfo{year}{2001}).

\bibitem[{\citenamefont{Sch{\"o}pe et~al.}(2006)\citenamefont{Sch{\"o}pe,
  Bryant, and van Megen}}]{schoepe:2006}
\bibinfo{author}{\bibfnamefont{H.~J.} \bibnamefont{Sch{\"o}pe}},
  \bibinfo{author}{\bibfnamefont{G.}~\bibnamefont{Bryant}}, \bibnamefont{and}
  \bibinfo{author}{\bibfnamefont{W.}~\bibnamefont{van Megen}},
  \bibinfo{journal}{Phys. Rev. E} \textbf{\bibinfo{volume}{74}},
  \bibinfo{pages}{060401(R)} (\bibinfo{year}{2006}).

\bibitem[{\citenamefont{Zaccarelli et~al.}(2009)\citenamefont{Zaccarelli,
  Valeriani, Sanz, Poon, Cates, and Pusey}}]{zaccarelli:2009}
\bibinfo{author}{\bibfnamefont{E.}~\bibnamefont{Zaccarelli}},
  \bibinfo{author}{\bibfnamefont{C.}~\bibnamefont{Valeriani}},
  \bibinfo{author}{\bibfnamefont{E.}~\bibnamefont{Sanz}},
  \bibinfo{author}{\bibfnamefont{W.~C.~K.} \bibnamefont{Poon}},
  \bibinfo{author}{\bibfnamefont{M.~E.} \bibnamefont{Cates}}, \bibnamefont{and}
  \bibinfo{author}{\bibfnamefont{P.~N.} \bibnamefont{Pusey}},
  \bibinfo{journal}{Phys. Rev. Lett.} \textbf{\bibinfo{volume}{103}},
  \bibinfo{pages}{135704} (\bibinfo{year}{2009}).

\bibitem[{\citenamefont{Pusey et~al.}(2009)\citenamefont{Pusey, Zaccarelli,
  Valeriani, Sanz, Poon, and Cates}}]{pusey:2009}
\bibinfo{author}{\bibfnamefont{P.~N.} \bibnamefont{Pusey}},
  \bibinfo{author}{\bibfnamefont{E.}~\bibnamefont{Zaccarelli}},
  \bibinfo{author}{\bibfnamefont{C.}~\bibnamefont{Valeriani}},
  \bibinfo{author}{\bibfnamefont{E.}~\bibnamefont{Sanz}},
  \bibinfo{author}{\bibfnamefont{W.~C.~K.} \bibnamefont{Poon}},
  \bibnamefont{and} \bibinfo{author}{\bibfnamefont{M.~E.} \bibnamefont{Cates}},
  \bibinfo{journal}{Phil. Trans. R. Soc. A} \textbf{\bibinfo{volume}{367}},
  \bibinfo{pages}{4993} (\bibinfo{year}{2009}).

\bibitem[{\citenamefont{Xu et~al.}(2010)\citenamefont{Xu, Sun, and
  An}}]{xu:2010a}
\bibinfo{author}{\bibfnamefont{W.-S.} \bibnamefont{Xu}},
  \bibinfo{author}{\bibfnamefont{Z.-Y.} \bibnamefont{Sun}}, \bibnamefont{and}
  \bibinfo{author}{\bibfnamefont{L.-J.} \bibnamefont{An}},
  \bibinfo{journal}{Eur. Phys. J. E} \textbf{\bibinfo{volume}{31}},
  \bibinfo{pages}{377} (\bibinfo{year}{2010}).

\bibitem[{\citenamefont{Valeriani et~al.}(2011)\citenamefont{Valeriani, Sanz,
  Zaccarelli, Poon, Cates, and Pusey}}]{valeriani:2011}
\bibinfo{author}{\bibfnamefont{C.}~\bibnamefont{Valeriani}},
  \bibinfo{author}{\bibfnamefont{E.}~\bibnamefont{Sanz}},
  \bibinfo{author}{\bibfnamefont{E.}~\bibnamefont{Zaccarelli}},
  \bibinfo{author}{\bibfnamefont{W.~C.~K.} \bibnamefont{Poon}},
  \bibinfo{author}{\bibfnamefont{M.~E.} \bibnamefont{Cates}}, \bibnamefont{and}
  \bibinfo{author}{\bibfnamefont{P.~N.} \bibnamefont{Pusey}},
  \bibinfo{journal}{J. Phys.: Condens. Matter} \textbf{\bibinfo{volume}{23}},
  \bibinfo{pages}{194117} (\bibinfo{year}{2011}).

\bibitem[{\citenamefont{Sanz et~al.}(2011)\citenamefont{Sanz, Valeriani,
  Zaccarelli, Poon, Pusey, and Cates}}]{sanz:2011}
\bibinfo{author}{\bibfnamefont{E.}~\bibnamefont{Sanz}},
  \bibinfo{author}{\bibfnamefont{C.}~\bibnamefont{Valeriani}},
  \bibinfo{author}{\bibfnamefont{E.}~\bibnamefont{Zaccarelli}},
  \bibinfo{author}{\bibfnamefont{W.~C.~K.} \bibnamefont{Poon}},
  \bibinfo{author}{\bibfnamefont{P.~N.} \bibnamefont{Pusey}}, \bibnamefont{and}
  \bibinfo{author}{\bibfnamefont{M.~E.} \bibnamefont{Cates}},
  \bibinfo{journal}{Phys. Rev. Lett.} \textbf{\bibinfo{volume}{106}},
  \bibinfo{pages}{215701} (\bibinfo{year}{2011}).

\bibitem[{\citenamefont{Franke et~al.}(2014)\citenamefont{Franke, Golde, and
  Sch{\"o}pe}}]{franke:2014}
\bibinfo{author}{\bibfnamefont{M.}~\bibnamefont{Franke}},
  \bibinfo{author}{\bibfnamefont{S.}~\bibnamefont{Golde}}, \bibnamefont{and}
  \bibinfo{author}{\bibfnamefont{H.~J.} \bibnamefont{Sch{\"o}pe}},
  \bibinfo{journal}{Soft Matter} \textbf{\bibinfo{volume}{10}},
  \bibinfo{pages}{5380} (\bibinfo{year}{2014}).

\bibitem[{\citenamefont{Sanz et~al.}(2014)\citenamefont{Sanz, Valeriani,
  Zaccarelli, Poon, Cates, and Pusey}}]{sanz:2014}
\bibinfo{author}{\bibfnamefont{E.}~\bibnamefont{Sanz}},
  \bibinfo{author}{\bibfnamefont{C.}~\bibnamefont{Valeriani}},
  \bibinfo{author}{\bibfnamefont{E.}~\bibnamefont{Zaccarelli}},
  \bibinfo{author}{\bibfnamefont{W.~C.~K.} \bibnamefont{Poon}},
  \bibinfo{author}{\bibfnamefont{M.~E.} \bibnamefont{Cates}}, \bibnamefont{and}
  \bibinfo{author}{\bibfnamefont{P.~N.} \bibnamefont{Pusey}},
  \bibinfo{journal}{Proc. Natl. Acad. Sci. USA} \textbf{\bibinfo{volume}{111}},
  \bibinfo{pages}{75} (\bibinfo{year}{2014}).

\bibitem[{\citenamefont{Pruppacher and Klett}(1978)}]{Pruppacher_Klett}
\bibinfo{author}{\bibfnamefont{H.~R.} \bibnamefont{Pruppacher}}
  \bibnamefont{and} \bibinfo{author}{\bibfnamefont{J.~D.} \bibnamefont{Klett}},
  \emph{\bibinfo{title}{Microphysics of clouds and precipitation}}
  (\bibinfo{publisher}{Reidel Publishing}, \bibinfo{address}{Dordrecht},
  \bibinfo{year}{1978}).

\bibitem[{\citenamefont{Hegg and Baker}(2009)}]{Hegg_Baker}
\bibinfo{author}{\bibfnamefont{D.~A.} \bibnamefont{Hegg}} \bibnamefont{and}
  \bibinfo{author}{\bibfnamefont{M.~B.} \bibnamefont{Baker}},
  \bibinfo{journal}{Rep. Prog. Phys.} \textbf{\bibinfo{volume}{72}},
  \bibinfo{pages}{056801} (\bibinfo{year}{2009}).

\bibitem[{\citenamefont{Gurganus et~al.}(2011)\citenamefont{Gurganus,
  Kostinski, and Shaw}}]{Shaw:2011}
\bibinfo{author}{\bibfnamefont{C.}~\bibnamefont{Gurganus}},
  \bibinfo{author}{\bibfnamefont{A.~B.} \bibnamefont{Kostinski}},
  \bibnamefont{and} \bibinfo{author}{\bibfnamefont{R.~A.} \bibnamefont{Shaw}},
  \bibinfo{journal}{J. Phys. Chem. Lett.} \textbf{\bibinfo{volume}{2}},
  \bibinfo{pages}{1449} (\bibinfo{year}{2011}).

\bibitem[{\citenamefont{Vega et~al.}(2005)\citenamefont{Vega, Abascal, Sanz,
  MacDowell, and McBride}}]{JPCM_Vega_ice_2005}
\bibinfo{author}{\bibfnamefont{C.}~\bibnamefont{Vega}},
  \bibinfo{author}{\bibfnamefont{J.~L.~F.} \bibnamefont{Abascal}},
  \bibinfo{author}{\bibfnamefont{E.}~\bibnamefont{Sanz}},
  \bibinfo{author}{\bibfnamefont{L.~G.} \bibnamefont{MacDowell}},
  \bibnamefont{and} \bibinfo{author}{\bibfnamefont{C.}~\bibnamefont{McBride}},
  \bibinfo{journal}{J. Phys.: Condens. Matter} \textbf{\bibinfo{volume}{17}},
  \bibinfo{pages}{S3283} (\bibinfo{year}{2005}).

\bibitem[{\citenamefont{Matsumoto et~al.}(2002)\citenamefont{Matsumoto, Saito,
  and Ohmine}}]{Ohmine:2002}
\bibinfo{author}{\bibfnamefont{M.}~\bibnamefont{Matsumoto}},
  \bibinfo{author}{\bibfnamefont{S.}~\bibnamefont{Saito}}, \bibnamefont{and}
  \bibinfo{author}{\bibfnamefont{I.}~\bibnamefont{Ohmine}},
  \bibinfo{journal}{Nature} \textbf{\bibinfo{volume}{416}},
  \bibinfo{pages}{409} (\bibinfo{year}{2002}).

\bibitem[{\citenamefont{Moore and Molinero}(2011)}]{Nature_Molinero_2011}
\bibinfo{author}{\bibfnamefont{E.~B.} \bibnamefont{Moore}} \bibnamefont{and}
  \bibinfo{author}{\bibfnamefont{V.}~\bibnamefont{Molinero}},
  \bibinfo{journal}{Nature} \textbf{\bibinfo{volume}{479}},
  \bibinfo{pages}{506} (\bibinfo{year}{2011}).

\bibitem[{\citenamefont{Molinero and Moore}(2009)}]{Molinero:2009}
\bibinfo{author}{\bibfnamefont{V.}~\bibnamefont{Molinero}} \bibnamefont{and}
  \bibinfo{author}{\bibfnamefont{E.~B.} \bibnamefont{Moore}},
  \bibinfo{journal}{J. Phys. Chem. B} \textbf{\bibinfo{volume}{113}},
  \bibinfo{pages}{4008} (\bibinfo{year}{2009}).

\bibitem[{\citenamefont{Li et~al.}(2011)\citenamefont{Li, Donadio, Russo, and
  Galli}}]{PCCP_Galli_2011}
\bibinfo{author}{\bibfnamefont{T.}~\bibnamefont{Li}},
  \bibinfo{author}{\bibfnamefont{D.}~\bibnamefont{Donadio}},
  \bibinfo{author}{\bibfnamefont{G.}~\bibnamefont{Russo}}, \bibnamefont{and}
  \bibinfo{author}{\bibfnamefont{G.}~\bibnamefont{Galli}},
  \bibinfo{journal}{Phys. Chem. Chem. Phys.} \textbf{\bibinfo{volume}{13}},
  \bibinfo{pages}{19807} (\bibinfo{year}{2011}).

\bibitem[{\citenamefont{Reinhardt and Doye}(2012)}]{JCP_Doye_2013}
\bibinfo{author}{\bibfnamefont{A.}~\bibnamefont{Reinhardt}} \bibnamefont{and}
  \bibinfo{author}{\bibfnamefont{J.~P.~K.} \bibnamefont{Doye}},
  \bibinfo{journal}{J. Chem. Phys.} \textbf{\bibinfo{volume}{136}},
  \bibinfo{pages}{054501} (\bibinfo{year}{2012}).

\bibitem[{\citenamefont{Li et~al.}(2013)\citenamefont{Li, Donadio, and
  Galli}}]{NatComm_Donadio_2013}
\bibinfo{author}{\bibfnamefont{T.}~\bibnamefont{Li}},
  \bibinfo{author}{\bibfnamefont{D.}~\bibnamefont{Donadio}}, \bibnamefont{and}
  \bibinfo{author}{\bibfnamefont{G.}~\bibnamefont{Galli}},
  \bibinfo{journal}{Nature Comm.} \textbf{\bibinfo{volume}{4}},
  \bibinfo{pages}{1887} (\bibinfo{year}{2013}).

\bibitem[{\citenamefont{Radhakrishnan and Trout}(2003)}]{Ravi_Trout_2003}
\bibinfo{author}{\bibfnamefont{R.}~\bibnamefont{Radhakrishnan}}
  \bibnamefont{and} \bibinfo{author}{\bibfnamefont{B.~L.} \bibnamefont{Trout}},
  \bibinfo{journal}{J. Am. Chem. Soc.} \textbf{\bibinfo{volume}{125}},
  \bibinfo{pages}{7743} (\bibinfo{year}{2003}).

\bibitem[{\citenamefont{Brukhno et~al.}(2008)\citenamefont{Brukhno, Anwar,
  Davidchack, and Handel}}]{JPCM_Handel_2008}
\bibinfo{author}{\bibfnamefont{A.~V.} \bibnamefont{Brukhno}},
  \bibinfo{author}{\bibfnamefont{J.}~\bibnamefont{Anwar}},
  \bibinfo{author}{\bibfnamefont{R.}~\bibnamefont{Davidchack}},
  \bibnamefont{and} \bibinfo{author}{\bibfnamefont{R.}~\bibnamefont{Handel}},
  \bibinfo{journal}{J. Phys.: Condens. Matter} \textbf{\bibinfo{volume}{20}},
  \bibinfo{pages}{494243} (\bibinfo{year}{2008}).

\bibitem[{\citenamefont{Bai and Li}(2005)}]{JCP_LI_2005}
\bibinfo{author}{\bibfnamefont{X.-M.} \bibnamefont{Bai}} \bibnamefont{and}
  \bibinfo{author}{\bibfnamefont{M.}~\bibnamefont{Li}}, \bibinfo{journal}{J.
  Chem. Phys.} \textbf{\bibinfo{volume}{122}}, \bibinfo{pages}{224510}
  (\bibinfo{year}{2005}).

\bibitem[{\citenamefont{Sanz et~al.}(2013)\citenamefont{Sanz, Vega, Espinosa,
  Caballero-Bernal, Abascal, and Valeriani}}]{JACS_Valeriani_2013}
\bibinfo{author}{\bibfnamefont{E.}~\bibnamefont{Sanz}},
  \bibinfo{author}{\bibfnamefont{C.}~\bibnamefont{Vega}},
  \bibinfo{author}{\bibfnamefont{J.~R.} \bibnamefont{Espinosa}},
  \bibinfo{author}{\bibfnamefont{R.}~\bibnamefont{Caballero-Bernal}},
  \bibinfo{author}{\bibfnamefont{J.~L.~F.} \bibnamefont{Abascal}},
  \bibnamefont{and}
  \bibinfo{author}{\bibfnamefont{C.}~\bibnamefont{Valeriani}},
  \bibinfo{journal}{J. Am. Chem. Soc.} \textbf{\bibinfo{volume}{135}},
  \bibinfo{pages}{15008} (\bibinfo{year}{2013}).

\bibitem[{\citenamefont{Quigley and Rodger}(2008)}]{Quigley_Rodger_2008}
\bibinfo{author}{\bibfnamefont{D.}~\bibnamefont{Quigley}} \bibnamefont{and}
  \bibinfo{author}{\bibfnamefont{P.~M.} \bibnamefont{Rodger}},
  \bibinfo{journal}{J. Chem. Phys.} \textbf{\bibinfo{volume}{128}},
  \bibinfo{pages}{154518} (\bibinfo{year}{2008}).

\bibitem[{\citenamefont{Haji-Akbari and
  Debenedetti}(2015)}]{PNAS_Debenedetti_2015}
\bibinfo{author}{\bibfnamefont{A.}~\bibnamefont{Haji-Akbari}} \bibnamefont{and}
  \bibinfo{author}{\bibfnamefont{P.~G.} \bibnamefont{Debenedetti}},
  \bibinfo{journal}{Proc. Natl. Acad. Sci. USA} \textbf{\bibinfo{volume}{112}},
  \bibinfo{pages}{10582} (\bibinfo{year}{2015}).

\bibitem[{\citenamefont{Moore and Molinero}(2010)}]{JCP_Molinero_2010}
\bibinfo{author}{\bibfnamefont{E.~B.} \bibnamefont{Moore}} \bibnamefont{and}
  \bibinfo{author}{\bibfnamefont{V.}~\bibnamefont{Molinero}},
  \bibinfo{journal}{J. Chem. Phys.} \textbf{\bibinfo{volume}{132}},
  \bibinfo{pages}{244504} (\bibinfo{year}{2010}).

\bibitem[{\citenamefont{Thapar and Escobedo}(2014)}]{thapar:2014}
\bibinfo{author}{\bibfnamefont{V.}~\bibnamefont{Thapar}} \bibnamefont{and}
  \bibinfo{author}{\bibfnamefont{F.~A.} \bibnamefont{Escobedo}},
  \bibinfo{journal}{Phys. Rev. Lett.} \textbf{\bibinfo{volume}{112}},
  \bibinfo{pages}{048301} (\bibinfo{year}{2014}).

\bibitem[{\citenamefont{Shah et~al.}(2011)\citenamefont{Shah, Santiso, and
  Trout}}]{shah_trout:2011}
\bibinfo{author}{\bibfnamefont{M.}~\bibnamefont{Shah}},
  \bibinfo{author}{\bibfnamefont{E.~E.} \bibnamefont{Santiso}},
  \bibnamefont{and} \bibinfo{author}{\bibfnamefont{B.~L.} \bibnamefont{Trout}},
  \bibinfo{journal}{J. Phys. Chem. B} \textbf{\bibinfo{volume}{115}},
  \bibinfo{pages}{10400} (\bibinfo{year}{2011}).

\bibitem[{\citenamefont{Raiteri and Gale}(2010)}]{raitieri:2010}
\bibinfo{author}{\bibfnamefont{P.}~\bibnamefont{Raiteri}} \bibnamefont{and}
  \bibinfo{author}{\bibfnamefont{J.~D.} \bibnamefont{Gale}},
  \bibinfo{journal}{J. Am. Chem. Soc.} \textbf{\bibinfo{volume}{132}},
  \bibinfo{pages}{17623} (\bibinfo{year}{2010}).

\end{thebibliography}
\end{document}